\documentclass[iop,revtex4]{emulateapj}
\slugcomment{Accepted for publication in ApJ}

\shorttitle{A Census of $z=7.0$ Ly$\alpha$ Emitters to 0.3 $L^*$}
\shortauthors{Ota et al.}

\begin{document}


%\title{A New Constraint on Reionization by a Census of $z=7.0$ Ly$\alpha$ Emitters with a Deep and Large Sample to 0.3 $L^*$: Evolution of Ly$\alpha$ Luminosity Function and Equivalent Width at $z\sim6$--7\altaffilmark{*}}
%\title{A New Constraint on Reionization from Evolution of the Ly$\alpha$ Luminosity Function at $z\sim6$--7 Probed by a Deep Census of $z=7.0$ Ly$\alpha$ Emitter Candidates to 0.3 $L^*$\footnote{Based on data collected at Subaru Telescope, which is operated by the National Astronomical Observatory of Japan.}}
%\title{A New Constraint on Reionization from Evolution of the Ly$\alpha$ Luminosity Function at $z\sim6$--7 Probed by a Deep Census of $z=7.0$ Ly$\alpha$ Emitter Candidates to 0.3 $L^*$\altaffilmark{*}}

\title{A New Constraint on Reionization from Evolution of the Ly$\alpha$ Luminosity Function at $z\sim6$--7 Probed by a Deep Census of $z=7.0$ Ly$\alpha$ Emitter Candidates to 0.3 $L^*$\altaffilmark{\dag}}

%\thanks{{\it Herschel} is an ESA space observatory with science instruments provided by European-led Principal Investigator consortia and with important participation from NASA.}


\author{Kazuaki Ota\altaffilmark{1,2,14}, Masanori Iye\altaffilmark{3}, Nobunari Kashikawa\altaffilmark{3,4}, Akira Konno\altaffilmark{5,6}, Fumiaki Nakata\altaffilmark{7}, Tomonori Totani\altaffilmark{5,8}, Masakazu A.R. Kobayashi\altaffilmark{9}, Yoshinobu Fudamoto\altaffilmark{10}, Akifumi Seko\altaffilmark{11}, Jun Toshikawa\altaffilmark{3}, Akie Ichikawa\altaffilmark{12,13}, Takatoshi Shibuya\altaffilmark{6}, Masafusa Onoue\altaffilmark{3,4}}

%Kazuhiro Shimasaku\altaffilmark{5,6},
%Masami Ouchi\altaffilmark{7,8}
%Tadayuki Kodama\altaffilmark{3,4}, 
%Ken-ichi Tadaki\altaffilmark{11}, 
%Tomoki Morokuma\altaffilmark{12},
%Linhua Jiang\altaffilmark{13}, 
%Takuya Fujiyoshi\altaffilmark{10},
%Akifumi Seko\altaffilmark{14},
%Hiroko Tamazawa\altaffilmark{7,15},
%Tomoe Takeuchi\altaffilmark{14},

%\author{Kazuaki Ota\altaffilmark{1,2,18}, Masanori Iye\altaffilmark{3}, Nobunari Kashikawa\altaffilmark{3,4}, Kazuhiro Shimasaku\altaffilmark{5,6}, Masami Ouchi\altaffilmark{7,8}, Tomonori Totani\altaffilmark{5,6}, Masakazu A.R. Kobayashi\altaffilmark{9}, Fumiaki Nakata\altaffilmark{10}, Tadayuki Kodama\altaffilmark{3,4}, Ken-ichi Tadaki\altaffilmark{11}, Tomoki Morokuma\altaffilmark{12}, Jun Toshikawa\altaffilmark{3}, Linhua Jiang\altaffilmark{13}, Takuya Fujiyoshi\altaffilmark{10}, Akira Konno\altaffilmark{5,7}, Yoshinobu Fudamoto\altaffilmark{11}, Akifumi Seko\altaffilmark{14}, Hiroko Tamazawa\altaffilmark{7,15}, Akie Ichikawa\altaffilmark{16,17}, Tomoe Takeuchi\altaffilmark{14}, Masafusa Onoue\altaffilmark{3,4}, Takatoshi Shibuya\altaffilmark{7}}

\affil{$^1$Kavli Institute for Cosmology, University of Cambridge, Madingley Road, Cambridge, CB3 0HA, UK; kota@ast.cam.ac.uk}
\affil{$^2$Cavendish Laboratory, University of Cambridge, 19 J.J. Thomson Avenue, Cambridge, CB3 0HE, UK}
\affil{$^3$National Astronomical Observatory of Japan, 2-21-1 Osawa, Mitaka, Tokyo, 181-8588, Japan}
\affil{$^4$The Graduate University for Advanced Studies, 2-21-1 Osawa, Mitaka, Tokyo, 181-8588, Japan}
\affil{$^5$Department of Astronomy, Graduate School of Science, The University of Tokyo, 7-3-1 Hongo, Bunkyo, Tokyo 113-0033, Japan}
\affil{$^6$Institute for Cosmic Ray Research, The University of Tokyo, 5-1-5 Kashiwanoha, Kashiwa, Chiba 277-8582, Japan}
\affil{$^7$Subaru Telescope, 650 North A'ohoku Place, Hilo, HI 96720, USA}
\affil{$^8$Research Center for the Early Universe, Graduate School of Science, The University of Tokyo, 7-3-1 Hongo, Bunkyo, Tokyo 113-0033, Japan}
\affil{$^9$Faculty of Natural Sciences, National Institute of Technology, Kure College, 2-2-11 Agaminami, Kure, Hiroshima 737-8506, Japan}
\affil{$^{10}$Observatoire de Gen\'eve, Chemin des Maillettes 51, CH-1290 Versoix, Switzerland}
\affil{$^{11}$Department of Astronomy, Kyoto University, Kitashirakawa-Oiwake-cho, Sakyo-ku, Kyoto 606-8502, Japan}
\affil{$^{12}$Research Center for Space and Cosmic Evolution, Ehime University, 2-5 Bunkyo-cho, Matsuyama, Ehime 790-8577, Japan}
\affil{$^{13}$Physics Department, Graduate School of Science \& Engineering, Ehime University, 2-5 Bunkyo-cho, Matsuyama, Ehime 790-8577, Japan}

\altaffiltext{}{---------------------------------------------------------}
\altaffiltext{\dag}{Based on data collected at Subaru Telescope, which is operated by the National Astronomical Observatory of Japan.}
%\altaffiltext{*}{Based on data collected at Subaru Telescope, which is operated by the National Astronomical Observatory of Japan.}
%\altaffiltext{*}{Based in part on data collected at Subaru Telescope, which is operated by the National Astronomical Observatory of Japan, observations made with the NASA/ESA {\it Hubble Space Telescope}, obtained from the Data Archive at the Space Telescope Science Institute, which is operated by the Association of Universities for Research in Astronomy, Inc.~under NASA contract NAS 5-26555 and observations made with the {\it Spitzer Space Telescope}, which is operated by the Jet Propulsion Laboratory, California Institute of Technology under a contract with NASA. {\it Herschel} is an ESA space observatory with science instruments provided by European-led Principal Investigator consortia and with important participation from NASA.}
\altaffiltext{14}{Kavli Institute Fellow}

%\altaffiltext{5}{Department of Astronomy, Graduate School of Science, University of Tokyo, 7-3-1 Hongo, Bunkyo-ku, Tokyo 113-0033, Japan}


\begin{abstract}
We detect 20 $z=7.0$ Ly$\alpha$ emitter (LAE) candidates to $L({\rm Ly}\alpha) \geq 2 \times 10^{42}$ erg s$^{-1}$ or $0.3$ $L^*_{z=7}$ and in $6.1\times 10^5$ Mpc$^3$ volume in the Subaru Deep Field and the Subaru/{\it XMM-Newton} Deep Survey field by 82 and 37 hours of Subaru Suprime-Cam narrowband NB973 and reddest optical $y$-band imaging. We compare their Ly$\alpha$ and UV luminosity functions (LFs) and densities and Ly$\alpha$ equivalent widths (EWs) to those of $z=5.7$, 6.6 and 7.3 LAEs from previous Suprime-Cam surveys. The Ly$\alpha$ LF (density) rapidly declines by a factor of $\times$1.5 (1.9) in $L({\rm Ly}\alpha)$ at $z=5.7$--6.6 (160 Myr), $\times$1.5 (1.6) at $z=6.6$--7.0 (60 Myr) at the faint end and $\times$2.0 (3.8) at $z=7.0$--7.3 (40 Myr). Also, in addition to the systematic decrease in EW at $z=5.7$--6.6 previously found, 2/3 of the $z=7.0$ LAEs detected in the UV continuum exhibit lower EWs than the $z=6.6$ ones. Moreover, while the UV LF and density do not evolve at $z=5.7$--6.6, they modestly decline at $z=6.6$--7.0, implying galaxy evolution contributing to the decline of the Ly$\alpha$ LF. Comparison of the $z=7.0$ Ly$\alpha$ LF to the one predicted by an LAE evolution model further reveals that galaxy evolution alone cannot explain all the decline of Ly$\alpha$ LF. If we attribute the discrepancy to Ly$\alpha$ attenuation by neutral hydrogen, the intergalactic medium transmission of Ly$\alpha$ photons at $z=7.0$ would be $T_{{\rm Ly}\alpha}^{\rm IGM} \leq 0.6$--0.7. It is lower (higher) than $T_{{\rm Ly}\alpha}^{\rm IGM}$ at $z=6.6$ (7.3) derived by previous studies, suggesting rapid increase in neutral fraction at $z > 6$.

\end{abstract}


%\keywords{cosmology: observations---early universe---galaxies: evolution---galaxies: formation}



\section{Introduction \label{Intro}}
Observations of distant objects in the first 1 Gyr after the Big Bang have been revealing cosmic reionization history. {\it WMAP} and {\it Planck} observations of cosmic microwave background (CMB) suggest that reionization occurred at $z=9.1$--11.7 and $z=7.4$--10.5, respectively \citep{Bennett13,Hinshaw13,Planck16a,Planck16b}. Meanwhile, Gunn-Peterson (GP) troughs found in $z \sim 6$ quasar (QSO) spectra imply that neutral hydrogen fraction of the universe at $z\sim6$ is $x_{\rm HI}^{z\sim6}\sim 0.01$--0.04 and reionization might have ended at $z \sim 6$ \citep{Fan06}. Also, analyses of $z\sim5.9$, 6.3 and 6.7 $\gamma$-ray burst (GRB) damping wing absorptions indicate that $x_{\rm HI}$ could increase with redshift with the constraints of $x_{\rm HI}^{z\sim5.9}\sim 0.06$, $x_{\rm HI}^{z\sim6.3}\leq 0.17$ and $x_{\rm HI}^{z\sim6.7} > 0.35$ and reionization might not be complete yet at $z>6$ \citep[][see also \citet{McQuinn08} for their estimate of $x_{\rm HI}^{z\sim6.3}\sim 0.5$ from the same GRB]{Totani06,Totani14,Totani16,Greiner09}.

Another probe of reionization is Ly$\alpha$ emitters (LAEs). Their Ly$\alpha$ luminosity function (LF) could decline as neutral hydrogen absorbs or scatters Ly$\alpha$ photons from LAEs \citep{Malhotra04}. According to observations of LAEs to date, Ly$\alpha$ LF does not evolve at $3 \lesssim z \lesssim 5.7$ \citep{Dawson07,Ouchi08} but significantly declines from $z=5.7$ to 6.6 \citep{Hu10,Ouchi10,Kashikawa11,Matthee15}. The Ly$\alpha$ LF evolves if LAEs evolve between $z=5.7$ and 6.6. However, \citet{Kashikawa11} also confirmed that the rest frame ultraviolet (UV) continuum LF of LAEs does not evolve between $z=5.7$ and 6.6, and thus the LAEs do not significantly evolve over these epochs. This is because the UV LF of LAEs can change only if LAEs evolve in number or luminosity (e.g., due to attenuation of their UV continua by dust), as the UV continuum is not absorbed by neutral hydrogen. \citet{Ono10} performed a spectral energy distribution (SED) fitting study of $z=5.7$ and $z=6.6$ LAEs in the samples of \citet{Ouchi08,Ouchi10} using their multiwavelength data and found that these LAE populations on average have negligible dust extinction. Hence, no evolution of UV LF of LAEs at $z=5.7$--6.6 suggests that the decline of the Ly$\alpha$ LF from $z=5.7$ to $z=6.6$ is not due to galaxy evolution but could be attributed to the attenuation of Ly$\alpha$ emission by neutral hydrogen. Comparing the amount of the decline with predictions from various reionization models \citep[e.g.,][]{Santos04,Furlanetto06,Dijkstra07,McQuinn07}, \citet{Ouchi10} and \citet{Kashikawa11} estimated the neutral fraction at $z=6.6$ to be $x_{\rm HI}^{z=6.6} < 0.4$. This supports the incomplete reionization at $z>6$ suggested by the QSO and GRB observation results. Moreover, \citet{Kashikawa11} found that the rest frame Ly$\alpha$ equivalent width (EW$_0$) distributions of $z\sim3$--5.7 LAEs are very similar but EW$_0$'s of $z=6.6$ LAEs are significantly lower. This also implies that Ly$\alpha$ emission might be attenuated by neutral hydrogen at $z=6.6$.  
%If this is the case, $W_{{\rm Ly}\alpha}^{\rm rest}$'s of $z=7$ LAEs could be even smaller than those of $z=6.6$ LAEs due to the larger $x_{\rm HI}$ at higher redshift. However, nobody has investigated this to date due to the difficulty in detecting statistically large number of $z=7$ LAEs. 

On the other hand, several theoretical models suggest that cosmic reionization state can be also probed by the spatial distribution of LAEs \citep[e.g.,][]{Furlanetto06,McQuinn07}. Attenuation of Ly$\alpha$ emission could modulate the sky distribution of LAEs and cause significant apparent clustering. This, if any, can be detected by investigating two point angular correlation function (ACF) of LAEs. \citet{Kashikawa06} investigated the ACF of 58 $z=6.6$ LAE photometric candidates distributed over 876 arcmin$^2$ sky area of the Subaru Deep Field \citep[SDF,][]{Kashikawa04} but did not detect any clustering signal. Meanwhile, \citet{Ouchi10} examined the ACF of 207 $z=6.6$ LAE candidates distributed over $\sim1$ deg$^2$ sky area of the Subaru/{\it XMM-Newton} Deep Survey field \citep[SXDS,][]{Furusawa08} and detected its clustering signal. Comparing the ACF with the predictions by reionization models \citep{Furlanetto06,McQuinn07}, they constrained the neutral fraction to be $x_{\rm HI}^{z=6.6} \lesssim 0.5$.    

Furthermore, possible increase in neutral fraction at $z>6$ has been also suggested by observations of Lyman break galaxies (LBGs). Some LBGs are known to exhibit strong Ly$\alpha$ emission (EW$_0 > 25$\AA) while some others do not. Fraction of Ly$\alpha$ emitting LBGs increases from $z\sim3$ to 6 but suddenly decreases from $z\sim6$ to 7--8 \citep{Stark10,Stark11,Pentericci11,Pentericci14,Ono12,Schenker12,Schenker14,Tilvi14,Caruana12,Caruana14,Treu12,Treu13,Furusawa16}. These studies pointed out that the reversal of the evolution trend at $z>6$ could be due to attenuation of Ly$\alpha$ emission by neutral hydrogen and rapid evolution of neutral fraction from $z\sim6$ to 7. Also, recent theoretical studies and simulations have investigated visibility of the Ly$\alpha$ emission line during reionization and predicted redshift evolution of Ly$\alpha$ fraction among galaxies by different reionization models \citep{Dijkstra11,Jensen13,Mesinger15,Choudhury15}. Comparisons of their predictions with the observed Ly$\alpha$ fractions among LBGs suggest that rapid evolution of neutral hydrogen is required to fully account for the large drop in Ly$\alpha$ fraction at $z>6$.

%%figure 1
\begin{figure}
\epsscale{1.17}
%\plotone{zyNB973_FDCCD_FilterResponse_Sky.eps}
\plotone{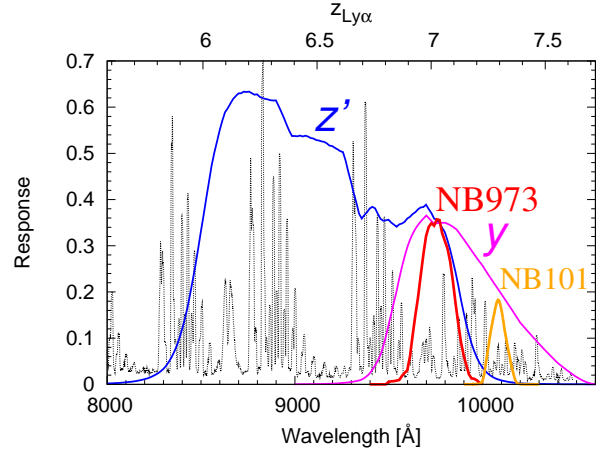}
\caption{Response curves (solid curves) of the Suprime-Cam $z'$, $y$ and NB973 bands used for our $z=7$ LAE study as well as the NB101 band used by \citet{Konno14} for their $z=7.3$ LAE survey. The response curves include the fully-depleted red-sensitive Hamamatsu CCD quantum efficiency, the reflection ratio of the telescope primary mirror, correction for the prime focus optics and transmission to the atmosphere (airmass $\sec z = 1.2$). The OH night sky lines are also overplotted with the dotted curve. The top axis indicates the redshift of Ly$\alpha$ emission corresponding to the wavelength at the bottom axis.\label{FilterTransmission}}
\end{figure}
%\vspace*{0.5cm}
%\vspace*{1cm}

Meanwhile, to investigate the possibility of incomplete reionization at $z=7.0$ (hereafter we simply say $z=7$), we also conducted Subaru Telescope Suprime-Cam \citep{Miyazaki02} and the narrowband NB973 \citep[$\lambda_c=9755$\AA, $\Delta\lambda_{\rm FWHM}=200$\AA, $z_{{\rm Ly}\alpha}=6.94$--7.10,][see Figure \ref{FilterTransmission}]{Iye06} imaging of the SDF and the SXDS \citep{Ota08,Ota10}. We reached the Ly$\alpha$ luminosity limits ($5\sigma$) of $L({\rm Ly}\alpha)_{\rm lim} \sim 1\times10^{43}$ erg s$^{-1}$ (SDF) and $\lesssim 9.2\times10^{42}$ erg s$^{-1}$ (the upper limit on the SXDS survey limit)\footnote[1]{Note that \citet{Ota10} very conservatively presented $L({\rm Ly}\alpha)_{\rm lim} \lesssim 9.2 \times10^{42}$ erg s$^{-1}$ as the upper limit on their $z=7$ LAE survey limit in SXDS by assuming that all the narrowband flux comes from the Ly$\alpha$ flux of an LAE; $F_{{\rm Ly}\alpha}=F_{\rm NB973}$. This is because there was no broadband image of SXDS covering the UV continuum redwards of the $z=7$ Ly$\alpha$ was available at that time. However, the narrowband flux also likely includes the UV continuum flux of an LAE; $F_{{\rm Ly}\alpha}+F_{\rm UV}=F_{\rm NB973}$. Considering the contribution from the UV continuum flux, the actual survey limit of \citet{Ota10} is deeper: $L({\rm Ly}\alpha)_{\rm lim} \simeq 4.1 \times10^{42}$ erg s$^{-1}$ if we use the equation (\ref{FLya_FNB_lambda_EW}) and adopt the Ly$\alpha$ EW threshold of EW$_0 =10$\AA~as a definition of an LAE. See Section \ref{LyaUVLumSFRLimit} for the details of estimation of $L({\rm Ly}\alpha)_{\rm lim}$ from a narrowband NB973 limiting magnitude.} but detected only one and three $z=7$ LAEs in SDF and SXDS, respectively \citep{Iye06,Ota08,Ota10}. We should have detected 8 $z=7$ LAEs in SDF to $L({\rm Ly}\alpha)_{\rm lim} = 1 \times10^{43}$ erg s$^{-1}$ and 52 $z=7$ LAEs in SXDS to $L({\rm Ly}\alpha)_{\rm lim} = 4.1 \times10^{42}$ erg s$^{-1}$ (see footnote 1) if there had been no evolution of the Ly$\alpha$ LF from $z=6.6$ to $z=7$ in SDF and SXDS\footnote[2]{We integrate the best-fit Schechter functions of the $z=6.6$ Ly$\alpha$ LFs in SDF and SXDS derived by \citet{Kashikawa11} and \citet{Ouchi10}, respectively, to our survey limits and in our survey volumes to estimate the expected detection numbers of LAEs.}. Although our survey limits were shallow and probed only the bright end of the $z=7$ Ly$\alpha$ LF, we found that the Ly$\alpha$ LF declines from $z=6.6$ to $z=7$ more rapidly than it does from $z=5.7$ to $z=6.6$, and neutral fraction could be higher at $z=7$ than $z=6.6$; $x_{\rm HI}^{z=7}\sim0.32$--0.64 and $x_{\rm HI}^{z=7}<0.63$ in SDF and SXDS, respectively. 
%We would have detected XX $z=7$ LAEs to $L_{{\rm Ly}\alpha} \sim XX\times10^{42}$ erg s$^{-1}$ if there had been no evolution of Ly$\alpha$ LF from $z=6.6$ to $z=7$, but we actually detected only three LAEs at $z=7$. 
%\footnote[2]{The decreases in the number and Ly$\alpha$ luminosity densities are $n_{z=6.6}/n_{z=5.7} \sim XX$, $n_{z=7}/n_{z=6.6} \sim XX$ and $\rho_{z=6.6}^{{\rm Ly}\alpha}/\rho_{z=5.7}^{{\rm Ly}\alpha} \sim XX$, $\rho_{z=7}^{{\rm Ly}\alpha}/\rho_{z=6.6}^{{\rm Ly}\alpha} \sim XX$ to $L_{{\rm Ly}\alpha} \sim XX\times10^{42}$ erg s$^{-1}$.}

This trend of more rapid or "accelerated" decline of Ly$\alpha$ LF (and thus possible accelerated increase in neutral fraction) at higher redshift was also confirmed by \citet{Konno14} at slightly higher redshift and to deeper survey limits. They used Subaru Suprime-Cam and the narrowband filter NB101 ($\lambda_c=10095$\AA, $\Delta\lambda_{\rm FWHM}=90$\AA, $z_{{\rm Ly}\alpha}=7.302\pm0.037$, see Figure \ref{FilterTransmission}) to detect three and four $z=7.3$ LAE candidates reaching $L({\rm Ly}\alpha)_{\rm lim} \sim 4.1\times10^{42}$ erg s$^{-1}$ and $2.4\times10^{42}$ erg s$^{-1}$ in SXDS and the Cosmic Evolution Survey \citep[COSMOS,][]{Scoville07} fields. They confirmed that the Ly$\alpha$ LF of LAEs declines from $z=6.6$ to 7.3 more rapidly than it does from $z=5.7$ to 6.6. 

\citet{Konno14} also compared the redshift evolution of the Ly$\alpha$ luminosity density of LAEs ($\rho_{{\rm Ly}\alpha}^{\rm LAE}$) over $z=5.7$--7.3 with that of the UV continuum luminosity density of LBGs ($\rho_{\rm UV}^{\rm LBG}$) over $z\sim6$--10 and found that $\rho_{{\rm Ly}\alpha}^{\rm LAE}$ modestly decreases from $z=5.7$ to $z=6.6$ and sharply decreases from $z=6.6$ to $z=7.3$ while $\rho_{\rm UV}^{\rm LBG}$ modestly decreases from $z\sim6$ to $z\sim8$ and sharply decreases from $z\sim8$ to $z\sim10$. Based on this result, they claimed that the sharp (accelerating) decrease in $\rho_{{\rm Ly}\alpha}^{\rm LAE}$ at $z>6.6$ is not due to galaxy (LAE) evolution but could be attributed to attenuation of Ly$\alpha$ emission of LAEs by neutral hydrogen. 

It should be noted that this conclusion assumes that evolution of LAEs and LBGs are the same at $z\sim6$--10. Despite some overlaps, they are two different galaxy populations selected by different photometric methods. LAEs are selected by detecting their Ly$\alpha$ emission using a combination of narrowband and broadband filters and thus tend to be bright in Ly$\alpha$ emission but faint in UV continuum while LBGs are selected by detecting their UV continua using broadband filters and thus inclined to be bright in UV continuum but not necessarily bright in Ly$\alpha$. Also, while high redshift (e.g., $z\sim5.7$--6.6) LAEs have almost negligible dust extinction \citep[e.g.,][]{Ono10}, some LBGs suffer significant dust attenuation in their UV continua \citep[e.g.,][]{Schaerer10,deBarros14}. Hence, the evolution of LAEs and LBGs may not necessarily be the same at $z\sim6$--10, and thus the evolution of $\rho_{\rm UV}^{\rm LBG}$ at $z\sim6$--10 might not reflect that of the UV continuum luminosity density of LAEs ($\rho_{\rm UV}^{\rm LAE}$) at those epochs. To examine whether the LAEs evolve or not at $z>6.6$, we have to derive the UV LF of LAEs to as a deep flux limit as possible.
%it should be noted that evolution of LAEs and LBGs, which are two different galaxy populations with some overlaps, may not necessarily be the same at $z>6.6$ and (see below). 

%Table 1 (deluxetable)
%\clearpage
\begin{turnpage}
\begin{deluxetable*}{cccccccl}
\tabletypesize{\scriptsize}
%\rotate
\tablecaption{Summary of Our Imaging Observations and Data\label{ImagingData}}
%\tablewidth{0pt}
%\tablewidth{510pt}
\tablewidth{0pt}
\tablehead{
\colhead{Field} & \colhead{Band} & \colhead{Exposure Time} &  \colhead{PSF Size$^{\rm a}$} & \colhead{Area$^{\rm b}$}        & \colhead{$m_0$$^{\rm c}$} & \colhead{$m_{\rm lim}$$^{\rm d}$} & \colhead{Observation Date and Note}\\
\colhead{(R.A., Decl. [J2000.0])}      & \colhead{}     & \colhead{(second)}      & \colhead{(arcsec)}          & \colhead{(arcmin$^2$)} & \colhead{(mag/count)}   & \colhead{(mag)}               & \colhead{} 
}
\startdata
SDF      & $B$   &  35700 & 0.98        &     & 34.780 & 28.45         & Public Data \citep{Kashikawa04}\\
13:24:38.9, $+$27:29:25.9         & $V$   &  20400 & 0.98        &     & 34.584 & 27.74         & Public Data\\
         & $R$   &  97200 & 1.15        &     & 34.109 & 28.38$^{\rm e}$ & Data from \citet{Toshikawa12}\\
         & $i'$  &  99720 & 0.93        &     & 34.148 & 27.80$^{\rm e}$ & Data from \citet{Toshikawa12}\\
         & $z'$  & 111240 & 0.97 (1.05) &     & 32.999 & 27.31$^{\rm e}$ & Data from \citet{Toshikawa12}\\
         & NB816 &  36000 & 0.98        &     & 32.880 & 26.63         & Public Data\\
         & NB921 &  53940 & 0.98        &     & 32.520 & 26.54         & Public Data\\
         & $y$   &  94515 & 0.79 (1.05) &     & 31.28  & 26.21         & Data from \citet{Ouchi09}$^{\rm f}$\\
         & NB973 & 211949 & 1.05        & 824 & 32.10  & 26.50         & 2004 Aug 12--15$^{\rm g}$, 2005 Mar 16--17$^{\rm g}$, 2007 May 9--10$^{\rm g}$\\
         &       &        &             &     &        &               & 2013 Jun 7--8, 2013 Nov 27, 2015 Feb 15--18\\
\hline
SXDS-C   & $B$   &  20700 & 0.80        &     & 34.723 & 28.09         & Public Data \citep{Furusawa08}\\
02:18:00.0, $-$05:00:00.0         & $V$   &  19140 & 0.80        &     & 33.639 & 27.78         & Public Data\\
         & $R$   &  14880 & 0.80        &     & 34.315 & 27.57         & Public Data\\
         & $i'$  &  38820 & 0.80        &     & 34.055 & 27.62         & Public Data\\
         & $z'$  &  13020 & 0.80 (1.05) &     & 33.076 & 26.70         & Public Data\\
         & NB816 &  17182 & 0.81        &     & 32.64  & 26.55         & Data from \citet{Ouchi08}\\
         & NB921 &  30000 & 0.81        &     & 32.04  & 26.2          & Data from \citet{Ouchi10}\\ 
SXDS-S   & $B$   &  19800 & 0.82        &     & 34.706 & 28.33         & Public Data\\
02:18:00.0, $-$05:25:00.0         & $V$   &  19260 & 0.82        &     & 33.643 & 27.75         & Public Data\\
         & $R$   &  13920 & 0.82        &     & 34.219 & 27.67         & Public Data\\
         & $i'$  &  18540 & 0.82        &     & 34.046 & 27.47         & Public Data\\
         & $z'$  &  11040 & 0.82 (1.05) &     & 32.258 & 26.39         & Public Data\\
         & NB816 &  14400 & 0.81        &     & 32.56  & 26.65         & Data from \citet{Ouchi08}\\
         & NB921 &  37138 & 0.83        &     & 32.04  & 26.2          & Data from \citet{Ouchi10}\\ 
SXDS-C+S & $y$   &  39660 & 1.05        &     & 31.69  & 25.85         & 2010 Oct 6--8$^{\rm h}$, 2015 Feb 15--18\\
02:18:00.0, $-$05:13:30.0         & NB973 &  83329 & 1.05        & 851 & 32.21  & 26.15         & 2008 Oct 25--26, 2008 Nov 30, 2013 Nov 27, 2014 Dec 20--21
%SXDS-C+S & $y$   &  39660 & 1.05        &     & 31.9(Tadaki)? or 31.57?(Ota)  & XX.X          & 2010 Oct 6--8$^{\rm g}$, 2015 Feb 15--18\\
\enddata
%% Text for table notes should follow after the \enddata but before
%% the \end{deluxetable}. Make sure there is at least one \tablenotemark
%% in the table for each \tablenotetext.
%\tablecomments{All the upper limits are $3\sigma$ at a resolution of $1\farcs5 \times 1\farcs2$ (See \textsection \ref{FIRcont_Prop} for details.)}
\tablecomments{}
\tablenotetext{a}{FWHM of PSFs. Those in parentheses are the FWHMs of PSFs convolved to be matched to that of the NB973 image in each field for the purpose of color measurements by aperture photometry (see Section \ref{Photometry} for details).}
\tablenotetext{b}{Effective areas of the NB973 images after removing the masked regions and the low S/N edges.}
\tablenotetext{c}{Magnitude zeropoint.}
\tablenotetext{d}{$3\sigma$ limiting magnitude measured in a $2''$ diameter aperture.}
\tablenotetext{e}{These are different from the magnitudes evaluated by \citet{Toshikawa12} as we measured the limiting magnitudes by ourselves. Our measurements in $R$ and $i'$ is similar to those of \citet{Toshikawa12} while our measurement in $z'$ is 0.2 mag deeper than that of \citet{Toshikawa12}.}
\tablenotetext{f}{This image was created by stacking $y$ band exposures taken with the MIT-LL CCDs and the Hamamatsu CCDs in Suprime-Cam.}
\tablenotetext{g}{These NB973 exposures were taken by using the MIT-LL CCDs while all the other NB973 exposures by the Hamamatsu CCDs.}
\tablenotetext{h}{This observation was conducted by \citet{Tadaki12}.}
\end{deluxetable*}
\end{turnpage}
%\clearpage

Although our previous $z=7$ LAE surveys and the \citet{Konno14} $z=7.3$ LAE survey might have caught a possible sign of rapidly increasing neutral fraction at $z>6.6$, there is still room for further and deliberate investigations before we come to this conclusion. In our previous $z=7$ LAE surveys, we confirmed that the Ly$\alpha$ LF of LAEs declines from $z=6.6$ to $z=7$ but the UV LF of LAEs does not \citep{Ota08,Ota10}. However, we probed only the bright ends of the LFs and thus do not know what the Ly$\alpha$ and UV LFs of $z=7$ LAEs entirely look like to their faint ends. 

Meanwhile, in the $z=7.3$ LAE survey, \citet{Konno14} confirmed that the Ly$\alpha$ LF of LAEs declines from $z=6.6$ to $z=7.3$ to fainter Ly$\alpha$ luminosity limits. However, they did not derive the Ly$\alpha$ EWs nor the UV LF of $z=7.3$ LAEs because any broadband image covering the UV continua of $z=7.3$ LAEs redwards of their Ly$\alpha$ was not available. They also selected photometric $z=7.3$ LAE candidates using the narrowband NB101 and $z'$ band. As seen in Figure \ref{FilterTransmission}, the $z'$ band is sensitive to the wavelengths completely blueward of $z=7.3$ Ly$\alpha$ ($\lambda \lesssim 10000$\AA) while NB101 is sensitive to the wavelengths 10050--10140\AA. Their LAE sample selection as well as estimates of Ly$\alpha$ luminosities of $z=7.3$ LAE candidates and their survey limits (in Ly$\alpha$ flux or luminosity) are based on NB101 and $z'$ band magnitudes. This selection method is more like a dropout technique and could selectively detect LBGs and LAEs with a bright UV continuum detectable even in the narrowband NB101. This may result in apparently extremely low number of detected LAEs. To estimate Ly$\alpha$ luminosities and survey limits more accurately and not to miss detecting any LAE candidates with faint UV continua with unbiased sampling of LAEs, deep broadband imaging covering wavelengths redwards of Ly$\alpha$ emission is necessary. Also, such a broadband image is indispensable to deriving the Ly$\alpha$ EWs and the UV LF of LAEs. 

To address all these issues, in this study, we have obtained very deep imaging of SDF and SXDS in NB973 as well as the reddest optical $y$ band\footnote[3]{This is the custom-made broadband filter for the Subaru Suprime-Cam originally called $z_R$ band and used by \cite{Shimasaku05} to detect $z\sim6$ LBGs. Hereafter we call it $y$ band for simplicity. Also see \citet{Ouchi09} for the details of this filter.} ($\lambda_c=9860$\AA, $\Delta\lambda_{\rm FWHM}=590$\AA, see Figure \ref{FilterTransmission}) sufficiently covering the UV continuum wavelengths redwards of $z=7$ Ly$\alpha$ to derive the $z=7$ Ly$\alpha$ and UV LFs as well as the $z=7$ Ly$\alpha$ EW distribution to the survey limit comparable to those of the previous Subaru $z=5.7$, $z=6.6$ and $z=7.3$ LAE surveys. Moreover, in the case of SDF, the $R$, $i'$ and $z'$ band images 0.4--0.7 mag deeper than the ones we used for our previous SDF $z=7$ LAE survey is now available \citep{Poznanski07,Graur11,Toshikawa12}. This also helps improve our selection of $z=7$ LAEs in more effectively removing foreground interlopers and detecting LAEs down to as faint Ly$\alpha$ emission as possible, which might be severely attenuated by neutral fraction. As we will show in the subsequent sections, we have detected 20 LAEs with the EW$_0 > 10$\AA~threshold and to $\sim 0.3 L_{z=7}^*$, where the $L_{z=7}^*$ is the characteristic Ly$\alpha$ luminosity of $z=7$ LAEs, and in the comoving volume of $6.1 \times 10^5$ Mpc$^3$ in SDF plus SXDS. This is the fairly large and deepest $z=7$ LAE sample ever obtained at this moment. We will use this sample to investigate reionization state at $z=7$.

On the other hand, shortly after our present study, \citet{Zheng17} very recently reported the first results from their ongoing large area sensitive survey of $z=6.9$ LAEs, LAGER (Lyman Alpha Galaxies in the Epoch of Reionization) project. Using the narrowband filter NB964 ($\lambda_c \sim 9642$\AA, $\Delta\lambda_{\rm FWHM} \sim 90$\AA, $z_{{\rm Ly}\alpha} \sim 6.89$--6.97) and the Dark-Energy Camera (DECam, 3 deg$^2$ field-of-view) on the NOAO/CTIO 4m Blanco telescope, they imaged the COSMOS field and detected 27 $z=6.9$ LAE candidates in the central 2 deg$^2$ region to $L({\rm Ly}\alpha) \sim 4 \times 10^{42}$ erg s$^{-1}$. They found that the Ly$\alpha$ LF and luminosity density significantly declines from $z=5.7$ and 6.6 to $z=6.9$ at $42.6 < \log L({\rm Ly}\alpha) < 43.2$ while there is an excess of number of the bright LAE candidates at $43.2 < \log L({\rm Ly}\alpha) < 43.6$ where the Ly$\alpha$ LF only very modestly changes among $z=5.7$, 6.6 and 6.9. Comparing the $z=6.9$ Ly$\alpha$ LF and the decline of Ly$\alpha$ luminosity density from $z=5.7$ to 6.9 with reionization models, they constrained the neutral fraction to be $x_{\rm HI}^{z=6.9} \sim 0.4$--0.6. This is consistent with the constraints $x_{\rm HI}^{z=7}\sim0.32$--0.64 and $x_{\rm HI}^{z=7.3}< 0.3$--0.8 obtained by our previous $z=7$ LAE surveys \citep{Ota08,Ota10} and the $z=7.3$ LAE survey by \citet{Konno14}, further supporting the rapid increase in neutral hydrogen in IGM with redshift at $z>6$. We will also compare our new constraint on reionization from the present work with the result from \citet{Zheng17}. 

This paper is organized as follows. In Section 2, we describe our observations and data reduction. Then, in Section 3, we perform photometry, select $z=7$ LAE candidates and estimate their physical properties. We derive the Ly$\alpha$ and UV LFs as well as the Ly$\alpha$ EWs of $z=7$ LAEs and compare them with those of $z=5.7$, 6.6 and 7.3 LAEs derived by the previous Subaru LAE surveys to the comparable survey limits in Section 4. Based on these results, we discuss the implications for galaxy evolution and cosmic reionization in Section 5. We summarize and conclude our study in Section 6. Throughout, we adapt AB magnitudes \citep{Oke74} and a concordance cosmology with $(\Omega_m, \Omega_{\Lambda}, h)=(0.3, 0.7, 0.7)$ consistent with the constraints by the recent {\it WMAP} and {\it Planck} observations \citep{Bennett13,Hinshaw13,Planck16a}, unless otherwise specified.  

\section{Observation and Data Reduction\label{Obs_Data}}
In our previous $z=7$ LAE surveys, we imaged the SDF \citep[$13^{\rm h}24^{\rm m}38.^{\rm s}9$, $+27^{\circ}29'25.''9$ (J2000),][]{Iye06,Ota08} and a part of the UKIDSS Ultra Deep Survey field \citep[UKIDSS-UDS,][]{Lawrence07} within the SXDS field \citep[$02^{\rm h}18^{\rm m}00.^{\rm s}00$, $-05^{\circ}13'30''$ (J2000), see Figure 2 of][]{Ota10} with Suprime-Cam (field of view is $34' \times 27'$) and the NB973 filter. Each field was observed by one pointing of Suprime-Cam. 

For the SDF survey, \citet{Iye06} and \citet{Ota08} used the public broadband $B$, $V$, $R$, $i'$, $z'$ and narrowband NB816 ($\lambda_{\rm c} = 8150$\AA, $\Delta \lambda_{\rm FWHM} = 120$\AA, $z_{{\rm Ly}\alpha}=5.65$--5.75) and NB921 ($\lambda_{\rm c} = 9196$\AA, $\Delta \lambda_{\rm FWHM} = 132$\AA, $z_{{\rm Ly}\alpha}=6.51$--6.62) images of SDF\footnote[4]{Available at http://soaps.nao.ac.jp/SDF/v1/index.html}. Currently, the SDF $R$, $i'$ and $z'$ images 0.4--0.7 mag deeper than the public images \citep{Poznanski07,Graur11,Toshikawa12} are available, and we use them in this study along with the public $B$, $V$, NB816 and NB921 images. 

Meanwhile, the entire SXDS field has an area of $\sim 1.3$ deg$^2$ and consists of five pointing of Suprime-Cam. They are called SXDS-C, SXDS-N, SXDS-S, SXDS-E, and SXDS-W with the central coordinates $02^{\rm h}18^{\rm m}00.^{\rm s}00$, $-05^{\circ}00'00''$ (J2000) of the SXDS-C corresponding to the center of the entire SXDS field. The deep public $B$, $V$, $R$, $i'$ and $z'$ images of SXDS\footnote[5]{Available at http://soaps.nao.ac.jp/SXDS/Public/DR1/index\_dr1.html} \citep{Furusawa08} are available for all the five subfields. Also, the deep narrowband NB816 and NB921 images of the SXDS were taken by \citet{Ouchi08,Ouchi09}. Moreover, the SXDS field has been observed with several different wavelengths from X-ray to radio. Hence, multiwavelength study of $z=7$ LAEs is also possible. For example, parts of the SXDS-C and the SXDS-S have been observed by the UKIDSS-UDS \citep{Lawrence07}, the {\it Spitzer Space Telescope} legacy (SpUDS; PI: J. Dunlop), the SEDS (PI G. Fazio) and the {\it Hubble Space Telescope} CANDELS \citep{Grogin11,Koekemoer11} surveys. However, the SXDS-C includes a few bright stars whose stellar halos contaminate large areas. Hence, to image the same area as the multiwavelength surveys and to avoid those bright stars, \citet{Ota10} observed an area between the SXDS-C and the SXDS-S in the NB973 filter by one pointing of the Suprime-Cam as shown in Figure 2 in their paper.

We extend our previous SDF and SXDS $z=7$ LAE surveys by substantially deepening the NB973 images and obtaining $y$ band images. Details of these NB973 and $y$ band imaging data as well as other band data used in this study are shown in Table \ref{ImagingData}. For our previous survey in SDF, \citet{Iye06} and \citet{Ota08} used 15 hr of NB973 imaging of SDF taken in 2005 with the MIT-Lincoln Laboratory (MIT-LL) CCDs installed in Suprime-Cam \citep{Miyazaki02}, which is about twice less sensitive to $z\sim7$ Ly$\alpha$ emission than the Hamamatsu fully depleted red-sensitive CCDs installed in Suprime-Cam since 2008 \citep{Kamata08}. In this study, we additionally use the NB973 imaging of SDF taken with either the MIT-LL or the Hamamatsu CCDs in 2004 (test observations in NB973 with short exposures), 2007, 2013, and 2015. The $y$ band imaging of SDF was taken with either the MIT-LL or the Hamamatsu CCDs in 2003, 2004, 2007 and 2009, and using these data, the final deep stacked image was created by \citet{Ouchi09} (see Table 1 of their paper for details of the SDF $y$ band image). The total integration time is 26.3 hr. \citet{Ouchi09} stacked $y$ band exposures taken with the MIT-LL and the Hamamatsu CCDs because the shapes of $y$ band total response curves are almost identical between the data of these CCDs. We also confirmed the same for the shapes of the NB973 filter response curves, and thus we combine the NB973 exposures taken with the both CCDs.   
 
Meanwhile, for our previous $z=7$ LAE survey in SXDS, \citet{Ota10} used 13 hr of NB973 imaging of SXDS taken in 2008. In this study, we additionally use NB973 imaging of SXDS newly taken in 2013 and 2014. The $y$ band imaging of the same part of the SXDS field was first taken in 2010 by \citet{Tadaki12} for different science objectives, and we took additional $y$ band imaging in 2013 and 2014 in order to deepen the data. All the NB973 and $y$ band exposures of SXDS were taken with the Hamamatsu CCDs in Suprime-Cam.

%We imaged the field centered at the $z=6.6141$ QSO J3035--3150 with Subaru Telescope Suprime-Cam in the broadbands $i'$ and $z'$ as well as the narrowband NB921. The observations were carried out at dark nights on 2013 November 27, and 2014 August 22--27. The sky conditions were partly clear/cloudy with the seeing of $\sim 0.''8$--$1.''3$ in 2013 and photometric with the seeing of $\sim 0.''5$--$0.''9$ in 2014. We took 240, 120--240 and 1200 second individual exposure frames with the $i'$, $z'$ and NB921 bands, respectively, using eight-point dithering patterns. 

Except for the final stacked $y$ band image of SDF created by \citet{Ouchi09}, we reduced the raw NB973 ($y$ band) exposures of SDF and SXDS (SXDS) to create the final stacked science images. We performed the data reduction using the software SDFRED2 \citep{Yagi02,Ouchi04} in the same standard manner as in \citet{Kashikawa04} and \citet{Ota08}, including bias subtraction, flat-fielding, distortion correction, measuring/matching of point spread functions (PSFs) between the 10 CCD chips in Suprime-Cam, sky subtraction and masking of the shadow of the auto guider probe. Then, the dithered exposure frames were matched and stacked. We did not use the individual exposures having full width at half maximum (FWHM) of PSF larger than $1.''0$ as they degrade the final stacked image while they do not help improve the depth. The integration times of the final stacked images are 58.9 hr for the SDF NB973 image, and 23.1 hr and 11.0 hr for the SXDS NB973 and $y$ band images, respectively. Meanwhile, images of the spectrophotometric standard stars GD71 and GD50 \citep{Oke90} taken in NB973 and $y$ band were used to calibrate the photometric zero points of the NB973 and $y$ band images. 

%% In this section, we use  the \subsection commnd to set off
%% a subsection.  \footnote is used to insert a footnote to the text.

%% figure 2
\begin{figure*}
\includegraphics[angle=0,scale=0.475]{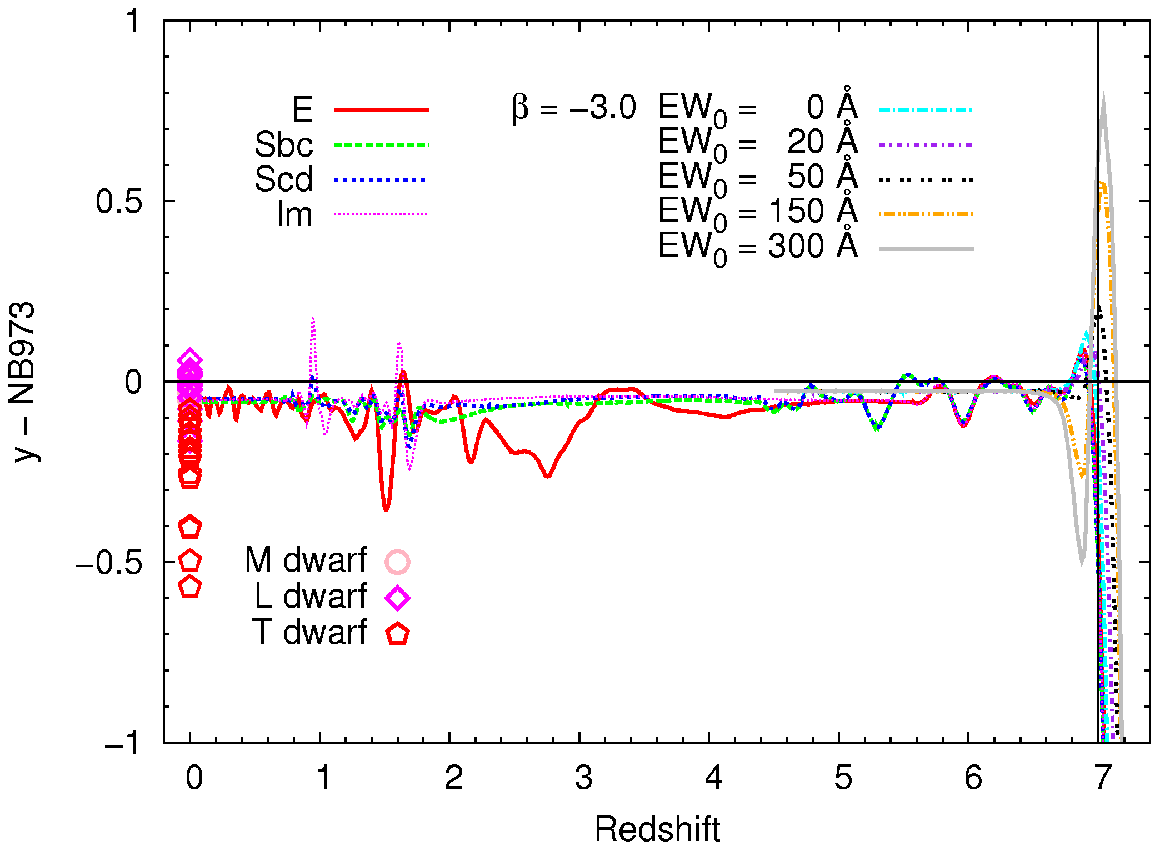}
\includegraphics[angle=0,scale=0.475]{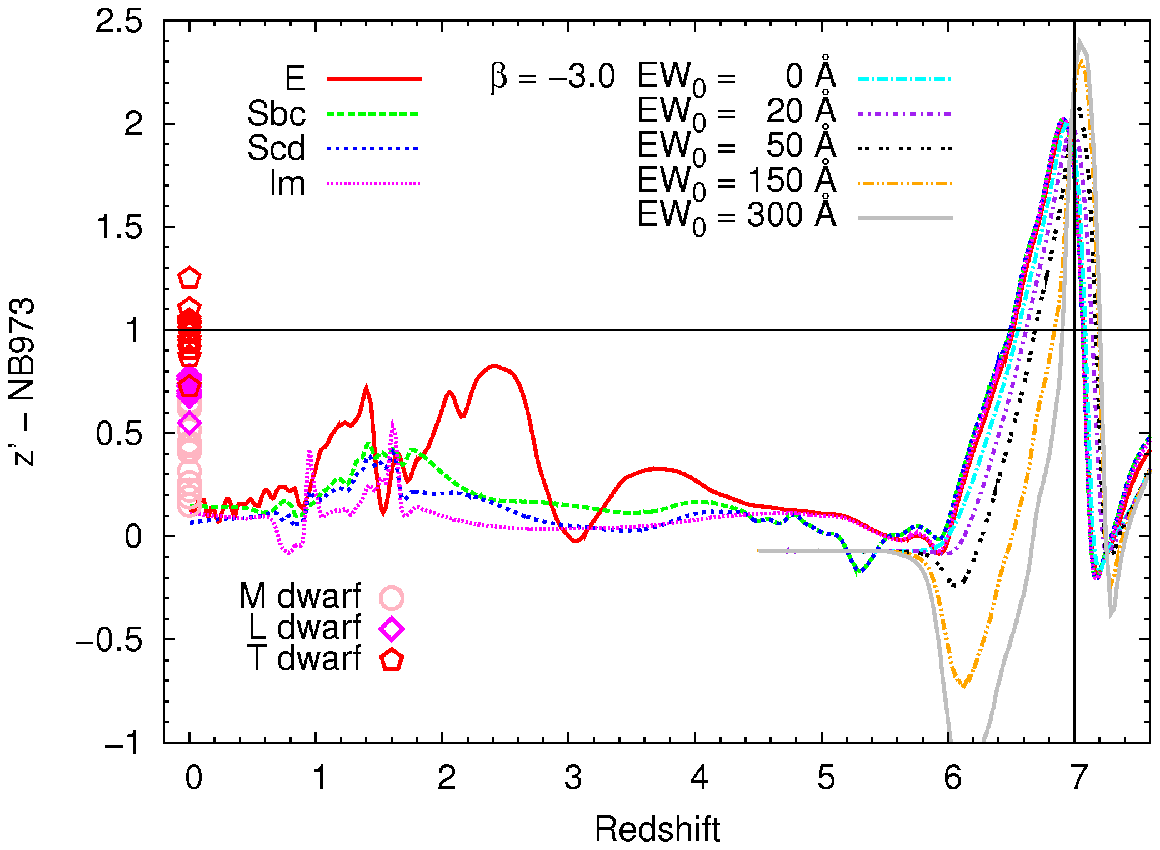}
\includegraphics[angle=0,scale=0.475]{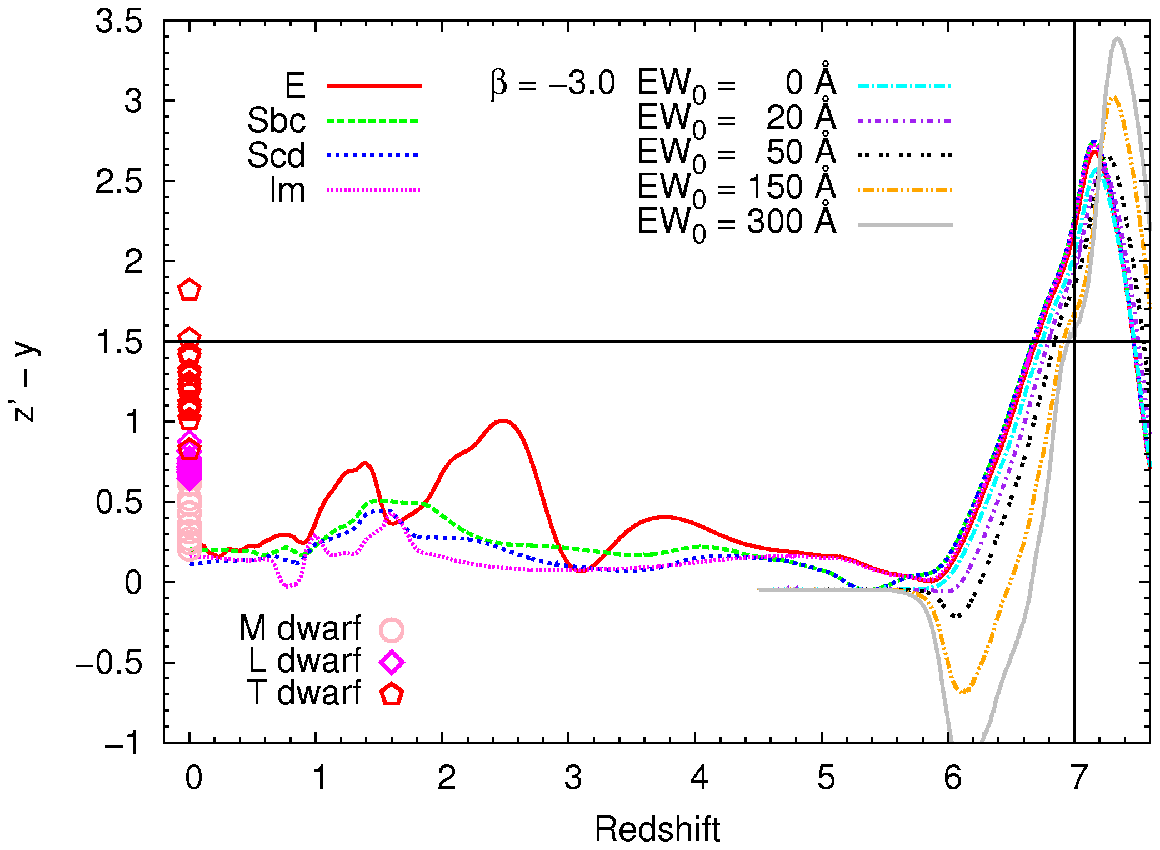}
\caption{$y-{\rm NB973}$, $z'-{\rm NB973}$ and $z'-y$ colors (Suprime-Cam $z'$, NB973 and $y$ bands) as a function of redshift of our model LAEs/LBGs, several types of galaxies and M/L/T type dwarf stars (left, middle and right panels, respectively). Here we plot colors of our model LAEs/LBGs with only a UV continuum slope $\beta=-3$ and several different rest frame Ly$\alpha$ EWs (EW$_0$) as an example because recent study of \citet{Ono10} found that $z\sim7$ LBGs and $z\sim5.7$ and $z\sim6.6$ LAEs have $\beta\sim-3$, but we have confirmed that these three colors do not change much with $\beta$. The colors of E (elliptical), Sbc, Scd and Im (irregular) galaxies were calculated using \citet{Coleman80} template spectra. Also, colors of M/L/T dwarfs (specifically, types M3--M9.5, L0--L9.5 and T0--T8) were calculated using their observed spectra provided by \citet{Burgasser04,Burgasser06a,Burgasser06b,Burgasser08,Burgasser10} and \citet{Kirkpatrick10} at the SpeX Prism Spectral Libraries (see footnote 6). The vertical lines show our target redshift $z=7$ while the horizontal lines indicate the $z=7$ LAE color selection criteria we finally adopted, $y-{\rm NB973}>0$ (see footnote 7 and Section \ref{LyaFaintObjects} for details), $z'-{\rm NB973}>1$ and $z'-y>1.5$.\label{Color_vs_redshift}}
\end{figure*}

\section{Photometry and LAE Candidate Selection\label{Phot_LAEselection}}
\subsection{Photometry and Object Catalogs\label{Photometry}}
In order to select $z=7$ LAEs in SDF and SXDS fields, we performed photometry to make the NB973-detected object catalogs. In selecting $z=7$ LAE candidates, along with the NB973 and $y$ band images, we used the public $B$, $V$, NB816 and NB921 band images of SDF \citep{Kashikawa04}, SDF $R$, $i'$ and $z'$ band images 0.4--0.7 mag deeper than the public images \citep{Poznanski07,Graur11,Toshikawa12} and the public $B$, $V$, $R$, $i'$ and $z'$ band images of SXDS \citep{Furusawa08} and the NB816 and NB921 images of SXDS taken by \citet{Ouchi08,Ouchi09}. Thus, we registered the NB973 and $y$ band images to the public images by using positions of the stellar objects commonly detected in these images. The deep SDF $R$, $i'$ and $z'$ images were also registered to the SDF public images. In the case of SXDS, the NB973 and $y$ band images were registered to the public SXDS-C and SXDS-S images to produce the four images we hereafter call NB973-SXDS-C, NB973-SXDS-S, $y$-SXDS-C and $y$-SXDS-S images, respectively. The NB816 and NB921 images of SXDS-C and SXDS-S were also registered to the public images of SXDS-C and SXDS-S, respectively. Then, we copied the astrometry of the public SDF and SXDS images to the NB973 and $y$ band images of SDF and SXDS, respectively \citep[see][for the details of astrometry of the SDF and SXDS images]{Kashikawa04,Furusawa08}. The SDF NB973 image and the SXDS NB973 and $y$ band images have the PSF FWHMs of $\simeq 1.''0$ as the worst PSF FWHM among the CCD chips in the individual exposures finally used for stacking was $\simeq 1.''0$ because we discarded the exposures having PSFs larger than $1.''0$ right before the PSF matching among the CCD chips during the data reduction. As seen in Sections \ref{LAE-Selections} and \ref{LyaFaintObjects}, we use $z'-y$, $z'-{\rm NB973}$ and $y-{\rm NB973}$ colors to select $z=7$ LAE candidates. To measure the colors accurately with a common aperture photometry, the PSF FWHMs of the $z'$, $y$ and NB973 images should be the same. Hence, we convolved the $z'$ and $y$ band images to have the same PSF FWHM as those of the NB973 images (see Table \ref{ImagingData}). 
%The SDF $y$ band image reduced by \citet{Ouchi09} was also convolved to have a PSF of $1.''0$. All the SDF $z'$, $y$ and NB973 images already had the same PSFs of $1.''0$.
%the $z'$ band images of SXDS-C and SXDS-S had the PSFs of $X.''XX$ and $X.''XX$, respectively, while all the SXDS NB973 and $y$ band images had the PSFs of $1.''0$. Hence, we convolved the PSFs of the SXDS-C and SXDS-S $z'$ band images to $1.''0$.

We used the SExtractor software version 2.8.6 \citep{BA96} for source detection and photometry. The Suprime-Cam CCDs have a pixel size of $0.''202$ pixel$^{-1}$. We considered an area larger than five contiguous pixels with a flux (mag arcsec$^{-2}$) greater than $2\sigma$ (two times the background rms) to be an object. Object detections were first made in the NB973 images, and then photometry was performed in the $B$, $V$, $R$, $i'$, $z'$, $y$, NB816, NB921 and NB973 images, using the double-image mode. We measured $2''$ aperture magnitudes of the detected objects with {\tt MAG\_APER} parameter and total magnitudes with {\tt MAG\_AUTO} (we also used {\tt MAG\_APER} $+$ aperture correction to estimate total magnitudes; see Section \ref{Aperture_Correction} for more details). The NB973-detected object catalogs were constructed for SDF and SXDS by combining the photometry in all the bands. Meanwhile, we also measured the limiting magnitudes of the images by placing $2''$ apertures in random blank positions excluding the low signal-to-noise ratio (S/N) regions near the edges of the images (see Section \ref{LAE-Selections} for the details of removing such edge regions). The limiting magnitudes of the images are shown in Table \ref{ImagingData}.

%\subsection{Selection of $z=7$ LAE Candidates \label{LAE-Selections}}
\subsection{Simulating Expected Colors of $z=7$ LAEs and Potential Contaminants\label{ColorSim}}
Figure \ref{FilterTransmission} shows that the bandpass of NB973 is located at the red side of but within the $z'$ band as well as at the slightly blue side of but almost in the middle of the $y$ band. If Ly$\alpha$ emission of an LAE is redshifted in the bandpass of NB973, it is expected to show significant excess in NB973 with respect to $z'$ and possibly $y$. Also, such an LAE would exhibit red $z'-y$ color due to its Ly$\alpha$ emission and Lyman break by significant intergalactic medium (IGM) absorption. These characteristics can be used to isolate $z=7$ LAEs from other objects. In our previous $z=7$ LAE surveys in SDF and SXDS, we photometrically selected $z=7$ LAE candidates mainly using $z'-{\rm NB973}$ color without $y$ band data \citep{Iye06,Ota08,Ota10}. Now we have deep $y$ band data, we consider using $z'-y$ as well as $z'-{\rm NB973}$ and/or $y-{\rm NB973}$ colors for the LAE selection. In order to determine these color criteria, we examined expected colors of LAEs. 

Figure \ref{Color_vs_redshift} shows $z'-y$, $z'-{\rm NB973}$ and $y-{\rm NB973}$ colors as a function of redshift of model LAEs as well as other types of galaxies and M/L/T type dwarf stars, which may be potential contaminants. We created model spectra of LAEs by assuming the power-law continuum $f_{\lambda} \propto \lambda^{\beta}$ with several different UV continuum slopes $\beta=-3$, $-2$, $-1$ and 0 and adding them the Ly$\alpha$ emission with rest-frame equivalent width of EW$_0 = 0$ (no emission), 20, 50, 150 and 300\AA. We did not assume any specific line profile or velocity dispersion of Ly$\alpha$ emission. Instead, we simply added the total line flux value to the spectra at 1216\AA. Then, we redshifted the spectra to $z=0$--8 and applied IGM absorption to them by using the prescription of \citet{Madau95}. Colors of these model LAEs were calculated using their redshifted spectra and response curves of the Suprime-Cam $z'$, $y$ and NB973 filters. While we assumed the UV continuum slopes $\beta=-3$, $-2$, $-1$ and 0 to cover the possible variety of LAEs and check their expected colors, it would be worth noting that \citet{Ono10} found $\beta\sim-3$ for $z\sim5.7$ and $z\sim6.6$ LAEs by performing an SED-fitting study. We show colors of the model LAEs with $\beta=-3$ for example in Figure \ref{Color_vs_redshift}.

For comparison, we also calculated the colors of E (elliptical), Sbc, Scd and Im (irregular) galaxies by using the filter response curves and the \citet{Coleman80} template spectra and also plot them in Figure \ref{Color_vs_redshift}. Moreover, we plot the colors of M/L/T type dwarf stars (specifically, types M3--M9.5, L0--L9.5 and T0--T8), which can be interlopers, in Figure \ref{Color_vs_redshift}. We calculated the colors by using the filter response curves and their actual spectra provided by \citet{Burgasser04,Burgasser06a,Burgasser06b,Burgasser08,Burgasser10} and \citet{Kirkpatrick10} at the SpeX Prism Spectral Libraries\footnote[6]{http://pono.ucsd.edu/{\textasciitilde}adam/browndwarfs/spexprism/library.html}. 

As seen in the color-redshift diagrams, $z\sim7$ LAEs are expected to show very red colors of $z'-y \gtrsim 1.5$, strong narrowband excess of $z'-{\rm NB973} \gtrsim 1.0$ but possibly either modest narrowband excess of $0<y-{\rm NB973}<1$ or even depression of $y-{\rm NB973} < 0$ to $-1$. Hence, at this stage we adopt $z'-y > 1.5$ and $z'-{\rm NB973} > 1.0$ colors as parts of our $z=7$ LAE selection criteria rather than $y-{\rm NB973}$ color\footnote[7]{
However, note that in Section \ref{LyaFaintObjects} we find/show that the objects detected by the color cuts $z'-y > 1.5$ and $z'-{\rm NB973} > 1.0$ (i.e. the $z=7$ LAE selection criteria (\ref{Criteria-1})) can be classified into either those with strong Ly$\alpha$ EWs (EW$_0>10$\AA) and $y-{\rm NB973} >0$ colors (considered $z=7$ LAEs) or those with extremely low/zero Ly$\alpha$ EWs (EW$_0 \simeq 0$--1\AA) and $y-{\rm NB973} <0$ colors (considered T-type dwarf stars or $z\sim 6.8$--7.1 LBGs) based on our analysis of Ly$\alpha$ and UV continuum fluxes of $z=7$ LAE candidates using the equation (\ref{Eqn_LyaUVLum}). Thus we will later adopt an additional color criterion $y-{\rm NB973} > 0$ to further remove contamination from T-type dwarf stars and LBGs (The left panel of Figure \ref{Color_vs_redshift} also shows that T-type dwarf stars have a color of $y-{\rm NB973} < 0$).}. Figure \ref{Color_vs_redshift} shows that low redshift galaxies are expected to be removed by these color criteria. On the other hand, by imposing these color criteria alone, we cannot completely eliminate T dwarfs as some of them exhibit colors of $z'-y \sim 1.5$--1.8 and $z'-{\rm NB973} \sim 1$--1.3. However, by imposing null detections in wavebands bluewards of $z\sim7$ Ly$\alpha$ (i.e., $B$, $V$, $R$, $i'$, NB816 and NB921) where fluxes of an LAE are absorbed by IGM, we could effectively remove dwarfs. For example, \citet{Taniguchi05} and \citet{Shimasaku06} selected 58 $z\sim6.6$ and 89 $z\sim5.7$ LAE candidates by using the Suprime-Cam filters and the color criteria analogous to ours and imposing null detections in wavebands bluewards of Ly$\alpha$ (less than $2\sigma$ in $B$ and $V$, $R - z' \geq 1.5$ and $z'-{\rm NB816} \geq 1.5$ for $z\sim5.7$ LAEs; less than $3\sigma$ in $B$, $V$ and $R$, $i' - z' > 1.3$ and $z'-{\rm NB921} > 1.0$ for $z\sim6.6$ LAEs). \citet{Kashikawa06,Kashikawa11} conducted spectroscopy of 66 out of 89 $z\sim5.7$ LAE candidates and 52 out of 58 $z\sim6.6$ LAE candidates in their studies and detected no dwarf stars. Hence, we expect that contamination of our sample by dwarfs would be very low or could be zero. Moreover, in Section \ref{LyaFaintObjects} we will show that the contamination from T-type dwarfs can be negligible by further adopting an additional color criterion $y-{\rm NB973} > 0$. 
%However, we will discuss effects of possible contamination, if any, on our subsequent results in this study in Section \ref{LyaFaintObjects}. 
%However, Note that in Section \ref{LyaFaintObjects} we find/show that the color cuts $z'-y > 1.5$ and $z'-{\rm NB973} > 1.0$ (i.e. the $z=7$ LAE selection criteria (\ref{Criteria-1})) actually detect $z=7$ LAE candidates with the narrowband depression $y-{\rm NB973} < 0$ based on our analysis of Ly$\alpha$ and UV continuum fluxes of $z=7$ LAE candidates using the equation (\ref{Eqn_LyaUVLum}). Thus we will later adopt an additional color criterion $y-{\rm NB973} > 0$ to further remove contamination from dwarf stars and LBGs.}. 

Meanwhile, another source of contamination is lower redshift LAEs and LBGs with EW$_0 =0$ at $6.5 \lesssim z < 6.9$ which could have $z'-y > 1.5$ and $z'-{\rm NB973} > 1.0$ colors as seen in Figure \ref{Color_vs_redshift}. Note that bandpasses of the NB973 and $y$ bands overlap with the wavelength range at the red edge of the $z'$ band as seen in Figure \ref{FilterTransmission}. This could cause LAEs and LBGs at $6.5 \lesssim z < 6.9$ to exhibit $z'-y > 1.5$ and $z'-{\rm NB973} > 1.0$ colors, if such galaxies have bright and steep UV continua. For example, \citet{Ota08} found that a $z=6.6$ LAE detected in NB921 by \citet{Taniguchi05} and spectroscopically identified by \citet{Kashikawa06} actually satisfied the color criterion $z' - {\rm NB973} > 1$ \citep[see Figure 3 and Table 1 in][]{Ota08}. Hence, the null detection in NB921 band, also previously adopted by \citet{Ota08}, is important to remove such lower redshift LAEs from the $z=7$ LAE selection. However, the rate of such contamination seems to be very low or zero. We examined $z' - {\rm NB973}$ and $z' - y$ colors and NB973 magnitudes of the 58 photometric $z\sim6.6$ LAE candidates in SDF \citep[the sample of][]{Taniguchi05,Kashikawa06,Kashikawa11}. We found that 7 out of them are detected in NB973 $< 26.2$ ($4\sigma$ limiting magnitude in SDF which is deeper than that in SXDS), of which only one satisfies $z' - {\rm NB973} > 1$ and none meets $z' - y >1.5$. 
%Hence, the criterion $z' - y >1.5$ can effectively remove lower redshift LAEs.
%In addition to this, the color criterion $z' - y > 1.5$ also helps remove lower redshift LAEs.

\subsection{Selection of $z=7$ LAE Candidates \label{LAE-Selections}}
Eventually, we used the following criteria (all the magnitudes are measured in a $2''$ aperture) to select $z=7$ LAE candidates.
\begin{eqnarray}
B > B_{2\sigma},~ V > V_{2\sigma},~ R > R_{2\sigma},~ i' > i'_{2\sigma}, \nonumber\\ 
{\rm NB816} > {\rm NB816}_{2\sigma},~ {\rm NB921} > {\rm NB921}_{2\sigma}, \nonumber\\ 
\left[(z' \leq z'_{2\sigma},~ z' - y > 1.5)~{\rm or}~(z' > z'_{2\sigma})\right],\nonumber\\
z' - {\rm NB973} > 1.0, \nonumber\\ 
{\rm NB973} \leq {\rm NB973}_{4\sigma}
\label{Criteria-1}
\end{eqnarray}
We limit our $z=7$ LAE samples in SDF and SXDS to NB973 $\leq {\rm NB973}_{4\sigma}$ where ${\rm NB973}_{4\sigma}$ is the $4\sigma$ limiting magnitude (NB973 $= 26.20$ and 25.83 in SDF and SXDS, respectively). The $B_{2\sigma}$, $V_{2\sigma}$, $R_{2\sigma}$, $i'_{2\sigma}$, ${\rm NB816}_{2\sigma}$, ${\rm NB921}_{2\sigma}$ and $z'_{2\sigma}$ are the $2\sigma$ limiting magnitudes in $B$, $V$, $R$, $i'$, NB816, NB921 and $z'$. As the fluxes of LAEs bluewards of the $z\sim7$ Ly$\alpha$ emission are absorbed by IGM, we impose null detections (fainter than $2\sigma$ limiting magnitudes) in $B$, $V$, $R$, $i'$, NB816 and NB921 bands. These criteria help remove contamination from dwarf stars and low redshift interlopers including LAEs/LBGs at $z\sim 6.5$--6.8 as discussed above. Figure \ref{Color_vs_redshift} shows that the criterion $z'-{\rm NB973} > 1.0$ corresponds to the rest frame Ly$\alpha$ EW threshold of EW$_0 \geq 0$ and can also detect $z\sim7$ LBGs with a zero or extremely weak Ly$\alpha$ emission (EW$_0=0$--10\AA) as contaminants. However, as described in Section \ref{LyaFaintObjects}, we can remove such interlopers by estimating their EW$_0$'s by using their NB973 and $y$-band total magnitudes and the equation (\ref{Eqn_LyaUVLum}). 

To select $z=7$ LAE candidates in the SDF and SXDS fields, we first masked blooming, smearing and halos of large bright stars, large galaxies, bad pixels and low S/N edge regions in the SDF and SXDS images by applying the public SDF and SXDS masking programs \citep{Kashikawa04,Furusawa08} to our NB973-detected object catalogs made in Section \ref{Photometry}. 

Then, we further applied the criteria (\ref{Criteria-1}) to the NB973-detected object catalogs. In this process, if $z'$ band magnitude of a source in the catalog is fainter than $1\sigma$ (i.e., $z'> z'_{1\sigma}$), we replaced it by $z'_{1\sigma}$. We found that the criteria (\ref{Criteria-1}) yielded a large number of objects, and most of them are located in the low S/N regions near the edges of the SDF and SXDS NB973 images which arose from the dithering at the time of observations and fringing peculiar to these images and were thus not masked by the public masking program. This implies that most of them could be noises that appear only in the NB973 images and not in the other wavebands. To examine if they are noises, we created the negative NB973 images by multiplying each pixel value by $-1$, performed source detection running SExtractor and limit the detected objects to NB973 $\leq {\rm NB973}_{4\sigma}$. The low S/N edge regions of the NB973 images where a large number of sources passed the criteria (\ref{Criteria-1}) was dominated with negative detections, which are considered noises. Hence, we removed the sources at these edge regions from the lists of objects that passed the LAE criteria (\ref{Criteria-1}). This left 25 and 14 objects in SDF and SXDS, respectively. 

%% figure 3
\begin{figure*}
\epsscale{1.17}
%\plottwo{SDF_Final_18z7LAEs_z-NB973_vs_NB973.eps}{SXDS_Final_6z7LAEs_z-NB973_vs_NB973.eps}
%\plottwo{z-NB973_vs_z-y_LAE_Beta-3_EW0-300_w_Final18z7LAEs_SDF.eps}{z-NB973_vs_z-y_LAE_Beta-3_EW0-300_w_Final6z7LAEs_SXDS.eps}
\plottwo{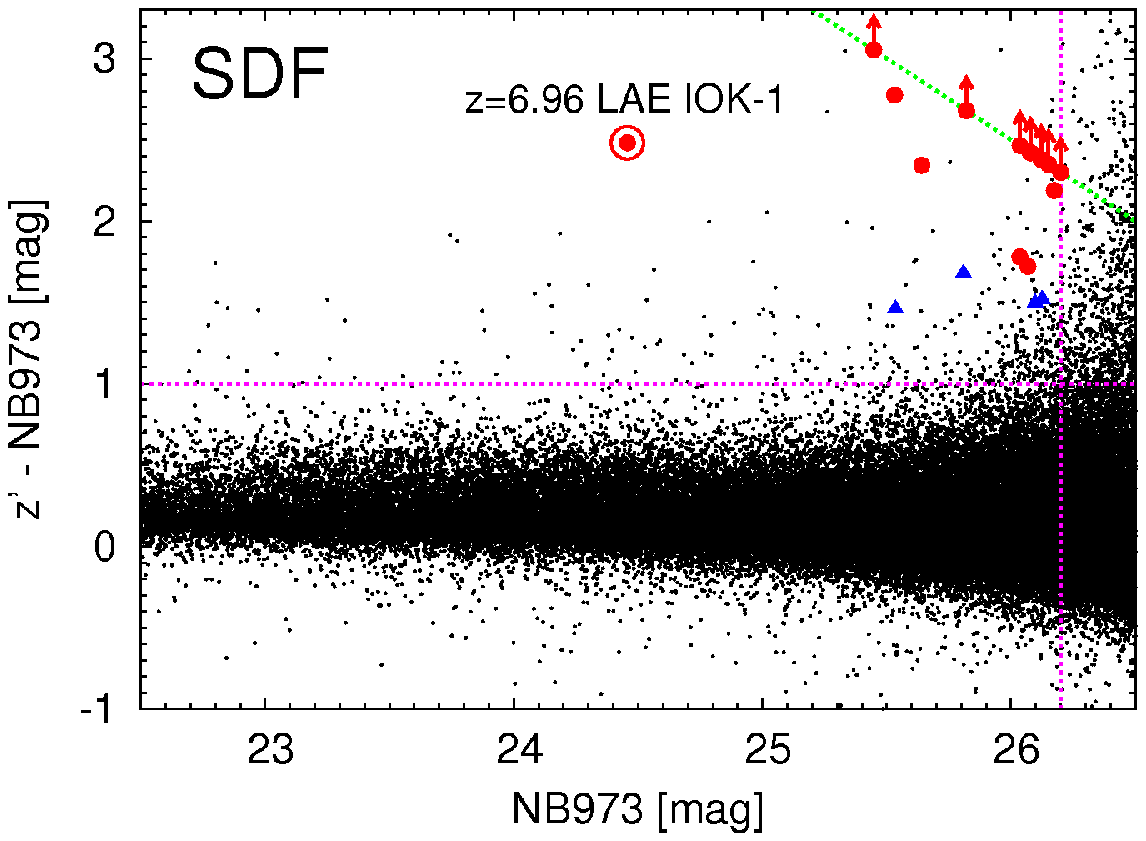}{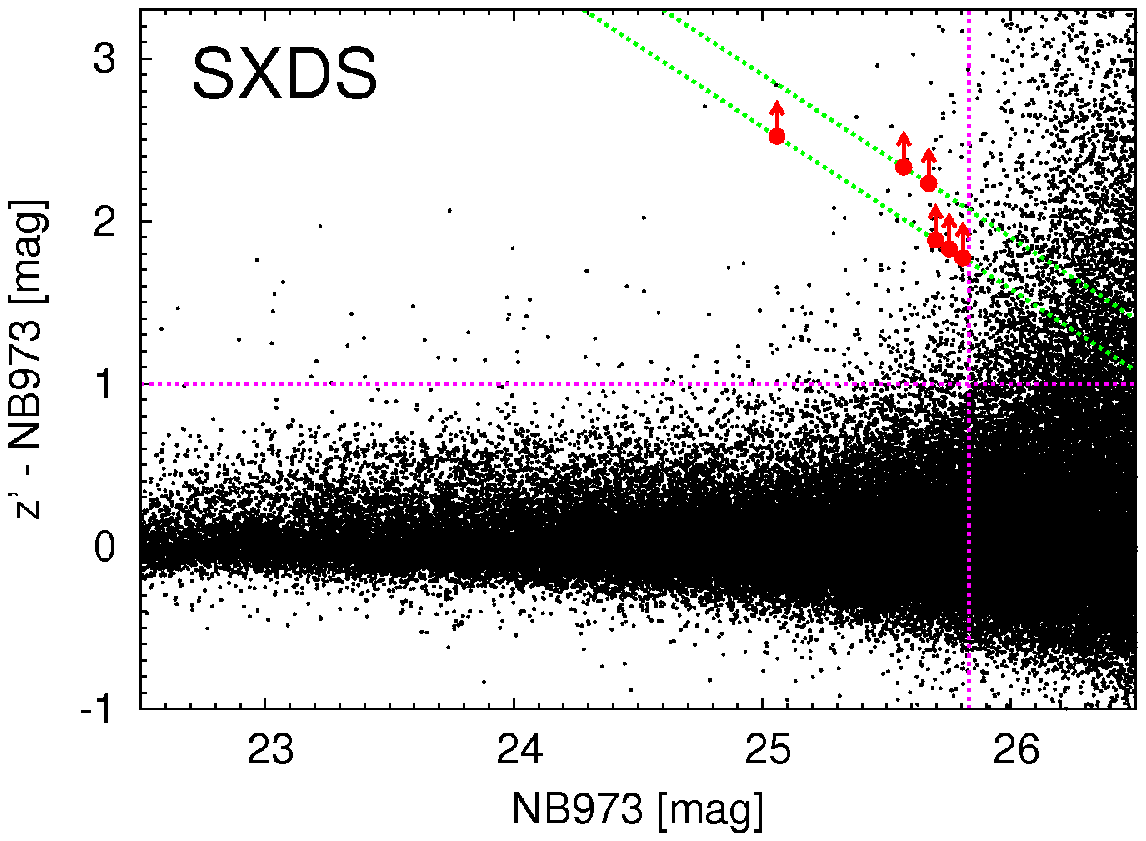}
\plottwo{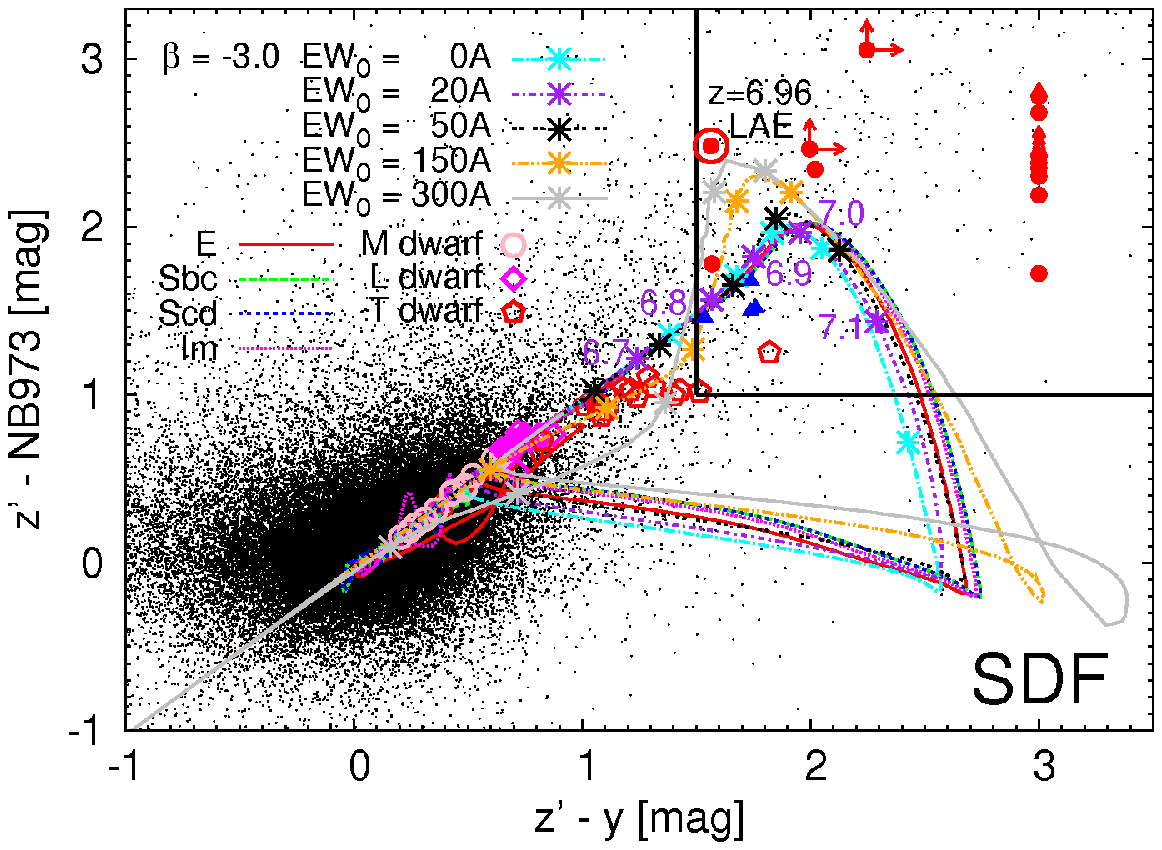}{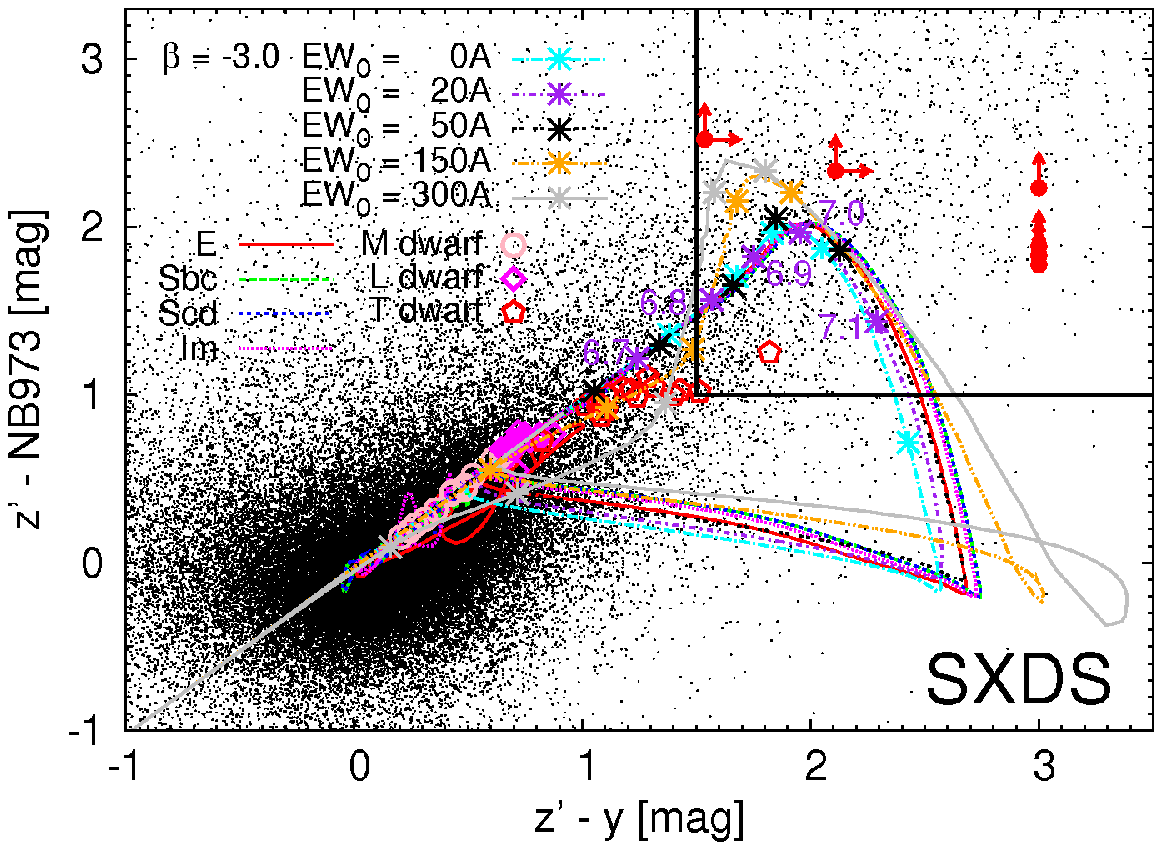}
\caption{Upper panels: $z' - {\rm NB973}$ color as a function of NB973 ($2''$ aperture) magnitude of all the objects detected in our SDF and SXDS NB973 images (shown by dots). The horizontal lines show a part of our $z=7$ LAE color selection criteria, $z'-{\rm NB973} > 1.0$. The vertical lines indicate the $4\sigma$ limiting magnitudes, ${\rm NB973} = 26.20$ and 25.83 in SDF and SXDS, respectively. The finally selected $z=7$ LAE candidates are denoted by the red filled circles with the arrows showing the $1\sigma$ limits on $z'-{\rm NB973}$ colors for those undetected in $z'$ band. The four objects with a $y-{\rm NB973}<0$ color and an extremely faint or zero Ly$\alpha$ flux in SDF finally removed from our $z=7$ LAE sample are shown by the blue triangles (see Tables \ref{z7LAECandidates} and \ref{Propertyz7LAECandidates} and Section \ref{LyaFaintObjects}). The $1\sigma$ limits on $z'-{\rm NB973}$ color for each NB973 magnitude are shown by the diagonal lines for SDF, SXDS-C and SXDS-S, respectively. Note that the $z'$ band depths in SXDS-C and SXDS-S, which our NB973 SXDS image consists of, are different (see Table \ref{ImagingData}). Lower panels: $z'-{\rm NB973}$ versus $z'-y$ color-color plot of all the objects detected in our SDF and SXDS NB973 images (shown by dots). The upper right rectangles surrounded by the solid lines indicate parts of our LAE selection criteria (\ref{Criteria-1}), $z'-y>1.5$ and $z'-{\rm NB973} > 1.0$. The selected $z=7$ LAE candidates are denoted by the filled circles with the arrows showing the $1\sigma$ limits on their colors. The LAE candidates undetected in $y$ band (fainter than $2\sigma$) are placed at $z'-y=3$ for presentation purpose. The four objects with a $y-{\rm NB973}<0$ color and an extremely faint or zero Ly$\alpha$ flux in SDF finally removed from our $z=7$ LAE sample are shown by the blue triangles. Using the same data, symbols and lines in Figure \ref{Color_vs_redshift}, we also plot the colors of M/L/T dwarfs (types M3--M9.5, L0--L9.5 and T0--T8) \citep[spectra from][]{Burgasser04,Burgasser06a,Burgasser06b,Burgasser08,Burgasser10,Kirkpatrick10}, $z=0$--8 E (elliptical), Sbc, Scd and Im (irregular) galaxies \citep[spectra of][]{Coleman80} as well as our model LAEs at $z=4.5$--8 with a UV slope $\beta=-3$ and several different rest frame Ly$\alpha$ EWs (EW$_0$). On the model LAE color evolution tracks, we denote by asterisks redshifts from $z=6.7$ to 7.1 by $\Delta z = 0.1$ step. We label them for the case of EW$_0=20$\AA~LAE model. In both upper and lower left panels, one of the LAE candidates in SDF, NB973-SDF-85821, previously spectroscopically confirmed as a $z=6.96$ LAE, IOK-1, by \citet{Iye06}, \citet{Ota08} and \citet{Ono12} is encircled and so labeled. \label{z-NB973_vs_NB973_and_z-NB973_vs_z-y}}
\end{figure*}

To further eliminate possible spurious sources, if any, we visually inspected $B$, $V$, $R$, $i'$, $z'$, $y$, NB816, NB921 and NB973 images of each of these sources. We especially removed obviously spurious sources such as columns of bad pixels, pixels saturated with bright stars, their blooming, smearing and halos (those not removed by the previous masking process), noise events of deformed shapes, and scattering pixels having anomalously large fluxes. Also, we removed objects seen very faintly in $B$, $V$, $R$, $i'$, NB816 and NB921 bands (i.e., wavebands bluewards of $z\sim7$ Ly$\alpha$) even though we imposed less than $2\sigma$ criterion on these bands as a part of the LAE selection criteria (\ref{Criteria-1}). After the visual inspection, we were left with 18 and 6 sources in SDF and SXDS, respectively. 

In Section \ref{LyaFaintObjects} (see the text there for details), we will further remove 4 objects in SDF whose estimated Ly$\alpha$ emission are extremely weak or zero (EW$_0 \leq 10$\AA). Hence, we consider 14 and 6 sources in SDF and SXDS to be our final $z=7$ LAE candidates. 

Our $z=7$ LAE criteria (\ref{Criteria-1}) successfully re-select a $z=6.96$ LAE, IOK-1, in SDF (dubbed NB973-SDF-85821 in this paper), which we previously spectroscopically confirmed \citep{Iye06,Ota08}. Moreover, the selected LAE candidates in SXDS successfully include two out of the three $z=7$ LAE candidates, NB973-SXDS-S-113268 and NB973-SXDS-S-66924 (dubbed NB973-SXDS-S-95993 and NB973-SXDS-S-66036, respectively, in this paper), which we selected previously without $y$ band \citep{Ota10}. However, we could not re-select the other $z=7$ LAE candidate NB973-SXDS-S-5729 \citet{Ota10} previously selected because this object is located slightly inside one of the low S/N edge regions we trimmed off our NB973 image. \citet{Ota10} also trimmed the similar low S/N edge region that determined by a negative image test similar to what we did in this study but our trimmed region is slightly larger than theirs. Nevertheless, the object is also detected in our deeper NB973 image and satisfies our LAE selection criteria (\ref{Criteria-1}) except for the criterion NB973 $\leq$ NB973$_{4\sigma}$. Its NB973 $2''$ aperture magnitude is 0.67 mag fainter than what \citet{Ota10} measured. This could be because the photometry of this object is affected by two factors. (i) Our NB973 image is deeper and has less positive noises (sky background residuals) than that of \citet{Ota10} at/around the position of the object. (ii) The object slightly partly blends with its neighbor. The object and/or neighbor might be a variable object and became fainter. \citet{Ota10} checked the magnitudes of this object in the NB973 images taken in October and November 2008 and confirmed no variability in one month interval. However, we created our current NB973 image by stacking not only exposures taken in 2008 but also those in 2013 and 2014. If the object is variable and became significantly fainter in 2013 and/or 2014, its magnitude measured in our NB973 image would be also fainter. Eventually, as we cannot clearly conclude whether this object can be a $z=7$ LAE and it is located within our trimmed region, we do not include it in our LAE sample.      

\subsection{Magnitudes, Colors and Images of the $z=7$ LAE Candidates \label{LAE-Photo-image}}
\begin{figure}
\epsscale{1.17}
%\plotone{SDF_18z7LAEs_BlackWhite-ReCont.eps}
\plotone{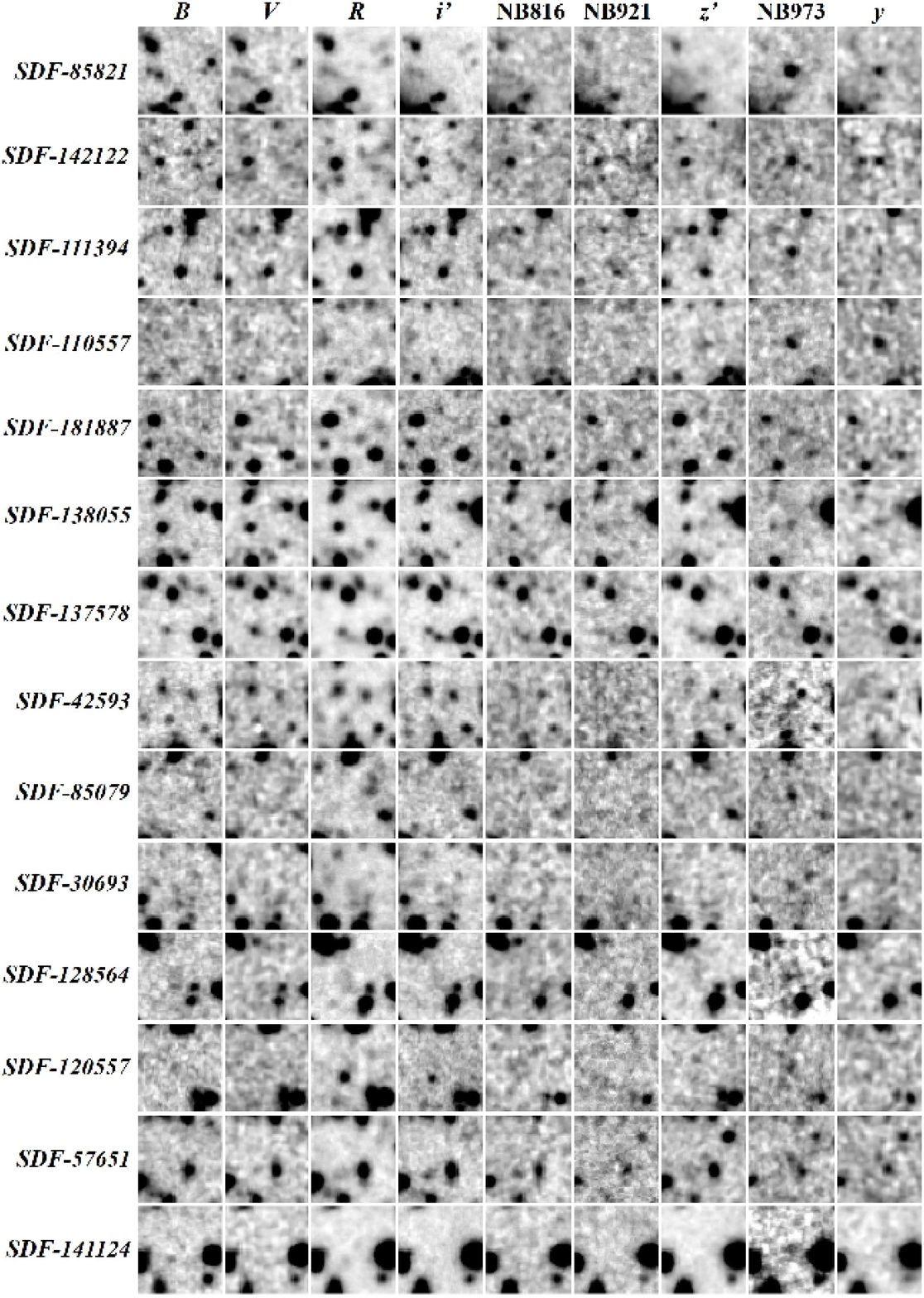}
\caption{Multi-waveband images of the 14 final $z=7$ LAE candidates in SDF. The size of each image is $10'' \times 10''$. North is up and east to the left. The topmost object NB973-SDF-85821 is a $z=6.96$ LAE, IOK-1, previously spectroscopically confirmed by \citet{Iye06}, \citet{Ota08} and \citet{Ono12}.\label{SDF_18z7LAEs_BlackWhite}}
\end{figure}

\begin{figure}
\epsscale{1.17}
\plotone{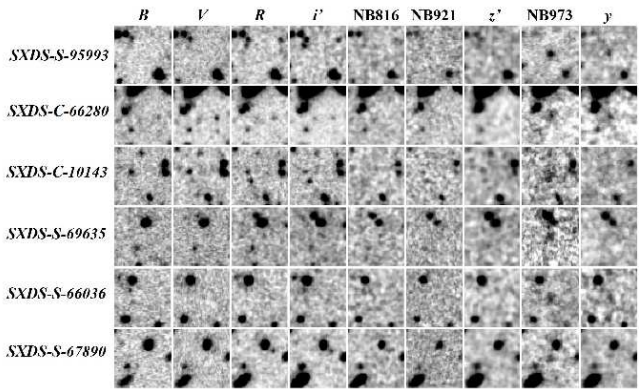}
\caption{Multi-waveband images of the six $z=7$ LAE candidates in SXDS. The size of each image is $10'' \times 10''$. North is up and east to the left.\label{SXDS_z7LAEs_BlackWhite}}
\end{figure}

The color-magnitude ($z'-{\rm NB973}$ versus NB973) and two color ($z'-{\rm NB973}$ versus $z'-y$) diagrams of our $z=7$ LAE candidates and all the NB973-detected objects are plotted in Figure \ref{z-NB973_vs_NB973_and_z-NB973_vs_z-y}. The $B$, $V$, $R$, $i'$, NB816, NB921, $z'$, NB973 and $y$ images of the $z=7$ LAE candidates in SDF and SXDS are shown in Figures \ref{SDF_18z7LAEs_BlackWhite} and \ref{SXDS_z7LAEs_BlackWhite}, respectively. The ID (dubbed based on SExtractor detection {\tt NUMBER} and our survey fields), coordinates, $z'$, NB973 and $y$ band magnitudes, colors and stellarities of the $z=7$ LAE candidates are listed in Table \ref{z7LAECandidates}. 

%%% figure 4 (skipped)
%\begin{figure*}
%\epsscale{1.17}
%\plottwo{z-NB973_vs_z-y_LAE_Beta-3_EW0-300_w_Final18z7LAEs_SDF.eps}{z-NB973_vs_z-y_LAE_Beta-3_EW0-300_w_Final6z7LAEs_SXDS.eps}
%\caption{(Left panel)  (Right panel) \label{z-NB973_vs_z-y}}
%\end{figure*}

%%figure 4
%\begin{figure*}
%\epsscale{0.9}
%%%%\plotone{SDF_18z7LAEs_BlackWhite.eps}
%\plotone{SDF_18z7LAEs_BlackWhite-ReCont.eps}
%\caption{Multi-waveband images of the 18 $z=7$ LAE candidates in SDF. The size of each image is $10'' \times 10''$. North is up and east to the left.\label{SDF_18z7LAEs_BlackWhite}}
%\end{figure*}
%%figure 4

%%figure 5
%\begin{figure*}
%\epsscale{0.9}
%\plotone{SXDS_6z7LAEs_BlackWhite.eps}
%\caption{Multi-waveband images of the six $z=7$ LAE candidates in SXDS. The size of each image is $10'' \times 10''$. North is up and east to the left.\label{SXDS_z7LAEs_BlackWhite}}
%\end{figure*}
%%figure 5

%Table 2 (deluxetable)
\begin{turnpage}
\begin{deluxetable*}{lccrrlrcrrrc}
\tabletypesize{\scriptsize}
%\rotate
\tablecaption{Photometry of the $z=7$ Ly$\alpha$ Emitter Candidates in SDF and SXDS\label{z7LAECandidates}}
\tablewidth{0pt}
%\tablewidth{510pt}
\tablehead{
\colhead{Object ID} & \colhead{R.A.(J2000)} & \colhead{Decl.(J2000)} & \colhead{$z'$} & \colhead{NB973} & \colhead{NB973$_{\rm total}$} &\colhead{$y$} &\colhead{$y_{\rm total}$$^{\rm d}$} & \colhead{$z'-y$} & \colhead{$z'-{\rm NB973}$} & \colhead{$y-{\rm NB973}$} & \colhead{stellarity}
}
\startdata
NB973-SDF-85821$^{\rm a}$ & 13:23:59.77 & $+$27:24:55.76 &    26.94 & 24.46 & 24.14         &    25.37$^{\rm c}$ & 25.27         & 1.57    & 2.48    & 0.91    & 0.54 \\      
NB973-SDF-142122        & 13:25:17.06 & $+$27:37:46.14 & $>$28.50 & 25.45 & 25.24         &    26.25$^{\rm c}$ & 26.22         & $>$2.25 & $>$3.05 & 0.80    & 0.45 \\      
NB973-SDF-111394        & 13:24:07.27 & $+$27:30:49.40 &  (28.31) & 25.53 & 25.37         &  (26.91)$^{\rm c}$ & 26.88         & 1.40    & 2.78    & 1.38    & 0.77 \\      
NB973-SDF-110557        & 13:24:50.43 & $+$27:30:37.66 &  (27.98) & 25.64 & 25.56         &    25.96$^{\rm c}$ & 25.93         & 2.02    & 2.34    & 0.32    & 0.25 \\      
NB973-SDF-181887        & 13:23:46.13 & $+$27:46:50.64 & $>$28.50 & 25.82 & 25.62$^{\rm b}$ & $>$27.40         & ---           & ---     & $>$2.68 & $>$1.58 & 0.44 \\      
NB973-SDF-138055        & 13:23:50.01 & $+$27:36:48.87 &  (27.82) & 26.04 & 25.52         &    26.25$^{\rm c}$ & 26.04         & 1.57    & 1.78    & 0.21    &  0.06 \\      
NB973-SDF-137578        & 13:25:06.62 & $+$27:36:42.45 & $>$28.50 & 26.04 & 25.84$^{\rm b}$ &    26.50$^{\rm f}$ & 26.40$^{\rm f}$ & $>$2.00 & $>$2.46 & 0.46    &  0.66 \\      
NB973-SDF-42593         & 13:25:25.95 & $+$27:15:29.78 &  (27.79) & 26.07 & 25.87$^{\rm b}$ &  (26.85)         & ---           & 0.94    & 1.72    & 0.78    &  0.00 \\      
NB973-SDF-85079         & 13:25:33.84 & $+$27:24:40.66 & $>$28.50 & 26.08 & 25.87         & $>$27.40         & ---           & ---     & $>$2.42 & $>$1.32 &  0.47 \\      
NB973-SDF-30693         & 13:25:05.89 & $+$27:13:11.62 & $>$28.50 & 26.08 & 26.06         & $>$27.40         & ---           & ---     & $>$2.42 & $>$1.32 &  0.37 \\      
NB973-SDF-128564        & 13:25:25.87 & $+$27:34:42.75 & $>$28.50 & 26.12 & 25.92$^{\rm h}$ & $>$27.40         & ---           & ---     & $>$2.38 & $>$1.28 &  0.01 \\      
NB973-SDF-120557        & 13:23:43.68 & $+$27:32:55.61 & $>$28.50 & 26.15 & 26.14         &  (27.29)         & ---           & $>$1.21 & $>$2.35 & 1.14    &  0.29 \\      
NB973-SDF-57651         & 13:25:19.67 & $+$27:18:25.77 &  (28.36) & 26.17 & 25.97$^{\rm b}$ &  (26.93)$^{\rm c}$ & 26.86$^{\rm e}$ & 1.43   & 2.19    & 0.76    &  0.04 \\      
NB973-SDF-141124        & 13:23:46.12 & $+$27:37:28.18 & $>$28.50 & 26.20 & 25.96         & $>$27.40         & ---            & ---    & $>$2.30 & $>$1.20 &  0.02 \\  
\hline
NB973-SXDS-S-95993      & 02:17:59.54 & $-$05:14:07.64 & $>$27.58 & 25.06 & 24.89         &    26.04$^{\rm c}$ & 25.86         & $>$1.54 & $>$2.52 & 0.98    &  0.85 \\
NB973-SXDS-C-66280      & 02:17:28.77 & $-$05:02:35.11 & $>$27.90 & 25.57 & 24.99         &    25.79$^{\rm f}$ & 25.69$^{\rm f}$ & $>$2.11 & $>$2.33 & 0.22    &  0.01 \\
NB973-SXDS-C-10143      & 02:17:42.08 & $-$05:12:21.06 & $>$27.90 & 25.67 & 25.47$^{\rm b}$ &  (26.96)         & ---           & $>$0.94 & $>$2.23 & 1.29    &  0.00 \\
NB973-SXDS-S-69635      & 02:17:04.23 & $-$05:18:08.41 & $>$27.58 & 25.70 & 25.50$^{\rm b}$ &  (26.94)         & ---           & $>$0.64 & $>$1.88 & 1.24    &  0.00 \\
NB973-SXDS-S-66036      & 02:17:57.86 & $-$05:18:47.42 & $>$27.58 & 25.75 & 25.55$^{\rm b}$ &  (26.53)$^{\rm c}$ & 26.27$^{\rm e}$ & $>$1.05 & $>$1.83 & 0.78    &  0.01 \\
NB973-SXDS-S-67890      & 02:19:02.80 & $-$05:18:29.63 & $>$27.58 & 25.81 & 25.72         & $>$27.05         & ---            & ---     & $>$1.77 & $>$1.24 &  0.00\\
\hline
\multicolumn{12}{c}{Objects with $y-{\rm NB973}<0$ and extremely faint or zero Ly$\alpha$ fluxes$^{\rm g}$}\\
\hline
NB973-SDF-109780        & 13:24:07.07 & $+$27:30:26.54 &    26.99 & 25.53 & 25.41         &    25.45$^{\rm c}$ & 25.22         & 1.54     &  1.46   & $-0.08$  &  0.68 \\      
NB973-SDF-155934        & 13:24:31.44 & $+$27:40:41.71 &    27.49 & 25.81 & 25.28         &    25.74$^{\rm c}$ & 25.15         & 1.75     &  1.68   & $-0.07$  &  0.73 \\ 
NB973-SDF-101846        & 13:25:09.62 & $+$27:28:32.16 &    27.59 & 26.10 & 25.89         &    25.85$^{\rm c}$ & 25.71         & 1.74     &  1.49   & $-0.25$  &  0.65 \\      
NB973-SDF-99670         & 13:23:49.40 & $+$27:28:00.27 &    27.64 & 26.13 & 26.11         &    25.88$^{\rm c}$ & 25.66         & 1.76     &  1.51   & $-0.25$  &  0.60 
\enddata
%% Text for table notes should follow after the \enddata but before
%% the \end{deluxetable}. Make sure there is at least one \tablenotemark
%% in the table for each \tablenotetext.
\tablecomments{Units of coordinate are hours: minutes: seconds (right ascension) and degrees: arcminutes: arcseconds (declination) using J2000.0 equinox. $z'$, NB973 and $y$ are all $2''$ aperture magnitudes ($2'' \simeq 2 \times$ FWHMs of PSFs of $z'$, NB973 and $y$ band images) while NB973$_{\rm total}$ and $y_{\rm total}$ are total magnitudes. The magnitudes between $1\sigma$ and $2\sigma$ levels are put in parentheses. Magnitudes are replaced by their $1\sigma$ limits if they are fainter than the limits. The LAE candidates in each field are listed in the order of increasing NB973 $2''$ aperture magnitude. The $z'-y$, $z'-{\rm NB973}$ and $y-{\rm NB973}$ colors were calculated from the $2''$ aperture magnitudes and/or their $1\sigma$ limits. The stellarity is the star/galaxy classifier index measured and given as {\tt CLASS\_STAR} parameter by SExtractor. It is 0 for a galaxy, 1 for a star, or any intermediate value for more ambiguous objects \citep{BA96}.}
\tablenotetext{a}{This object is a $z=6.96$ LAE, IOK-1, previously spectroscopically confirmed by \citet{Iye06}, \citet{Ota08} and \citet{Ono12}.}
\tablenotetext{b}{These total NB973 magnitudes were estimated by applying the aperture correction ($-0.2$ mag for NB973, see Section \ref{Aperture_Correction}) to their $2''$ aperture NB973 magnitudes because these LAE candidates have bright close neighbors and/or slightly blend with another object (values 1 and/or 2 in SExtractor {\tt FLAGS} parameter), which biases the {\tt MAG\_AUTO} measurements by SExtractor. All the other LAE candidates neither have any bright close neighbor nor blend with any objects (value 0 in {\tt FLAGS}), and thus we adopt the {\tt MAG\_AUTO} measured by SExtractor as their total magnitudes.}
\tablenotetext{c}{These LAE candidates are visually seen and detected by the SExtractor single image photometry in the $y$ band images while all the others are either unseen and undetected or visually faintly seen but undetected by the SExtractor single image photometry in the $y$ band images (see Figures \ref{SDF_18z7LAEs_BlackWhite} and \ref{SXDS_z7LAEs_BlackWhite} and footnote f).}
\tablenotetext{d}{For $y$ band total magnitude, $y_{\rm total}$, we adopt {\tt MAG\_AUTO} measured by the single image photometry on the $y$ band images by SExtractor rather than the double-image mode with the NB973 images as the latter's {\tt MAG\_AUTO} measurement does not always estimate total magnitudes accurately while the former does for unblended objects. The $y_{\rm total}$'s of the objects not visually seen and/or undetected by SExtractor single image mode in $y$ band are left blank.}
\tablenotetext{e}{These LAE candidates blend with other objects in NB973 but do not ({\tt FLAGS} $=0$) in $y$ band.}
\tablenotetext{f}{These LAE candidates are visually faintly seen but undetected by the SExtractor single image photometry in the $y$ band images. However, their $2''$ aperture $y$ band magnitudes measured by the SExtractor double image photometry in the NB973 and the $y$ band images are $>2\sigma$ ($2.3\sigma$ in NB973-SDF-137578 and $3.2\sigma$ in NB973-SXDS-C-66280). Hence, we applied the aperture correction ($-0.1$ mag for $y$ band, see Section \ref{Aperture_Correction}) to their $2''$ aperture $y$ band magnitudes to estimate their $y_{\rm total}$'s.}
\tablenotetext{g}{These four objects have colors of $y-{\rm NB973}<0$ and turn out to exhibit extremely faint or zero Ly$\alpha$ fluxes ($f({\rm Ly}\alpha) \lesssim 1.7 \times 10^{-18}$ erg s$^{-1}$ cm$^{-2}$ and EW$_0 \lesssim 0.17$\AA, see Table \ref{Propertyz7LAECandidates}) as a result of calculating $f_{\rm line}$ and $f_c$ by the equation (\ref{Eqn_LyaUVLum}). Hence, we consider them $z\sim7$ LBGs with a bright UV continuum and extremely faint or no Ly$\alpha$ emission (T-type dwarf stars can exhibit colors of $z'-y>1.5$, $z'-{\rm NB973}>1$ and $y-{\rm NB973}<0$, but given the low stellarities of these four objects, they are likely LBGs) and remove them from our $z=7$ LAE sample.}
\tablenotetext{h}{This object has SExtractor {\tt FLAGS} $=0$ but blends with faint background noises near it. This causes the SExtractor to consider it a larger single object and measure its {\tt MAG\_AUTO} with a very large aperture including the noises. This results in overestimate of NB973$_{\rm total}$. Hence, we instead applied the aperture correction to its $2''$ aperture NB973 magnitude to estimate its NB973$_{\rm total}$.}
%\tablenotetext{b}{They were also previously detected as $z=7$ LAE candidates, NB973-SXDS-S-113268 and NB973-SXDS-S-66924, respectively, by \citet{Ota10}.}
\end{deluxetable*}
\end{turnpage}

The stellarity is the star/galaxy classifier index measured for each LAE candidate and given as {\tt CLASS\_STAR} parameter by SExtractor. It is 0 for a galaxy, 1 for a star, or any intermediate value for more ambiguous objects \citep{BA96}. One of the LAE candidates, NB973-SXDS-S-95993, has somewhat high value 0.85 although we cannot tell whether it is a $z=7$ LAE or a dwarf star until we obtain its spectrum. On the other hand, all the other LAE candidates have relatively to considerably low stellarities of 0.0--0.77. This supports our earlier argument that our LAE selection criteria would include very low or zero contamination by dwarfs. 

\subsubsection{Total Magnitudes and Aperture Corrections for Blended Sources \label{Aperture_Correction}}
Meanwhile, the total NB973 magnitudes listed in Table \ref{z7LAECandidates} were measured by the {\tt MAG\_AUTO} parameter of SExtractor for the $z=7$ LAE candidates unblended with any other objects. In this case, SExtractor returns a {\tt FLAGS} value of 0 to the objects detected in NB973. 12 out of the 20 LAE candidates in SDF and SXDS are unblended with {\tt FLAGS} $=0$ in the NB973 images except for one LAE candidate NB973-SDF-128564 which has {\tt FLAGS} $=0$ but blends with faint noise sources near it. \citet{Konno14} pointed out that {\tt MAG\_AUTO} magnitude measurements could be biased in the case of faint objects near a limiting magnitude. To examine this, we also estimated total magnitudes of the 12 unblended LAE candidates by performing multi-aperture photometry (between $2''$--$5''$ with small steps and not including other objects inside), plotting curves of magnitude versus aperture and measuring their plateaus. The total magnitudes measured by {\tt MAG\_AUTO} and the multi-aperture photometry were in good agreement for all the unblended LAE candidates. Hence, we conclude that we can safely adopt {\tt MAG\_AUTO} as total magnitudes for the unblended LAE candidates in the case of our NB973 images. On the other hand, in addition to the LAE candidate NB973-SDF-128564 mentioned above, the remaining 7 LAE candidates have bright close neighbors and/or slightly blend with another object (values 1 and/or 2 in SExtractor {\tt FLAGS} parameter; see the footnote b in Table \ref{z7LAECandidates}), which biases the {\tt MAG\_AUTO} measurements. We estimated their total NB973 magnitudes by applying the aperture correction of $-0.2$ mag to their $2''$ aperture magnitudes. We obtained this correction factor by measuring the differences between total NB973 magnitudes (either {\tt MAG\_AUTO} or multi-aperture photometry) and $2''$ aperture NB973 magnitudes of the 12 unblended LAE candidates and taking their median value. We did not use isolated stellar objects to estimate the aperture correction factor because shapes of the LAE candidates are not necessarily similar to those of stellar objects as their stellarities indicate in Table \ref{z7LAECandidates}. 

Meanwhile, for $y$ band total magnitude, we basically adopt {\tt MAG\_AUTO} measured by the single image photometry on the $y$ band images by SExtractor rather than the double-image mode with the NB973 and $y$ band images. This is because the latter's {\tt MAG\_AUTO} measurement does not always estimate total magnitudes accurately (i.e., does not agree with $y$ band total magnitudes estimated by multi-aperture photometry) while the former does for unblended objects. 8 out of 20 LAE candidates are detected by the SExtractor single image photometry on the $y$ band images. They do not blend with any other objects in the $y$ band images (confirmed by both {\tt FLAGS} $=0$ in the $y$-band images and visual inspection) although some of them blend with other objects in the NB973 images. We also confirmed that the {\tt MAG\_AUTO}'s of these unblended LAE candidates agree well with their total $y$-band magnitudes estimated by the multi-aperture photometry method. On the other hand, there are two other LAE candidates, NB973-SDF-137578 and NB973-SXDS-C-66280, which are visually faintly seen in the $y$-band images but undetected by the SExtractor $y$-band single image photometry. However, their $2''$ aperture $y$ band magnitudes measured by the SExtractor double image photometry with the NB973 and the $y$-band images are $>2\sigma$ significance ($2.3\sigma$ in NB973-SDF-137578 and $3.2\sigma$ in NB973-SXDS-C-66280). Hence, we applied the aperture correction of $-0.1$ mag to their $2''$ aperture $y$ band magnitudes to estimate their total $y$ band magnitudes. We obtained this correction factor by measuring the differences between total $y$-band magnitudes (either {\tt MAG\_AUTO} or multi-aperture photometry) and $2''$ aperture $y$-band magnitudes of the 8 unblended LAE candidates detected by the $y$-band single image photometry and taking their median value. %These edge regions coincide with the locations where most of the sources selected with the $z\sim7$ LAE candidate criteria (\ref{Criteria-1}) distribute. Hence, we trimmed these edge regions off the original NB973 images. Then, running SExtractor on these NB973 images as well as other waveband images with the same edge regions trimmed, we constructed the NB973-detected object catalogs again and applied the criteria (\ref{Criteria-1}) to them.

\subsection{Final Survey Area and Volume\label{Area_Volume}}
In the process of our LAE candidate selection, we masked blooming, smearing and halos of large bright stars, large galaxies and bad pixels as well as removed noise sources at the low S/N edge regions of the NB973 images (see the grey shaded regions of Figures \ref{SDF_SkyDist} and \ref{SXDS_SkyDist} for the cases of SDF and SXDS, respectively). As a result, total effective areas of the SDF and SXDS images used to select $z=7$ LAE candidates are 824 and 851 arcmin$^2$, respectively. The comoving distance along the line of sight corresponding to the redshift range $6.94 \leq z \leq 7.11$ for LAEs covered by NB973 filter is 58 Mpc. Therefore, we have probed comoving volumes of $\sim 3.0 \times 10^5$ Mpc$^3$ and $\sim 3.1 \times 10^5$ Mpc$^3$ in SDF and SXDS, respectively, for our $z\sim7$ LAE selection.
%To remove the noises in the process of the LAE candidate selection, we trimmed the low S/N edge regions off the NB973 image.

\subsection{Detection Completeness and LAE Number Counts\label{Completeness}}
What fraction of real objects in an image we can reliably detect by photometry depends on the magnitudes and blending of objects. The fraction usually decreases as magnitude becomes fainter due to difficulty in detecting fainter objects. Also, the detectability of target objects is affected by their blending with neighboring objects in projection. To examine what fraction of objects in the NB973 images of SDF and SXDS the SExtractor can detect or fails to detect to fainter magnitude, we measured the detection completeness of our photometry as it is used to correct the number of detected LAEs when we derive their number counts, Ly$\alpha$ LF, UV LF and Ly$\alpha$ EW distribution. 

%%figure 6
\begin{figure}
\epsscale{1.17}
%\plotone{SmallerSize_f6_AVE10_Blend_0p5NB973_Completeness_SDF_and_SXDS_ASSOC.eps}
\plotone{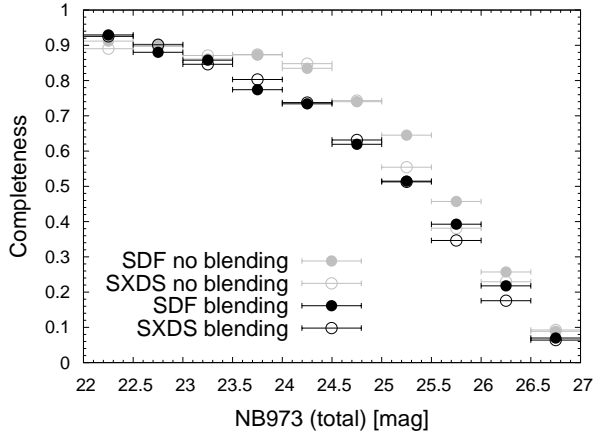}
\caption{Detection completeness of the NB973 images of SDF and SXDS per 0.5 mag bin (in total NB973 magnitude). Effects of blending of LAEs themselves and with other objects imaged in the fields were taken into account on the calculation of the completeness (black filled and open circles). For comparison, we also measure and plot the completeness without considering source blending by grey filled and open circles.\label{NB973Completeness}}
\end{figure}
%\vspace*{0.5cm}
%\vspace*{1cm}

Using the IRAF task {\tt starlist} and considering $z=7$ LAEs to be point sources, we first created $\sim 10,000$ artificial point sources with the same PSFs as the real objects and with random but uniform spatial and magnitude distributions, avoiding coordinates in the masked and low S/N edge regions in the NB973 images and ranging from 20 to 27 mag. We spread them over the NB973 images by using the IRAF task {\tt mkobject} allowing them to blend with themselves and real objects. Then the SExtractor was run for source detections in exactly the same way as our actual photometry. 

We extracted the input artificial objects from all the detected objects by the cross-identification based on their positions in the NB973 images. The artificial objects whose differences between their original magnitudes (those generated by IRAF {\tt starlist}) and measured magnitudes (SExtractor {\tt MAG\_AUTO}'s) are $> 0.5$ mag ($|m_{\rm AUTO} - m_{\rm org}| > 0.5$) tend to partly to completely blend with other objects or regions where sky backgrounds were locally oversubtracted, while those with the magnitude difference $\leq 0.5$ mag ($|m_{\rm AUTO} - m_{\rm org}| \leq 0.5$) are either isolated ($\gtrsim 96$\%) or slightly blend with other objects (only $\sim 2$--4\%), which can be detected and deblended by SExtractor. Hence, we define artificial objects with the magnitude difference of $\leq 0.5$ mag as detected and derive the completeness per 0.5 mag bin. 

Finally, we calculated the ratio of the number of the detected artificial objects (in every 0.5 mag bin in {\tt MAG\_AUTO}) to that of the created ones (in every 0.5 mag bin in {\tt starlist} magnitude) to obtain the detection completeness. We repeated this procedure ten times and averaged the obtained completeness for each of the SDF and SXDS NB973 images. The result is shown in Figure \ref{NB973Completeness}, which allows us to infer the actual number of $z=7$ LAEs from their detected number and measured total NB973 magnitudes. The completeness is $\sim 22$\% and $\sim 35$\% at our LAE detection limits of NB973 $={\rm NB973}_{4\sigma}$ in SDF and SXDS, respectively. 

To see how much effect the source blending has on the completeness, we also repeated the same completeness measurements but spreading artificial objects over the blank regions avoiding real objects in the NB973 images (we avoided the locations with distances shorter than 3/2 of FWHMs of real objects). The result is also shown in Figure \ref{NB973Completeness}. At the magnitude range of NB973 $=$ 24.0--26.5 that magnitudes of our $z=7$ LAE candidates span, the completenesses with source blending are worse by $\sim 4$\%--12\% than those without source blending. Hence, the effect of source blending is small and does not significantly affect the derivation of LAE number count and LFs.

\begin{figure*}
\epsscale{1.17}
\plottwo{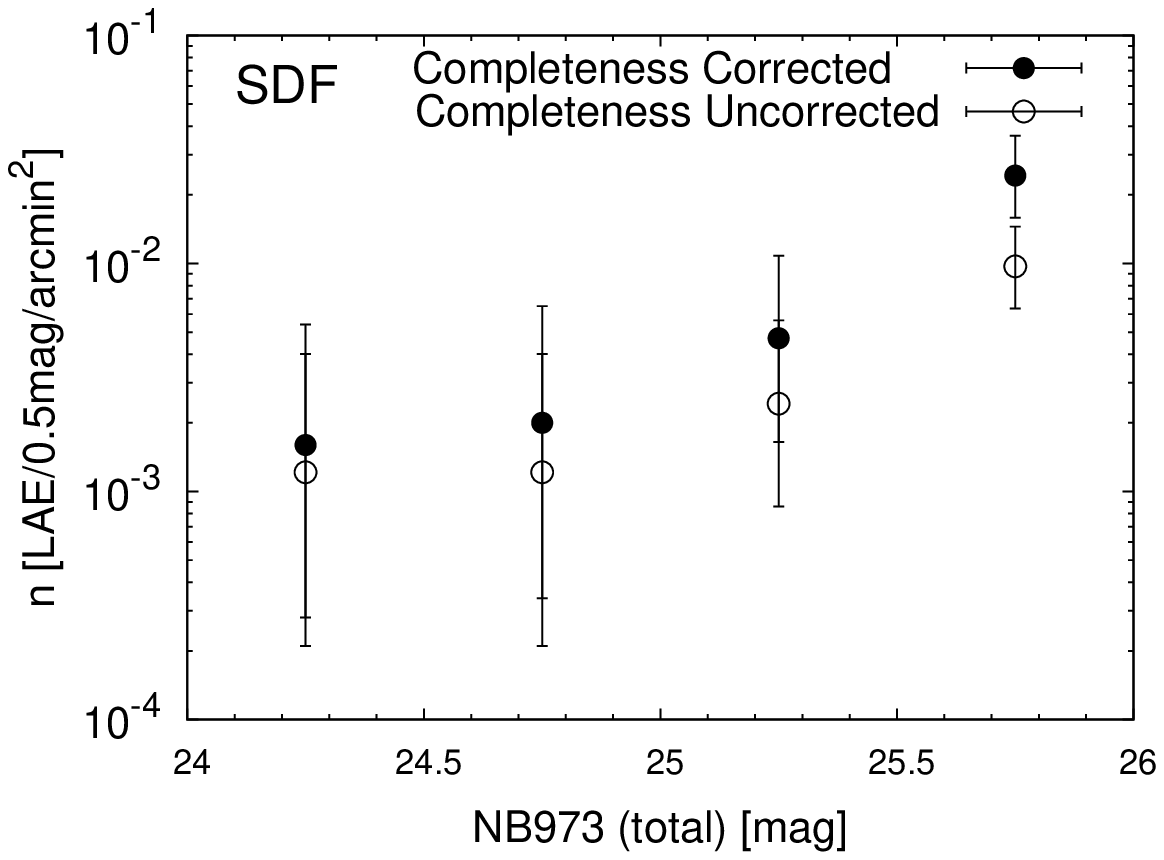}{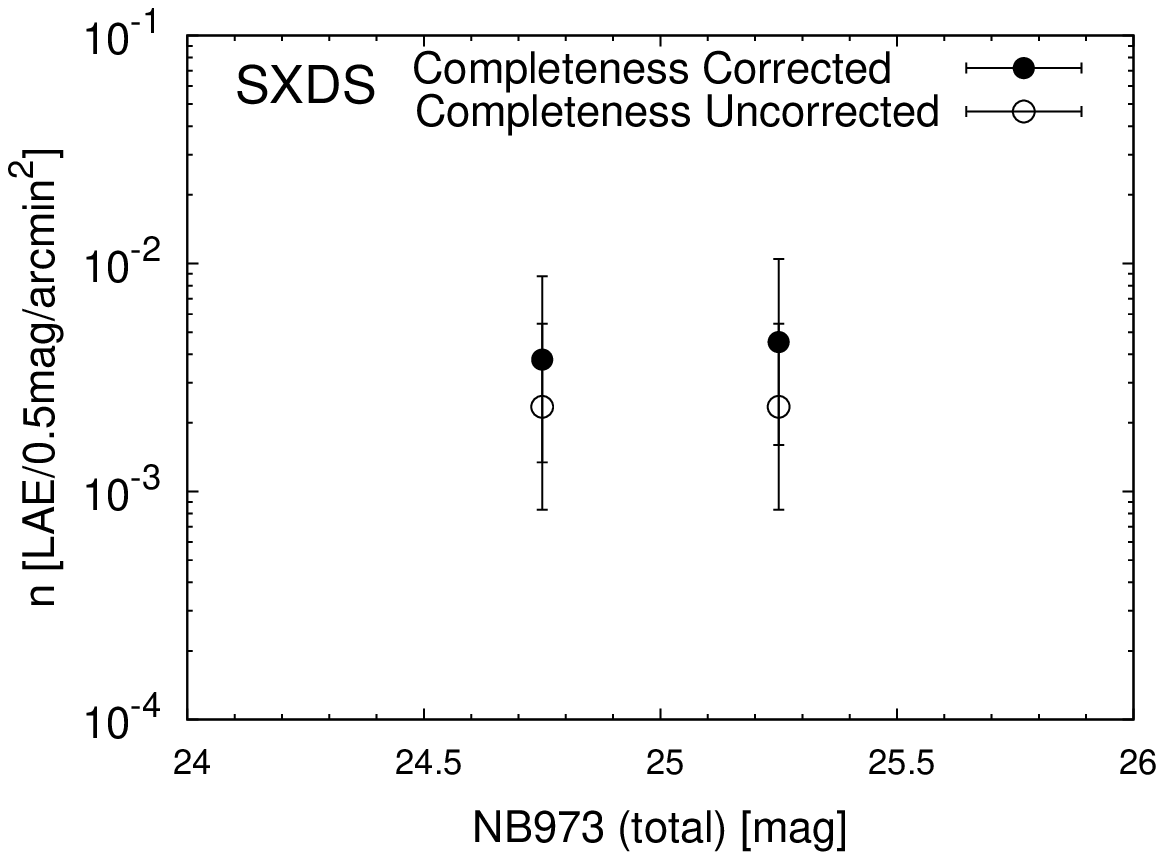}
\caption{The number counts of the $z=7$ LAE candidates in SDF and SXDS corrected (filled circles) and uncorrected (open circles) for detection completeness. The errors include Poisson errors for small number statistics quoted from \citet{Gehrels86}.\label{NC}}
\end{figure*}

In Figure \ref{NC}, we plot the number counts of the $z=7$ LAE candidates in SDF and SXDS corrected and uncorrected for detection completeness (we use the completeness allowing source blending for the correction). The detection completeness with source blending is also used to correct the Ly$\alpha$ and UV LFs as well as Ly$\alpha$ EW distribution of $z=7$ LAEs derived in Sections \ref{LyaLF}, \ref{UVLF} and \ref{subsectionLyaEWdist}.

%%figure 7
%\vspace*{0.5cm}
%\vspace*{1cm}
%log$L$(Ly$\alpha$) $=$ 43.2--43.4
%log$L$(Ly$\alpha$) $=$ 43.0--43.2  
%log$L$(Ly$\alpha$) $=$ 42.8--43.0
%log$L$(Ly$\alpha$) $=$ 42.6--42.8

\subsection{Estimating Ly$\alpha$ and UV Continuum Luminosities, Star Formation Rates and Survey Flux Limits\label{LyaUVLumSFRLimit}}
As the NB973 and $y$-band cover $z\sim7$ Ly$\alpha$ emission and the UV continuum redwards of it, we can estimate Ly$\alpha$ and UV continuum luminosities of the $z=7$ LAE candidates. Based on the NB973 and $y$-band total magnitudes of the $z=7$ LAE candidates, we photometrically derived their Ly$\alpha$ and UV continuum luminosities using the same method as the one \citet{Kashikawa11} used. The advantages of using their method for our study is as follows. \citet{Kashikawa11} used NB921 (NB816) and $z'$-band magnitudes of the $z\sim6.5$ ($z\sim5.7$) LAEs in their sample to photometrically estimate their Ly$\alpha$ and UV continuum luminosities as these bands cover $z\sim6.5$ ($z\sim5.7$) Ly$\alpha$ emission and UV continuum redwards of it. \citet{Kashikawa11} compared the Ly$\alpha$ line fluxes of 45 (54) spectroscopically identified $z\sim6.5$ ($z\sim5.7$) LAEs estimated photometrically from NB921 (NB816) and $z'$ with those measured from their spectra, and confirmed that they are in fairly good agreement (see Figures 5 and 6 in their paper). Because they are the largest spectroscopic $z\sim6.5$ and $z\sim5.7$ LAE samples ever constructed including LAEs with bright to faint Ly$\alpha$ luminosities, the validity of the method was statistically proved highly reliable. Also, the NB921 band is located in the middle of $z'$-band wavelength range while our NB973 band is analogously located in the middle of $y$-band wavelength range as seen in Figure \ref{FilterTransmission}. Hence, we can apply the same method to robustly and analogously estimate Ly$\alpha$ and rest frame UV continuum luminosities of $z=7$ LAEs. 

\citet{Kashikawa11} used the following formula to estimate the Ly$\alpha$ line flux ($f_{\rm line}$ in erg s$^{-1}$ cm$^{-2}$) and the rest frame UV continuum flux density ($f_c$ in erg s$^{-1}$ cm$^{-2}$ Hz$^{-1}$ in observer's frame) from narrowband (NB) and broadband (BB) magnitudes, $m_{\rm NB}$ and $m_{\rm BB}$.
\begin{equation}
m_{\rm NB,BB} + 48.6 = -2.5\log\frac{\int^{\nu_{{\rm Ly}\alpha}}_0 (f_c + f_{\rm line})T_{\rm NB,BB}d\nu/\nu}{\int T_{\rm NB,BB}d\nu/\nu} 
\label{Eqn_LyaUVLum}
\end{equation}
Here, $\nu_{{\rm Ly}\alpha}$ is the observed frequency of Ly$\alpha$, and $T_{\rm NB}$ and $T_{\rm BB}$ are the transmission bandpasses of the NB and BB filters as a function of observed frequency, respectively. We use the NB973 and $y$-band total magnitudes of each $z=7$ LAE candidate for $m_{\rm NB}$ and $m_{\rm BB}$. We adopt $\nu_{{\rm Ly}\alpha} = 3.08 \times 10^{14}$ Hz, the central frequency of the NB973 band, assuming the LAEs are at $z=7.02$. We use the NB973 and $y$-band response curves (see Figure \ref{FilterTransmission}) for $T_{\rm NB}$ and $T_{\rm BB}$, respectively. As in \citet{Kashikawa11}, we assume that SED of LAEs has a constant $f_c$ (i.e. flat continuum), $\delta$-function Ly$\alpha$ emission profile (i.e. flux value of $f_{\rm line}$ at $\nu_{{\rm Ly}\alpha}$ and 0 otherwise) and zero flux at the wavelength bluewards of Ly$\alpha$ due to the IGM absorption. If $f_{\rm line} \leq 0$, we set $f_{\rm line}$ to 0. Also, if an LAE candidate is not detected in $y$-band, we use $y$-band $1\sigma$ limiting magnitudes 27.40 (SDF) and 27.05 (SXDS) for $m_{\rm BB}$. If $f_c \leq 0$ (UV continuum is not detected), we estimate upper limit on $f_c$ directly from the $1\sigma$ $y$-band magnitudes. The estimated Ly$\alpha$ and rest frame UV continuum luminosities of the $z=7$ LAE candidates (converted from $f_{\rm line}$ and $f_c$ or $f_c$ limit) are listed in Table \ref{Propertyz7LAECandidates}. From these luminosities, we derive rest frame Ly$\alpha$ EWs, EW$_0$, and rest frame UV continuum apparent and absolute magnitudes, $m_{\rm cont}$ and $M_{\rm UV}$, and also list them in Table \ref{Propertyz7LAECandidates}.

We convert the Ly$\alpha$ line and UV continuum luminosities into star formation rates (SFRs), SFR(Ly$\alpha$) and SFR(UV). For SFR(Ly$\alpha$), we use the following relation derived from the Kennicutt equation \citep{Kennicutt98} with the case B recombination theory \citep{Brocklehurst71},
\begin{equation}
{\rm SFR}({\rm Ly}\alpha) = 9.1 \times 10^{-43} L({\rm Ly}\alpha) M_{\odot} {\rm yr}^{-1}.
\label{Eqn_SFR_Lya}
\end{equation}
For SFR(UV), we use the following relation \citep{Kennicutt98,Madau98},
\begin{equation}
{\rm SFR}({\rm UV}) = 1.4 \times 10^{-28} L_{\nu}({\rm UV}) M_{\odot} {\rm yr}^{-1}.
\label{Eqn_SFR_UV}
\end{equation}
These SFRs of the $z=7$ LAE candidates are also listed in Table \ref{Propertyz7LAECandidates}.
%As the NB973 and $y$-band cover $z\sim7$ Ly$\alpha$ emission and UV continuum redwards of it, we can estimate Ly$\alpha$ and UV continuum luminosities, $L_{{\rm Ly}\alpha}$ and $L_{\nu,{\rm UV}}$ (or UV continuum absolute magnitude $M_{\rm UV}$), of the $z=7$ LAE candidates.

Meanwhile, using the equation (\ref{Eqn_LyaUVLum}), we can translate our survey limiting magnitudes (NB973 $= 26.2$ and 25.8 at $4\sigma$ in SDF and SXDS, respectively) into Ly$\alpha$ line flux and luminosity limits, $f$(Ly$\alpha$)$_{\rm lim}$ and $L$(Ly$\alpha$)$_{\rm lim}$. To do this, we have to fix the rest frame Ly$\alpha$ EW to EW$_{0, {\rm lim}}$ that serves as the threshold for detecting $z=7$ LAEs because our LAE color criteria (\ref{Criteria-1}) can select objects down to EW$_0 =0$ and because $f$(Ly$\alpha$)$_{\rm lim}$ and $L$(Ly$\alpha$)$_{\rm lim}$ depend on $y$-band magnitude (the brighter the $y$-band magnitude is, the fainter these limits are). By fixing the Ly$\alpha$ EW, the $y$-band magnitude and $f_c$ are also correspondingly fixed. We compare the previous Subaru Suprime-Cam $z=5.7$--7.3 LAE narrowband surveys in Table \ref{LAESurveys}. Especially, the EW thresholds of $z=5.7$ and $z=6.6$ LAE surveys in SDF and SXDS are EW$_{0, {\rm lim}} =7$, 10, 14 and 27\AA~\citep{Taniguchi05,Shimasaku06,Ouchi08,Ouchi10}. Hence, we adopt a comparable limit of EW$_{0, {\rm lim}} = 10$\AA~above which an object selected by the color criteria (\ref{Criteria-1}) is considered a $z=7$ LAE candidate. 

%Table 3 (deluxetable)
\begin{turnpage}
\begin{deluxetable*}{lcccrrrrr}
\tabletypesize{\scriptsize}
%\rotate
\tablecaption{Properties of the $z=7$ Ly$\alpha$ Emitter Candidates Estimated from NB973 and $y$ Band Photometry\label{Propertyz7LAECandidates}}
\tablewidth{0pt}
%\tablewidth{510pt}
\tablehead{
\colhead{Object ID} & \colhead{$f({\rm Ly}\alpha$)$^{\rm a}$} & \colhead{$L({\rm Ly}\alpha$)} & \colhead{SFR(Ly$\alpha$)} & \colhead{EW$_0$$^{\rm b}$} & \colhead{$m_{\rm cont}$$^{\rm c}$} & \colhead{$M_{\rm UV}$$^{\rm d}$} & \colhead{$L_{\nu}$(UV)$^{\rm e}$} & \colhead{SFR(UV)}\\
 & \colhead{(10$^{-17}$ erg s$^{-1}$ cm$^{-2}$)} & \colhead{(10$^{42}$ erg s$^{-1}$)} & \colhead{($M_{\odot}$yr$^{-1}$)} & \colhead{(\AA)} & \colhead{(mag)} & \colhead{(mag)} & \colhead{(10$^{29}$ erg s$^{-1}$ Hz$^{-1}$)} & \colhead{($M_{\odot}$yr$^{-1}$)} 
}
\startdata
NB973-SDF-85821 (IOK-1)$^{\rm f}$  & 2.8	&	16	&	14.8	&	43	&	25.4	&     $-21.6$	&	1.8	&	25.1    \\
NB973-SDF-142122	 &	1.7	&	10	&	9.2	&	146.5	&	27.22	&     $-19.75$	&	0.34	&	4.8	\\
NB973-SDF-111394$^{\rm g}$ &	1.9	&	11	&	9.8	&    $>$184.4	&    $>$27.40	&   $>$$-19.56$	&    $<$0.29	&    $<$4.0     \\
NB973-SDF-110557	 &	0.80	&	4.7	&	4.2	&	17.5	&	25.75	&     $-21.22$	&	1.3	&	18.4	\\
NB973-SDF-181887$^{\rm g}$ &	1.6	&	9.0	&	8.2	&    $>$154.8	&    $>$27.40	&   $>$$-19.56$	&    $<$0.29	&    $<$4.0	\\
NB973-SDF-138055	 &	0.99	&	5.7	&	5.2	&	27.3	&	26.01	&     $-20.95$	&	1.0	&	14.5	\\
NB973-SDF-137578         &	0.76	&	4.4	&	4.0	&	30.8	&	26.42	&     $-20.55$	&       0.71	&	10.0	\\
NB973-SDF-42593$^{\rm g}$  &	1.2	&	6.8	&	6.2	&    $>$116.9	&    $>$27.40	&  $>$$-19.56$	&    $<$0.29	&    $<$4.0	\\
NB973-SDF-85079$^{\rm g}$  &	1.2	&	6.8	&	6.2	&    $>$116.9	&    $>$27.40	&  $>$$-19.56$	&    $<$0.29	&    $<$4.0	\\
NB973-SDF-30693$^{\rm g}$  &	0.94	&	5.4	&	4.9	&    $>$93.4	&    $>$27.40	&  $>$$-19.56$	&    $<$0.29	&    $<$4.0	\\
NB973-SDF-128564$^{\rm g}$ &	1.1	&	6.4	&	5.8	&    $>$110.3	&    $>$27.40	&  $>$$-19.56$	&    $<$0.29	&    $<$4.0	\\
NB973-SDF-120557$^{\rm g}$ &	0.85	&	4.9	&	4.5	&    $>$84.6	&    $>$27.40	&  $>$$-19.56$	&    $<$0.29	&    $<$4.0	\\
NB973-SDF-57651          &	0.85	&	4.9	&	4.5	&	94.6	&	27.52	&     $-19.44$	&	0.26	&	3.6	\\
NB973-SDF-141124$^{\rm g}$ &	1.1	&	6.1	&	5.6	&    $>$105.2	&    $>$27.40	&  $>$$-19.56$	&    $<$0.29	&    $<$4.0	\\
\hline
NB973-SXDS-S-95993	   &	2.4	&	14	&	12.7	&	138.7	&	26.81	&     $-20.15$	&	0.49	&	6.9	\\
NB973-SXDS-C-66280         &	1.9	&	11	&	9.9	&	47.3	&	25.91	&     $-21.05$	&       1.1	&	15.8	\\
NB973-SXDS-C-10143$^{\rm g}$ &	1.7	&	9.9	&	9.1	&    $>$123.5	&    $>$27.05	&  $>$$-19.92$	&    $<$0.40    &     $<$5.6	\\
NB973-SXDS-S-69635$^{\rm g}$ &	1.7	&	9.6	&	8.7	&    $>$119.3	&    $>$27.05	&  $>$$-19.92$	&    $<$0.40	&     $<$5.6	\\
NB973-SXDS-S-66036	   &	1.1	&	6.6	&	6.0	&	50.5	&	26.53	&     $-20.43$	&	0.64	&	9.0	\\
NB973-SXDS-S-67890$^{\rm g}$ &	1.3	&	7.4	&	6.7	&    $>$92.0	&    $>$27.05	&  $>$$-19.92$	&     $<$0.40	&     $<$5.6     \\
\hline
\multicolumn{9}{c}{Objects with $y-{\rm NB973}<0$ and an extremely faint or zero Ly$\alpha$ flux$^{\rm h}$}\\
\hline
NB973-SDF-109780	&	0.03	&	0.17	&	0.2	&	0.25	&	24.71	&     $-22.26$	&	3.4	&	48.0	\\
NB973-SDF-155934	&	0.17	&	0.97	&	0.9	&	1.34	&	24.66	&     $-22.30$	&	3.6	&	50.1	\\
NB973-SDF-101846	&	0.032	&	0.19	&	0.2	&	0.42	&	25.20	&     $-21.76$	&	2.2	&	30.5	\\
NB973-SDF-99670$^{\rm i}$ &       0.0     &       0.0	&       0.0	&	0.00	&	25.06	&     $-21.90$	&	2.5	&	34.7
\enddata
%% Text for table notes should follow after the \enddata but before
%% the \end{deluxetable}. Make sure there is at least one \tablenotemark
%% in the table for each \tablenotetext.
%\tablecomments{All the upper limits are $3\sigma$ at a resolution of $1\farcs5 \times 1\farcs2$ (See \textsection \ref{FIRcont_Prop} for details.)}
\tablecomments{}
\tablenotetext{a}{Ly$\alpha$ line flux, $f_{\rm line}$, calculated by the equation (\ref{Eqn_LyaUVLum}).}
\tablenotetext{b}{Rest frame Ly$\alpha$ equivalent width estimated from $f_{\rm line}$ and rest frame UV continuum flux density $f_c$ (observer's frame) calculated by the equation (\ref{Eqn_LyaUVLum}).}
\tablenotetext{c}{Apparent UV continuum magnitude calculated from $f_c$.}
\tablenotetext{d}{Absolute UV continuum magnitude calculated from $m_{\rm cont}$.}
\tablenotetext{e}{Rest frame UV continuum luminosity per unit frequency calculated from $f_c$.}
\tablenotetext{f}{This object is a $z=6.96$ LAE, IOK-1, previously spectroscopically confirmed by \citet{Iye06}, \citet{Ota08} and \citet{Ono12}. All the physical quantities listed here (except for $L_{\nu}$(UV) and SFR(UV)) of this LAE were taken from \citet{Ono12} who measured them based on their spectroscopy of this LAE by Keck Telescope DEIMOS. We converted $m_{\rm cont}$ to $L_{\nu}$(UV) and SFR(UV). These physical quantities of IOK-1 are also consistent with those in \citet{Jiang13} who especially measured the rest frame UV continuum magnitude using their deep {\it Hubble Space Telescope} Wide Field Camera 3 image of IOK-1.}
\tablenotetext{g}{These 11 objects exhibit $f_c <0$ and $y > y_{2\sigma}$ (6 of them show $y > y_{1\sigma}$). Only one (NB973-SDF-111394) is very faintly visually seen in the $y$ band image. Thus we consider them undetected in rest frame UV continuum whether they are visually seen or not in $y$ band. In the case of being visually seen in $y$ band, we consider all the $y$ band flux comes from Ly$\alpha$ emission alone. Hence, we use the $1\sigma$ magnitudes ($y = y_{1\sigma}$) of the SDF and SXDS $y$ band images to estimate the conservative upper limits on their rest frame UV continuum flux densities $f_c$ and then their limits on EW$_0$, $m_{\rm cont}$, $M_{\rm UV}$, $L_{\nu}$(UV) and SFR(UV).}
%Hence, their properties related to rest frame UV continuum cannot be inferred.
\tablenotetext{h}{These four objects have colors of $y-{\rm NB973}<0$ and turn out to exhibit extremely faint or zero Ly$\alpha$ fluxes ($f({\rm Ly}\alpha) \lesssim 1.7 \times 10^{-18}$ erg s$^{-1}$ cm$^{-2}$ and EW$_0 \lesssim 0.17$\AA) as a result of calculating $f_{\rm line}$ and $f_c$ by the equation (\ref{Eqn_LyaUVLum}). Hence, we consider them $z\sim7$ LBGs with a bright UV continuum and extremely faint or no Ly$\alpha$ emission and remove them from our $z=7$ LAE sample.}
\tablenotetext{i}{This object has $f_{\rm line} < 0$, and thus we consider it undetected in Ly$\alpha$ ($f_{\rm line}$ set to 0).}
\end{deluxetable*}
\end{turnpage}

Finally, fixing the NB973 magnitude to our survey limiting magnitude in SDF or SXDS and changing the $y$-band magnitude as a free parameter in the equation (\ref{Eqn_LyaUVLum}), we find the $y$-band magnitudes and thus $f_{\rm line}$'s and $f_c$'s that lead to EW$_0 =10$\AA. From these $f_{\rm line}$'s, we obtain $f$(Ly$\alpha$)$_{\rm lim} = 3.4 \times 10^{-18}$ and $4.7 \times 10^{-18}$ erg s$^{-1}$ cm$^{-2}$ ($4\sigma$) and $L$(Ly$\alpha$)$_{\rm lim} = 2.0 \times 10^{42}$ and $2.7 \times 10^{42}$ erg s$^{-1}$ ($4\sigma$) for SDF and SXDS, respectively. 
%In the case of $5\sigma$ limiting magnitudes (NB973 $= 25.9$ and 25.6 in SDF and SXDS), they are $f$(Ly$\alpha$)$_{\rm lim} = 4.3 \times 10^{-18}$ and $5.9 \times 10^{-18}$ erg s$^{-1}$ cm$^{-2}$, and $L$(Ly$\alpha$)$_{\rm lim} = 2.5 \times 10^{42}$ and $3.4 \times 10^{42}$ erg s$^{-1}$ for SDF and SXDS, respectively. 

For consistency check, we also estimate $f$(Ly$\alpha$)$_{\rm lim}$ and $L$(Ly$\alpha$)$_{\rm lim}$ from our NB973 limiting magnitudes in SDF or SXDS and EW$_{0, {\rm lim}} = 10$\AA~in the same method as the equations (5)--(8) in \citet{Ota12}. In this method, we assume that an NB973 flux $F_{\rm NB}$ comes from Ly$\alpha$ line and UV continuum fluxes, $F_{{\rm Ly}\alpha}$ and $F_{\rm cont}$.
\begin{equation}
F_{\rm NB}=F_{{\rm Ly}\alpha}+F_{\rm cont}
\label{FNB_FLya_Fcont}
\end{equation}
A Ly$\alpha$ flux is related to a UV continuum flux density $f_{\lambda,{\rm cont}}$ with an observed Ly$\alpha$ equivalent width EW$_{\rm obs}$. 
\begin{equation}
F_{{\rm Ly}\alpha} = {\rm EW}_{\rm obs}f_{\lambda,{\rm cont}}
\label{FLya_EW_Fcont}
\end{equation}
As the fluxes bluewards of Ly$\alpha$ is mostly absorbed by neutral IGM, the $f_{\lambda,{\rm cont}}$ is approximately the UV continuum flux divided by the wavelength from Ly$\alpha$ to the red edge of the NB973 bandpass ($\lambda_{\rm NB}^{\rm max} - \lambda_{{\rm Ly}\alpha}^{\rm obs}$). 
\begin{equation}
f_{\lambda,{\rm cont}} = F_{\rm cont}/(\lambda_{\rm NB}^{\rm max} - \lambda_{{\rm Ly}\alpha}^{\rm obs})
\label{Fcont_lambda}
\end{equation}
where $\lambda_{\rm NB}^{\rm max} = 9855$\AA, and we assume that Ly$\alpha$ is at $z=7.02$ (center of the NB973 bandpass), i.e. $\lambda_{{\rm Ly}\alpha}^{\rm obs} = (1+z)1216$\AA~= 9755\AA. Solving the equations (\ref{FNB_FLya_Fcont})--(\ref{Fcont_lambda}) for $F_{{\rm Ly}\alpha}$, we have 
\begin{equation}
F_{{\rm Ly}\alpha} = F_{\rm NB}/\{1+(\lambda_{\rm NB}^{\rm max} - \lambda_{{\rm Ly}\alpha}^{\rm obs})/{\rm EW}_{\rm obs}\}.
\label{FLya_FNB_lambda_EW}
\end{equation}
Substituting EW$_{\rm obs}=(1+z){\rm EW}_{0, {\rm lim}}=(1+z)10$\AA $=$ 80.2\AA~and $F_{\rm NB}=7.6 \times 10^{-18}$ and $1.1 \times 10^{-17}$ erg s$^{-1}$ cm$^{-2}$ (total NB973 fluxes corresponding to the limiting magnitudes NB973 $= 26.2$ and 25.8 at $4\sigma$ in SDF and SXDS) into this equation (\ref{FLya_FNB_lambda_EW}), we estimate our limiting Ly$\alpha$ fluxes to be $f$(Ly$\alpha$)$_{\rm lim} = F_{{\rm Ly}\alpha} = 3.4 \times 10^{-18}$ and $4.7 \times 10^{-18}$ erg s$^{-1}$ cm$^{-2}$. These values are consistent with (exactly same as) the $f$(Ly$\alpha$)$_{\rm lim}$'s estimated above using the equation (\ref{Eqn_LyaUVLum}).

We compare our survey limits with those of the previous deep Subaru Suprime-Cam $z=5.7$, 6.6 and 7.3 LAE narrowband surveys \citep{Taniguchi05,Shimasaku06,Kashikawa06,Kashikawa11,Ouchi08,Ouchi10,Konno14} in Table \ref{LAESurveys}. Our survey limits are comparable to those of the previous Subaru surveys and allow a fair comparison of the $z=7$ Ly$\alpha$ LF we derive with those at $z=5.7$, 6.6 and 7.3 from bright to faint end.

%Table 4 (deluxetable)
\begin{deluxetable*}{llccccccc}
\tabletypesize{\scriptsize}
%\rotate
\tablecaption{Comparison and Summary of the Subaru Suprime-Cam $z=5.7$--7.3 LAE Narrowband Surveys\label{LAESurveys}}
\tablewidth{0pt}
%tablewidth{510pt}
\tablehead{
Field & $z$$^{\rm a}$ & EW$_{0, {\rm lim}}$$^{\rm b}$ & $f$(Ly$\alpha$)$_{\rm lim}$$^{\rm c}$ & $L$(Ly$\alpha$)$_{\rm lim}$$^{\rm d}$ & Area         & Volume          & $N_{\rm LAE}$$^{\rm e}$ & References$^{\rm f}$\\
      &     & (\AA)                  & (10$^{-18}$erg s$^{-1}$ cm$^{-2}$)   & (10$^{42}$ erg s$^{-1}$)            & (arcmin$^2$) & (10$^5$ Mpc$^3$) &             &
}
\startdata
SDF    & 5.7  & 10         & 2.2$^{\rm i}$ ($5\sigma$) & 0.8$^{\rm i}$ & 725  & 1.8 & 54--93 & S06, K06, K11 \\
       & 6.56 & 7          & 4.5$^{\rm i}$ ($5\sigma$) & 2.2$^{\rm i}$ & 875  & 2.2 & 45--60 & T05, K06, K11 \\
       & 7.02 & 10$^{\rm g}$ & 3.4 ($4\sigma$)         & 2.0          & 824  & 3.0 & 14     & This study \\
       & 7.27 & 0$^{\rm h}$  & 9.1$^{\rm i}$ ($5\sigma$) & 5.7$^{\rm i}$ & 855  & 2.96 & 1     & S12 \\
\hline 
SXDS   & 5.7  & 27         & 8.4 ($5\sigma$)         & 3.0          & 3722 & 9.2 & 401    & Ou08 \\
       & 6.56 & 14         & 5.0 (3.6--$4.3\sigma$)  & 2.5          & 3238 & 8.0 & 207    & Ou10 \\
       & 7.02 & 10$^{\rm g}$ & 4.7 ($4\sigma$)         & 2.7          & 851  & 3.1 & 6      & This study \\
       & 7.27 & 0$^{\rm h}$  & 19$^{\rm i}$ ($5\sigma$)  & 12$^{\rm i}$   & 863 & 2.98 & 2     & S12 \\
       & 7.3  & 0$^{\rm h}$  & 6.5 ($5\sigma$)         & 4.1          & 790  & 1.2 & 3      & K14 \\
\hline
COSMOS & 7.3  & 0$^{\rm h}$  & 3.8 ($5\sigma$)         & 2.4          & 840  & 1.3 & 4      & K14
\enddata
%% Text for table notes should follow after the \enddata but before
%% the \end{deluxetable}. Make sure there is at least one \tablenotemark
%% in the table for each \tablenotetext.
%\tablecomments{All the upper limits are $3\sigma$ at a resolution of $1\farcs5 \times 1\farcs2$ (See \textsection \ref{FIRcont_Prop} for details.)}
\tablecomments{}
\tablenotetext{a}{Redshift of Ly$\alpha$ corresponding to the central wavelength of the narrowband filter used by each survey.}
\tablenotetext{b}{Rest frame Ly$\alpha$ equivalent width threshold adopted by each survey.}
\tablenotetext{c}{Ly$\alpha$ line flux limit and significance of the limiting narrowband magnitudes of each survey.}
\tablenotetext{d}{Ly$\alpha$ line luminosity limit of each survey.}
\tablenotetext{e}{The number of LAEs detected by each survey. For the cases of $z=5.7$ and $z=6.56$ surveys in SDF, LAEs serendipitously found by spectroscopy are also included.}
\tablenotetext{f}{The abbreviations for the references are S06: \citet{Shimasaku06}, K06: \citet{Kashikawa06}, K11: \citet{Kashikawa11}, T05: \citet{Taniguchi05}, S12: \citet{Shibuya12}, Ou08: \citet{Ouchi08}, Ou10: \citet{Ouchi10} and K14: \citet{Konno14}.}
\tablenotetext{g}{Our $z=7$ LAE color criteria (\ref{Criteria-1}) can select objects down to EW$_0 =0$. However, we adopt EW$_{0, {\rm lim}} =10$\AA~as the definition of LAEs. Using this EW, our limiting NB973 magnitudes and the equation (\ref{Eqn_LyaUVLum}), we estimate our $f$(Ly$\alpha$)$_{\rm lim}$ and $L$(Ly$\alpha$)$_{\rm lim}$ (see Section \ref{LyaUVLumSFRLimit} for the details).}
\tablenotetext{h}{These surveys' LAE color criteria select objects down to EW$_0=0$. We have taken the Ly$\alpha$ fluxes and luminosities of spectroscopically observed objects as $f$(Ly$\alpha$)$_{\rm lim}$ and $L$(Ly$\alpha$)$_{\rm lim}$ for S12 $z=7.27$ LAE survey. For the $z=7.3$ LAE survey, K14 used their $z'$ band and narrowband limiting magnitudes to estimate their $f$(Ly$\alpha$)$_{\rm lim}$ and $L$(Ly$\alpha$)$_{\rm lim}$.}
\tablenotetext{i}{These are the spectroscopically measured Ly$\alpha$ fluxes and luminosities of the faintest LAE in each survey.}
\end{deluxetable*}

%%figure 8
\begin{figure}
\epsscale{1.17}
\plotone{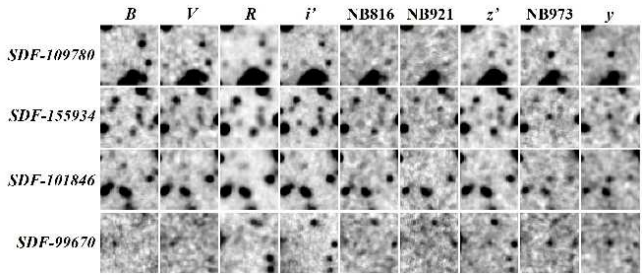}
\caption{Multi-waveband images of the four objects in SDF with a $y-{\rm NB973}<0$ color and an extremely faint or zero Ly$\alpha$ flux finally removed from our $z=7$ LAE sample (see Tables \ref{z7LAECandidates} and \ref{Propertyz7LAECandidates} and Section \ref{LyaFaintObjects}). They can be $z\simeq 7$ LBGs. The size of each image is $10'' \times 10''$. North is up and east to the left.\label{SDF_4zLBGs_BlackWhite}}
\end{figure}

\subsection{Final LAE Sample after Removing Objects with an Extremely Faint or Zero Ly$\alpha$ Flux\label{LyaFaintObjects}}
In Table \ref{Propertyz7LAECandidates}, we list Ly$\alpha$ and rest frame UV continuum properties of the 18 and 6 $z=7$ LAE candidates in SDF and SXDS we selected by using the color criteria (\ref{Criteria-1}). These properties were derived from their total NB973 and $y$-band magnitudes by using the equation (\ref{Eqn_LyaUVLum}). We found that 4 out of the 18 $z=7$ LAE candidates in SDF have extremely faint or zero Ly$\alpha$ fluxes, $f$(Ly$\alpha$) $=0$--$1.7 \times 10^{-18}$ erg s$^{-1}$ cm$^{-2}$, while no such object is found in SXDS. These objects have rest frame Ly$\alpha$ EWs of EW$_0=0$--1.3\AA, which are much lower than our EW threshold of EW$_{0, {\rm lim}}=10$\AA. On the other hand, they all have colors of $y-{\rm NB973} < 0$ (see Table \ref{z7LAECandidates}) and very bright rest frame UV continua (see Table \ref{Propertyz7LAECandidates}). 

In Section \ref{ColorSim}, we carefully examined possible $y-{\rm NB973}$, $z'-{\rm NB973}$ and $z'-y$ colors of LAEs, LBGs and potential contaminants (low-$z$ galaxies and M/L/T dwarfs) in Figure \ref{Color_vs_redshift}. We did not impose any $y-{\rm NB973}$ color criterion when selecting $z=7$ LAE candidates by using the color criteria (\ref{Criteria-1}) because Figure \ref{Color_vs_redshift} left panel shows that a $z\sim7$ LAE may possibly exhibit either $y-{\rm NB973} >0$ or $y-{\rm NB973} <0$ color. However, the equation (\ref{Eqn_LyaUVLum}) and Tables \ref{z7LAECandidates} and \ref{Propertyz7LAECandidates} suggest that objects with an extremely faint or zero Ly$\alpha$ flux and a very bright UV continuum would have $y-{\rm NB973} <0$ colors. Also, Figure \ref{Color_vs_redshift} suggests that the color criteria (\ref{Criteria-1}) could also select LBGs with no or very faint Ly$\alpha$ emission and T type dwarfs as contaminants. Moreover, Figure \ref{Color_vs_redshift} also indicates that T-type dwarf stars have colors of $y-{\rm NB973} <0$. However, stellarities of the four objects in question are not so high as seen in Table \ref{z7LAECandidates}. Hence, we consider them $z\simeq 7$ LBG candidates with a bright UV continuum and an extremely faint or no Ly$\alpha$ emission and remove them from our $z=7$ LAE sample (they are not noises as they are detected in $z'$, NB973 and $y$ bands; see Table \ref{z7LAECandidates} and Figure \ref{SDF_4zLBGs_BlackWhite}). Imposing this additional criterion $y-{\rm NB973} > 0$ reduces the contamination by LBGs and completely removes the T-type dwarfs from our $z=7$ LAE sample. Eventually, we are left with 14 and 6 $z=7$ LAE candidates in SDF and SXDS, which we consider our final sample (see Figures \ref{SDF_18z7LAEs_BlackWhite} and \ref{SXDS_z7LAEs_BlackWhite} for their images). 

\section{Result\label{Result}}
\subsection{Ly$\alpha$ Luminosity Function\label{LyaLF}}
\subsubsection{Ly$\alpha$ Luminosity Function of the $z=7.0$ LAE Candidates\label{LyaLF_z7LAEs}}
\begin{figure*}
\includegraphics[angle=0,scale=0.69]{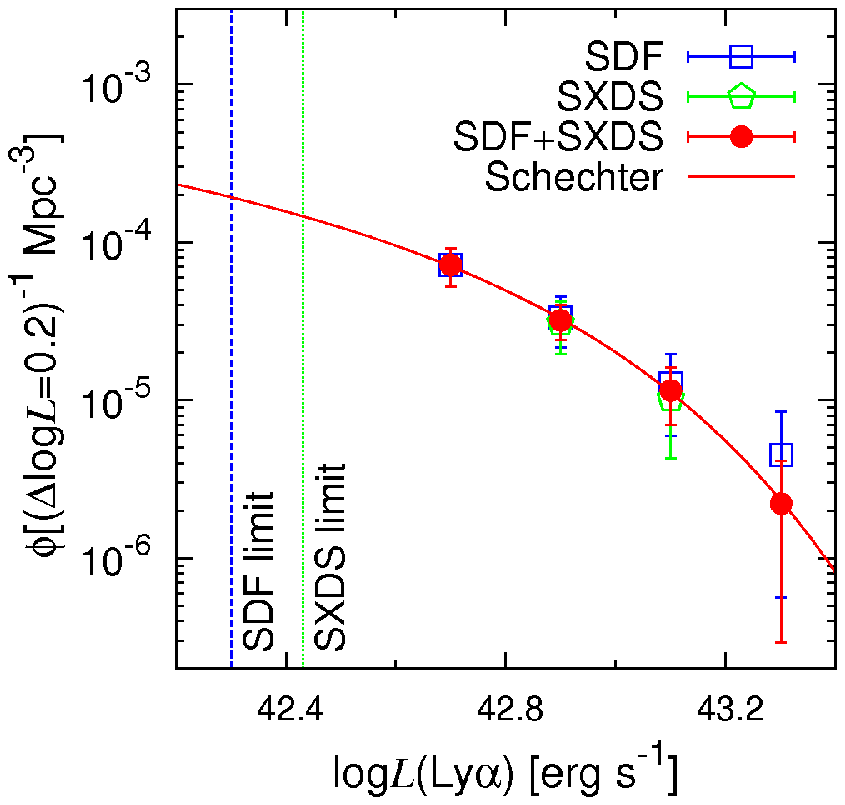} 
\includegraphics[angle=0,scale=0.69]{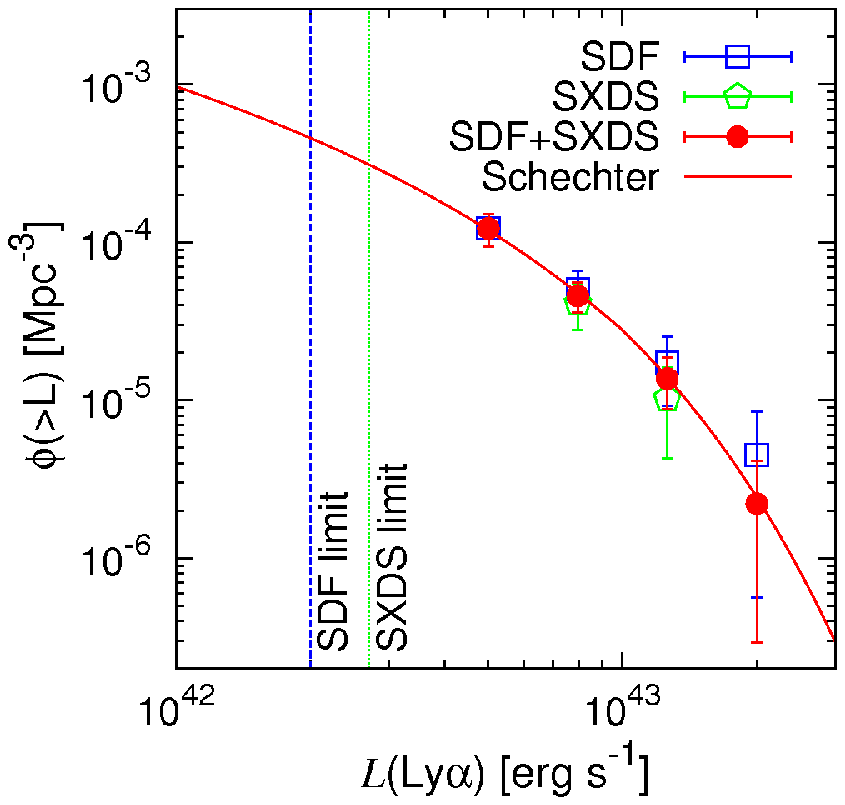}
\includegraphics[angle=0,scale=0.69]{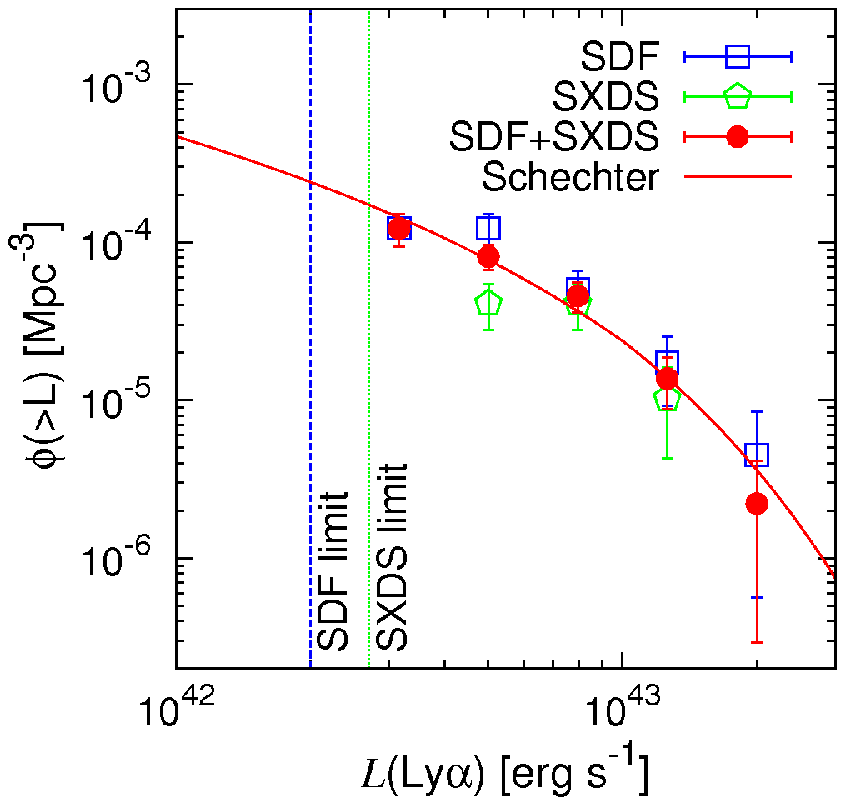}
\caption{The Ly$\alpha$ LFs of $z=7$ LAEs in SDF (open square) and SXDS (open pentagon). The total Ly$\alpha$ LF derived by combining the SDF and SXDS LAE samples is shown by filled circle. The error includes both Poisson errors and cosmic variance. The solid line is the Schechter function best-fitted to the total Ly$\alpha$ LF. The vertical dashed and dotted lines show our survey limits in SDF and SXDS, respectively. (Left) The differential Ly$\alpha$ LFs. (Middle) The cumulative Ly$\alpha$ LFs that exclude the faintest bin (within the survey limit) of each of the SDF and SXDS LFs in which no LAE is detected. (Right) The cumulative Ly$\alpha$ LFs that include the faintest bins in which no LAE is detected.\label{plottingLyaLF}}
\end{figure*}

With Ly$\alpha$ luminosities of the $z=7$ LAE candidates estimated in Section \ref{LyaUVLumSFRLimit}, we derive their Ly$\alpha$ LFs in SDF and SXDS to our survey limits and show them in Figure \ref{plottingLyaLF}. We estimate the number density of LAEs with the so-called $"$classical method$"$ by simply dividing the observed differential or cumulative number of LAEs in each Ly$\alpha$ luminosity bin by the SDF and SXDS effective survey volumes calculated in Section \ref{Area_Volume} by multiplying the FWHM of the NB973 filter by the survey areas. We have also assumed that each LAE has Ly$\alpha$ emission located at the center of the NB973 filter bandpass and a flat UV continuum when calculating its Ly$\alpha$ luminosity using the equation (\ref{Eqn_LyaUVLum}) in Section \ref{LyaUVLumSFRLimit}. The classical method has been widely used to derive observed Ly$\alpha$ LFs by many previous studies including \citet{Ouchi08,Ouchi10}, \citet{Kashikawa06,Kashikawa11} and \citet{Konno14} that derived the Ly$\alpha$ LFs of $z=5.7$, 6.6 and 7.3 LAEs detected by the Suprime-Cam narrowbands NB816, NB921 and NB101, respectively. As we compare our $z=7$ Ly$\alpha$ LF to their $z=5.7$, 6.6 and 7.3 ones below, we also use the classical method to derive the $z=7$ Ly$\alpha$ LF for consistency and fair comparison.

As pointed out by the previous LAE studies \citep{Shimasaku06,Ouchi08,Ouchi10,Konno14}, the classical method has two possible uncertainties in deriving a Ly$\alpha$ LF. (1) Ly$\alpha$ luminosities of LAEs having the same narrowband magnitude vary with redshift within a narrowband bandpass. Hence, Ly$\alpha$ luminosities of some LAEs could be over/underestimated. (2) Redshift distribution of LAEs within a narrowband bandpass depends on the Ly$\alpha$ EW distribution of LAEs. \cite{Shimasaku06} and \citet{Ouchi08}, who detected and studied $z=3.1$ and 3.7 LAEs and/or $z=5.7$ LAEs with Suprime-Cam narrowband filters, investigated these uncertainties with Monte Carlo simulations. They created mock catalogs of LAEs using a set of the Schechter function parameters ($\phi^*$, $L^*$, $\alpha$) and the dispersion of Ly$\alpha$ EW distribution that they assumed is Gaussian. Then, they uniformly distributed these LAEs in comoving volumes over the redshift ranges corresponding to their narrowband filter bandpasses. They $"$observed$"$ these LAEs in their narrowbands and broadbands to be the same as real band response. Then, they selected LAEs by the same criteria as the ones used to select real LAEs and derived their number densities and color distributions. They compared these results with the observed number densities and color distributions. Conducting this simulation over a wide range of the parameters, they derived the best-fit Schechter parameters with $\chi^2$ minimization, which are the Ly$\alpha$ LFs with no bias caused by the uncertainties (1) and (2). They compared these LFs with the Ly$\alpha$ LFs derived by the classical method and found that they are consistent (see Figure 11 of \cite{Shimasaku06} and Figures 16--18 of \citet{Ouchi08}). They concluded that the uncertainties (1) and (2) are negligible and/or cancel out each other and that the classical method is a good estimate of a Ly$\alpha$ LF. Because of this reason, \citet{Ouchi10} and \citet{Konno14} also used the classical method to derive their $z=6.6$ and $z=7.3$ Ly$\alpha$ LFs, and so do we.

For the error of the number density in each bin of the $z=7$ Ly$\alpha$ LFs we derived in Figure \ref{plottingLyaLF}, we include Poisson errors for small number statistics and cosmic variance estimated in the same way as in \citet{Ota08,Ota10}. Namely, we use column 2 in Tables 1 and 2 in \citet{Gehrels86} for the Poisson errors. For the cosmic variance $\sigma_v$ estimate, \citet{Ota08,Ota10} used the relation, $\sigma_v = b \sigma_{\rm DM}$, adopting a bias parameter of $b=3.4\pm1.8$ derived from the sample of 515 $z\sim5.7$ LAEs detected by \citet{Ouchi05} in the entire SXDS field and the dark matter variance $\sigma_{\rm DM}=0.044$ at $z=7$ obtained using the analytic cold dark matter model \citep{Sheth99,MoWhite02} and their survey volumes in SDF and SXDS. In this study, we use a bias parameter of $b=3.6\pm0.7$ derived more recently from the sample of 207 $z\sim6.6$ LAEs detected by \citet{Ouchi10} in the entire SXDS field. As our SDF and SXDS survey volumes are almost the same as those of \citet{Ota08,Ota10}, we adopt the same $\sigma_{\rm DM}$ value as they used. This gives the cosmic variance of $\sigma_v \sim 0.16$ for each of SDF and SXDS. Also, we correct the number density and the error for the detection completeness estimated in Section \ref{Completeness} and shown in Figure \ref{NB973Completeness} by number weighting according to the NB973 magnitude. 

In the similar way, we also derive the $z=7$ Ly$\alpha$ LF from the total sample combining the $z=7$ LAE candidates in SDF and SXDS and show it in Figure \ref{plottingLyaLF}. The cosmic variance included in the error of this total Ly$\alpha$ LF is estimated to be $\sigma_{\rm tot} \sim 0.11$ by combining the cosmic variances of SDF and SXDS as a volume weighted average as follows \citep{Moster11}.
\begin{equation}
\sigma_{\rm tot}^2 = \frac{\Sigma_i (\alpha_1 \alpha_2)_i^2 \sigma_i^2}{\left[\Sigma_i (\alpha_1 \alpha_2)_i\right]^2} 
\end{equation}
where $\alpha_1$ and $\alpha_2$ are the angular dimensions of each survey field.

%%figure 9

The left panel in Figure \ref{plottingLyaLF} shows the differential Ly$\alpha$ LF of the $z=7$ LAE candidates and our survey limits in SDF and SXDS, $\log L({\rm Ly}\alpha)_{\rm lim}$ (erg s$^{-1}$) $=$ 42.30 and 42.43, respectively. We notice that we do not detect any $z=7$ LAE candidates at the Ly$\alpha$ luminosity ranges of $\log L({\rm Ly}\alpha)$ (erg s$^{-1}$) = 42.3--42.6 in SDF and 42.43--42.8 in SXDS although these luminosity ranges are close to but still within our survey limits. There are two possibilities. We underestimate our limiting Ly$\alpha$ luminosities when converted from the NB973 limiting magnitudes, and actual sensitivities are somewhat shallower than our estimates, even though the survey limits estimated by our two independent methods agreed (see Section \ref{LyaUVLumSFRLimit}). Another possibility is that we do not actually detect any LAEs at these luminosity ranges even though we have really reached the very deep limits. For example, \citet{Matthee15} modeled evolution of a Ly$\alpha$ LF with neutral IGM and suggested that the faint end of the Ly$\alpha$ LF could be suppressed as Ly$\alpha$ emissions of fainter LAEs are more preferentially suppressed by neutral IGM. We cannot distinguish between these two possibilities from the current data alone. Hence, we derive two different cumulative Ly$\alpha$ LFs of the $z=7$ LAE candidates: (1) the LF excluding the faintest Ly$\alpha$ luminosity bins within our survey limits where no LAE candidate is detected and (2) the LF including these faintest bins. They are presented in the middle and the right panels of Figure \ref{plottingLyaLF}, respectively.
 
To investigate the derived $z=7$ Ly$\alpha$ LFs in more details, we fit the Schechter function \citep{Schechter76}
\begin{equation}
\phi(L)dL = \phi^*\left(\frac{L}{L^*}\right)^{\alpha}\exp\left(\frac{-L}{L^*}\right)d\left(\frac{L}{L^*}\right)
\end{equation}
to each of the three different total $z=7$ Ly$\alpha$ LFs in the three panels of Figure \ref{plottingLyaLF} by treating the characteristic luminosity $L^*$ and the normalization $\phi^*$ as free parameters, fixing the faint end slope to $\alpha=-1.5$ (to facilitate the comparison to previous studies) and minimizing $\chi^2$. In Figure \ref{plottingLyaLF} and Table \ref{Best-fitSchechter}, we show the best-fit Schechter functions and the best-fit parameters. The best-fit $L^*$ and $\phi^*$ values are slightly different among the three $z=7$ Ly$\alpha$ LFs but consistent with each other within the fitting errors. Our SDF and SXDS survey limits reach 0.22--0.36 $L^*$ and 0.29--0.49 $L^*$, respectively, and probe the $z=7$ Ly$\alpha$ LFs to the very faint end.

%Table 5 
\begin{deluxetable*}{llccccc}
\tabletypesize{\scriptsize}
%\rotate
\tablecaption{Best-fit Schechter Parameters for the Ly$\alpha$ LFs and the Number and Ly$\alpha$ Luminosity Densities of LAEs at $z=5.7$, 6.6, 7 and 7.3\label{Best-fitSchechter}}
%\tablewidth{0pt}
\tablewidth{530pt}
\tablehead{
$z$ & Reference/Ly$\alpha$ LF & $L^*_{{\rm Ly}\alpha}$ & $\phi^*$ & $n_{{\rm Ly}\alpha}^{\rm obs}$$^{\rm d}$ & $\rho_{{\rm Ly}\alpha}^{\rm obs}$$^{\rm d}$ & $\rho_{{\rm Ly}\alpha}^{\rm tot}$$^{\rm e}$ \\
     &                                             & (10$^{42}$ erg s$^{-1}$) & (10$^{-4}$ Mpc$^{-3}$)  & (10$^{-4}$ Mpc$^{-3}$)              & (10$^{39}$ erg s$^{-1}$ Mpc$^{-3}$)         & (10$^{39}$ erg s$^{-1}$ Mpc$^{-3}$) 
}
\startdata
5.7 & \citet{Ouchi08}                              & 6.8$_{-2.1}^{+3.0}$       & 7.7$_{-3.9}^{+7.4}$      & 6.8$_{-3.1}^{+5.2}$$^{\rm f}$          & 3.6$_{-1.7}^{+3.1}$$^{\rm f}$                & 9.2$_{-3.7}^{+6.6}$$^{\rm f}$\\
5.7 & \citet{Kashikawa11}                          & 10.5$_{-1.4}^{+1.6}$      & 2.8$_{-0.6}^{+0.6}$      & 4.1$_{-0.8}^{+0.8}$                  & 2.4$_{-0.5}^{+0.5}$                         & 5.0$_{-1.0}^{+1.0}$\\
\hline
6.6 & \citet{Ouchi10}                              & 4.4$_{-0.6}^{+0.6}$       & 8.5$_{-2.2}^{+3.0}$      & 4.1$_{-0.8}^{+0.9}$$^{\rm f}$          & 1.9$_{-0.4}^{+0.5}$$^{\rm f}$                & 6.6$_{-0.8}^{+1.0}$$^{\rm f}$\\
6.6 & \citet{Kashikawa11}                          & 5.8$_{-1.2}^{+1.5}$       & 5.2$_{-1.9}^{+3.1}$      & 3.7$_{-1.4}^{+2.2}$                  & 1.9$_{-0.7}^{+1.1}$                         & 5.2$_{-1.9}^{+3.1}$ \\
\hline
7.0 & Differential LF$^{\rm a}$             & 5.5$_{-2.0}^{+0.6}$      & 3.6$_{-2.0}^{+1.3}$      & 2.4$_{-1.4}^{+0.8}$     & 1.2$_{-0.7}^{+0.4}$                         & 3.5$_{-2.0}^{+1.2}$\\
7.0 & Cum LF excl.~0 bins$^{\rm b}$  & 6.3$_{-1.9}^{+2.6}$      & 4.1$_{-2.3}^{+6.4}$      & 3.3$_{-1.8}^{+5.2}$     & 1.7$_{-0.9}^{+2.6}$                         & 4.5$_{-2.5}^{+7.0}$\\
7.0 & Cum LF incl.~0 bins$ ^{\rm c}$ & 9.3$_{-3.0}^{+4.2}$      & 1.4$_{-0.7}^{+1.4}$      & 1.8$_{-0.9}^{+1.8}$     & 1.1$_{-0.5}^{+1.1}$                         & 2.3$_{-1.1}^{+2.3}$\\
\hline
7.3 & \citet{Konno14}                              & 2.7$_{-1.2}^{+8.0}$       & 3.7$_{-3.3}^{+17.6}$     & 0.76$_{-0.68}^{+4.8}$                &  0.31$_{-0.12}^{+0.19}$$^{\rm f}$              & 1.8$_{-1.1}^{+3.8}$$^{\rm f}$
\enddata
%% Text for table notes should follow after the \enddata but before
%% the \end{deluxetable}. Make sure there is at least one \tablenotemark
%% in the table for each \tablenotetext.
\tablecomments{All the best-fit parameters ($L^*_{{\rm Ly}\alpha}$ and $\phi^*$) were obtained by each author or us by fixing the faint end slope to $\alpha=-1.5$.}
\tablenotetext{a}{The Schechter function was fitted to the total differential Ly$\alpha$ LF of $z=7$ LAEs. This is the LF shown by the red line in the left panel of Figure \ref{plottingLyaLF}.}
\tablenotetext{b}{The Schechter function was fitted to the total cumulative Ly$\alpha$ LF of $z=7$ LAEs that excludes the faintest SDF and SXDS bins in which no LAE is detected. This is the LF shown by the red line in the middle panel of Figure \ref{plottingLyaLF}.}
\tablenotetext{c}{The Schechter function was fitted to the total cumulative Ly$\alpha$ LF of $z=7$ LAEs that includes the faintest SDF and SXDS bins in which no LAE is detected. This is the LF shown by the red line in the right panel of Figure \ref{plottingLyaLF}.}
\tablenotetext{d}{Number and Ly$\alpha$ luminosity densities of LAEs obtained by integrating each best-fit Schechter function down to the observed Ly$\alpha$ luminosity limit $\log L({\rm Ly}\alpha)$ (erg s$^{-1}$) $=$ 42.4.}
\tablenotetext{e}{Total Ly$\alpha$ luminosity densities of LAEs obtained by integrating each best-fit Schechter function down to $L({\rm Ly}\alpha)=0$.}
\tablenotetext{f}{These densities were calculated by the authors of the corresponding references in column 2. All the other densities listed in this table are calculated by us.}
\end{deluxetable*}

\subsubsection{Comparison with the $z=5.7$, 6.6 and 7.3 Ly$\alpha$ LFs Mostly Based on Photometric LAE Candidates\label{LyaLF_Comp1}}
\begin{figure}
\epsscale{1.17}
%\plotone{Cropped_SDF_SXDS_z7LAE_LyaLF_Differential_PoissonSqrt_noZeroLAEpoints_z5p7_z6p6_z7p3_LyaLFs.eps}
%\plotone{Cropped_SDF_SXDS_z7LAE_LyaLF_Differential_PoissonSqrt_noZeroLAEpoints_z5p7_z6p6_z7p3_LyaLFs_3LAECorr.eps}
\plotone{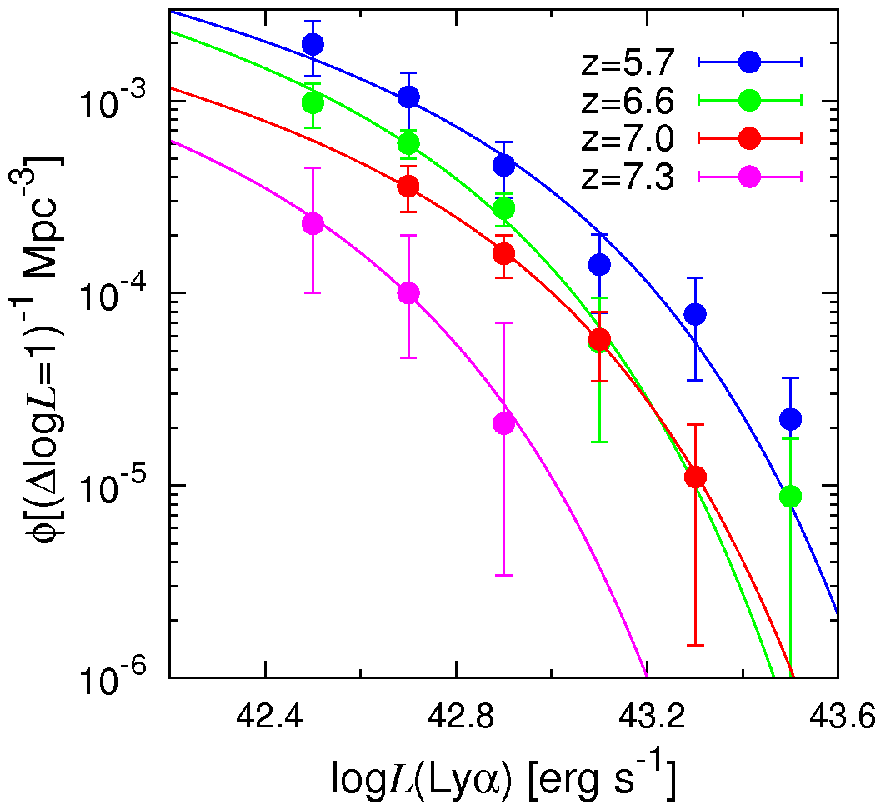}
\caption{Comparison of the differential Ly$\alpha$ LFs of LAEs at $z=5.7$, 6.6, 7.0 and 7.3 (color coded circles) and their best-fit Schechter functions (color coded curves). The $z=5.7$, 6.6 and 7.3 LFs are taken from \citet{Ouchi08}, \citet{Ouchi10} and \citet{Konno14}, respectively. The errors include Poisson errors and cosmic variance. These four LFs are mostly based on the photometric LAE candidates with some or no fractions confirmed by spectroscopy.\label{LyaLFs_at_z5p7-7p3}}
\end{figure}

In Figure \ref{LyaLFs_at_z5p7-7p3}, we plot the total differential $z=7$ Ly$\alpha$ LF and the best-fit Schechter function with $\alpha=-1.5$ and compare them with the differential Ly$\alpha$ LFs at $z=5.7$, 6.6 and 7.3 and their best-fit Schechter functions with $\alpha=-1.5$ derived by the previous Subaru LAE surveys \citep{Ouchi08,Ouchi10,Konno14}. The $z=7$ Ly$\alpha$ LF entirely shows a significant deficit from the $z=5.7$ LF from the bright to faint end beyond the errors including Poisson errors and cosmic variance. The $z=7$ Ly$\alpha$ LF also exhibits a significant deficit from the $z=6.6$ LF at the fainter end beyond the errors, but these LFs are consistent at the bright end. Moreover, the $z=7.3$ LF entirely displays a considerable deficit from the $z=7$ LF beyond the errors, which appears to be more significant than the deficits between $z=5.7$ and 6.6 and between $z=6.6$ and 7. This might support the accelerated evolution of the Ly$\alpha$ LF suggested by \citet{Konno14}.

%%figure 10

%%figure 11
\begin{figure}
\epsscale{1.17}
\plotone{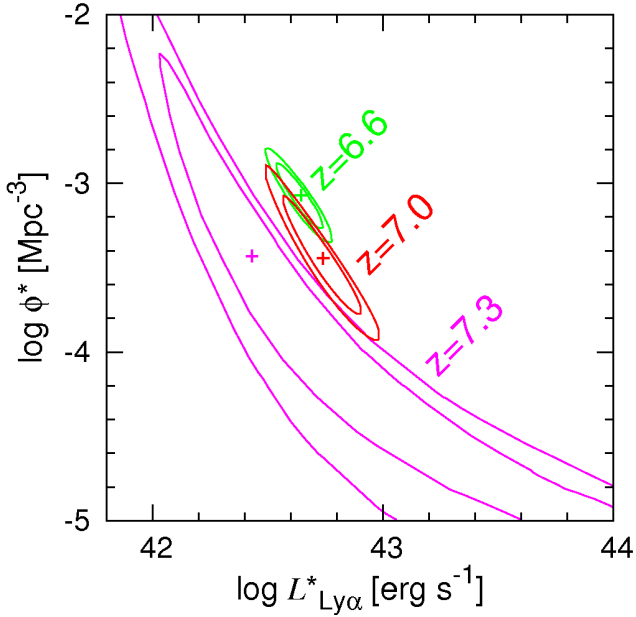}
\caption{Error contours of the $\phi^*$ and $L^*$ parameters of the Schechter functions fitted to the Ly$\alpha$ LFs at $z=6.6$, 7 and 7.3 shown in Figure \ref{LyaLFs_at_z5p7-7p3}. The plus symbols, inner and outer curves are the best-fit $\phi^*$ and $L^*$ values, the 68\% and 90\% confidence level contours. The $z=6.6$ and $z=7.3$ data are taken from \citet{Ouchi10} and \citet{Konno14}.\label{Phi_vs_L_z5p7-7p3}}
\end{figure}

To examine degree of the difference or the evolution of the Ly$\alpha$ LF among $z=6.6$, 7 and 7.3 more quantitatively, we derive the error contours of the $L^*$ and $\phi^*$ of the Schechter function fitted to the $z=7$ differential Ly$\alpha$ LF and compare them to those of $z=6.6$ and 7.3 Ly$\alpha$ LFs derived by \citet{Ouchi10} and \citet{Konno14} in Figure \ref{Phi_vs_L_z5p7-7p3}. The $z=7$ Ly$\alpha$ LF is different from the $z=7.3$ one at 90\% confidence level while the $z=7$ LF is different from the $z=6.6$ one less significantly but at 68\% confidence level. This suggests that the Ly$\alpha$ LF evolves modestly from $z=6.6$ to 7 and more rapidly from $z=7$ to 7.3 as implied from Figure \ref{LyaLFs_at_z5p7-7p3}. There are three possible explanations for the apparent non-evolution of the Ly$\alpha$ LF between $z=6.6$ and $z=7$ at the bright end. (1) Field-to-field variance: As our $z=7$ LAE sample consists of the LAEs from the sky area (two Suprime-Cam pointings) smaller than those of the $z=6.6$ LAE sample (6 Suprime-Cam pointings), the bright end of the $z=7$ Ly$\alpha$ LF might suffer some degree of field-to-field variance in the detected number of LAEs. (2) Completeness: Our estimate of the NB973 detection completeness considers an object detected by SExtractor with change in magnitude more than 0.5 mag due to noise and/or object blending to be a non-detection while \citet{Ouchi10}'s estimate of the NB921 ($z=6.6$ LAE) detection completeness does not. Hence, our completeness estimate for $z=7$ LAE detections is stricter than that for $z=6.6$ LAE detections, resulting in the relatively larger correction of the detected number of $z=7$ LAEs than that of $z=6.6$ LAEs. (3) Located in ionized bubbles: The LAEs in the two brightest Ly$\alpha$ luminosity bins of the $z=7$ Ly$\alpha$ LF are themselves UV-bright LAEs detected in the UV continuum and/or have a $z\sim7$ LBG candidate as an immediate neighbor. As such LAEs and LBGs would be stronger ionizing sources, their surroundings could be largely ionized, allowing higher transmission of Ly$\alpha$ photons (see Section \ref{SkyDist} and Figures \ref{SDF_SkyDist} and \ref{SXDS_SkyDist} for more details).

\subsubsection{Comparison with the $z=5.7$ and 6.6 Ly$\alpha$ LFs Mostly Based on Spectroscopically Confirmed LAEs\label{LyaLF_Comp2}}
\begin{figure}
\epsscale{1.17}
%\plotone{Cropped_Upper_Lower_SDF_SXDS_z7LAE_LyaLF_Cumulative_PoissonSqrt_z5p7_z6p6.eps}
%\plotone{Cropped_Upper_Lower_SDF_SXDS_z7LAE_LyaLF_Cumulative_PoissonSqrt_z5p7_z6p6_3LAECorr.eps}
\plotone{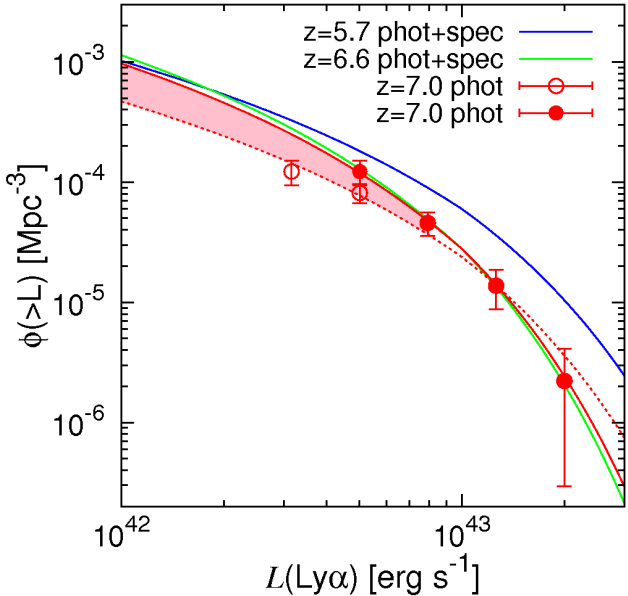}
\caption{Comparison of the cumulative Ly$\alpha$ LFs of LAEs at $z=5.7$, 6.6 and 7. The blue and green solid curves are the Schechter functions best-fitted to the $z=5.7$ and 6.6 Ly$\alpha$ LFs derived by \citet{Kashikawa11}, respectively. The filled circles and the red solid curve are the $z=7$ Ly$\alpha$ LF excluding the faintest bins where no LAE is detected and its best-fit Schechter function taken from the middle panel of Figure \ref{plottingLyaLF}. The open circles and the red dashed curve are the $z=7$ Ly$\alpha$ LF including the faintest bins where no LAE is detected and its best-fit Schechter function taken from the right panel of Figure \ref{plottingLyaLF}. The errors include Poisson errors and cosmic variance. The pink shaded region shows the difference between the two $z=7$ Ly$\alpha$ LFs at their faint ends which can be considered the possible range of the $z=7$ Ly$\alpha$ LF at faint end.\label{LyaLFs_at_z5p7-7_in_SDF}}
\end{figure}

The $z=5.7$ and 6.6 Ly$\alpha$ LFs compared in Figure \ref{LyaLFs_at_z5p7-7p3} are mostly based on the photometric LAE samples with some fraction confirmed by spectroscopy despite their large sample sizes drawn from the large sky area of the entire SXDS plus SDF by \citet{Ouchi08,Ouchi10}. While these LFs have large advantage in terms of very robust statistics, they might include some degree of contamination. Meanwhile, although limited to a smaller sky area of only SDF, \citet{Kashikawa11} carried out extensive spectroscopy campaigns of photometric $z=5.7$ and 6.6 LAE candidates and identified 54 (45) real $z=5.7$ (6.6) LAEs, which are 70\% (81\%) of their photometric samples. Then, they derived $z=5.7$ and 6.6 Ly$\alpha$ LFs based on their samples consisting of the large fractions of spectroscopically confirmed LAEs and small fractions of remaining photometric candidates. While these LFs have disadvantage in statistics compared to the LFs from \citet{Ouchi08,Ouchi10}, they suffer significantly less amount of contaminations because of the very large fractions of spectroscopically confirmed LAEs. Hence, we also compare our $z=7$ Ly$\alpha$ LF to the $z=5.7$ and 6.6 Ly$\alpha$ LFs (their best-fit Schechter functions) derived by \citet{Kashikawa11} in Figure \ref{LyaLFs_at_z5p7-7_in_SDF}. As \citet{Kashikawa11} derived their LFs cumulatively, in Figure \ref{LyaLFs_at_z5p7-7_in_SDF} we plot our two types of cumulative $z=7$ Ly$\alpha$ LFs excluding or including the faintest bins where no LAE is detected and their best-fit Schechter functions taken from the middle and the right panels of Figure \ref{plottingLyaLF}. If we exclude the faintest bins from the $z=7$ Ly$\alpha$ LF, it looks entirely very similar to the $z=6.6$ Ly$\alpha$ LF while it shows a significant deficit from the $z=5.7$ Ly$\alpha$ LF at the bright end. On the other hand, if we include the faintest bins in the $z=7$ Ly$\alpha$ LF, it exhibits a deficit from the $z=6.6$ Ly$\alpha$ LF at the faint end and entirely from the $z=5.7$ Ly$\alpha$ LF. 

%%figure 12

%%figure 13
\begin{figure*}
\epsscale{1.17}
%\plottwo{Cropped_1sigma_3sigma_ErrorContours_Schechter_Cumulative_z7LyaLF+Kashikawa2011LyaLF_noZeroLAEpoints.eps}{Cropped_1sigma_3sigma_ErrorContours_Schechter_Cumulative_z7LyaLF+Kashikawa2011LyaLF_withZeroLAEpoints.eps}
\plottwo{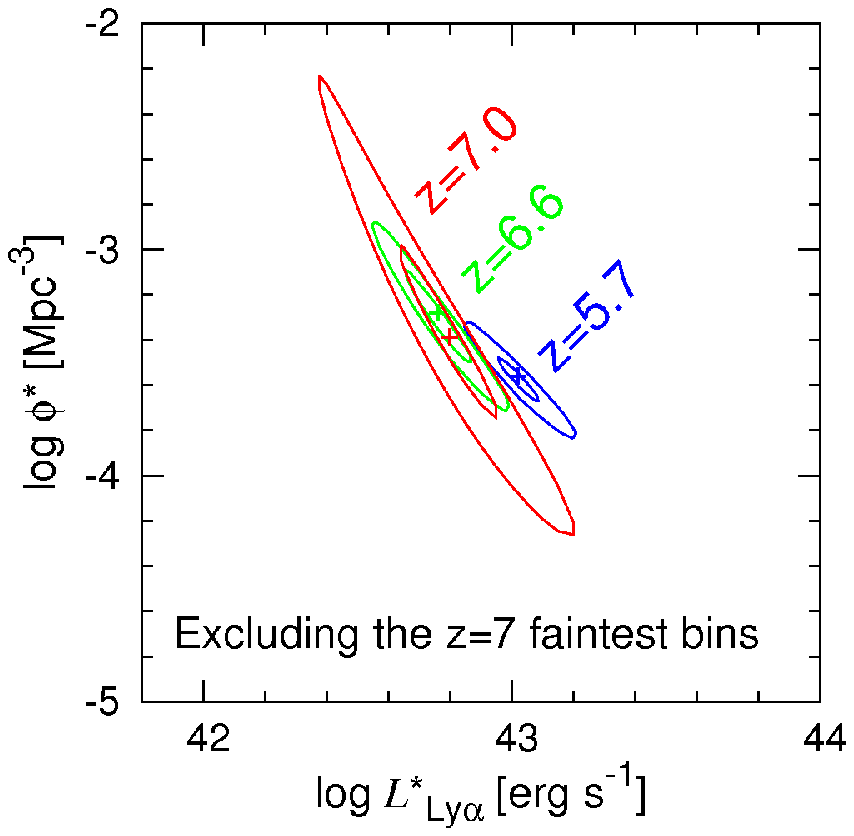}{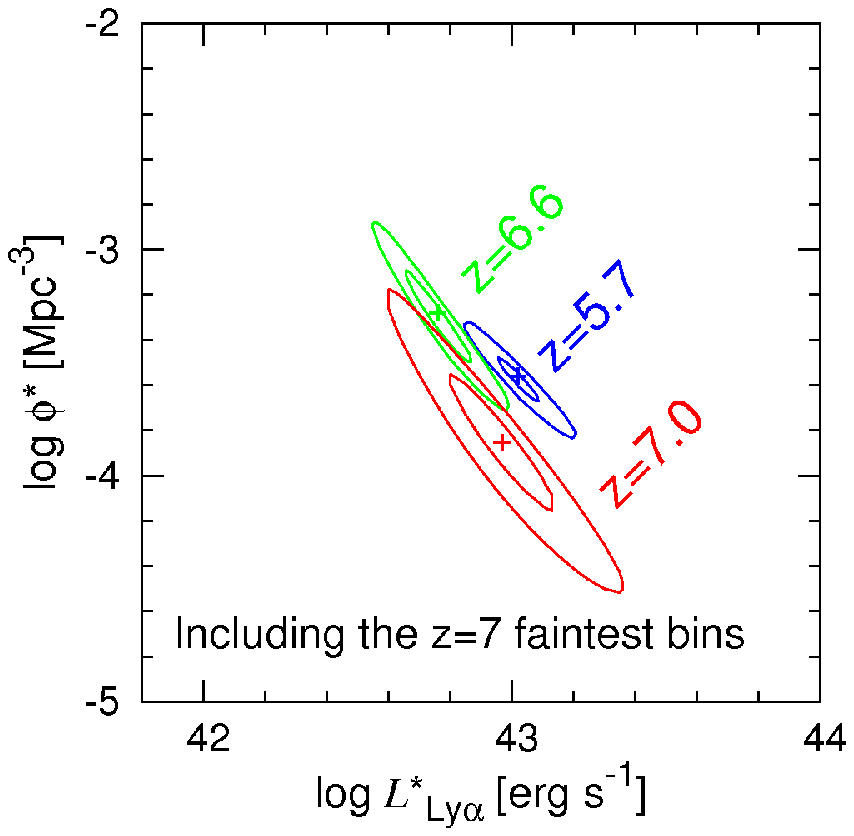}
\caption{Error contours of $\phi^*$ and $L^*$ parameters of the Schechter functions fitted to the $z=5.7$ and 6.6 cumulative Ly$\alpha$ LFs derived by \citet{Kashikawa11} and the $z=7$ cumulative Ly$\alpha$ LFs that exclude (left panel) or include (right panel) the faintest bins where no LAE is detected (the two $z=7$ cumulative Ly$\alpha$ LFs shown in the middle and right panels of Figure \ref{plottingLyaLF}). The plus symbols, inner and outer curves are the best-fit $\phi^*$ and $L^*$ values, the $1\sigma$ and $3\sigma$ confidence level contours.\label{Phi_vs_L_Cum}}
\end{figure*}

To look into this in more quantitatively, we derive the error contours of the $L^*$ and $\phi^*$ of the Schechter functions fitted to the two $z=7$ cumulative Ly$\alpha$ LFs and compare them to those of $z=5.7$ and 6.6 Ly$\alpha$ LFs derived by \citet{Kashikawa11} in Figure \ref{Phi_vs_L_Cum}. In the case of the $z=7$ Ly$\alpha$ LF excluding the faintest bins where no LAE is detected, the $z=7$ error contours at $1\sigma$ and $3\sigma$ confidence levels completely overlap with the $z=6.6$ ones while the $z=7$ error contour at $3\sigma$ confidence level barely overlaps with that of the $z=5.7$ Ly$\alpha$ LF. Meanwhile, in the case of the $z=7$ Ly$\alpha$ LF including the faintest bins, the $z=7$ error contour at $3\sigma$ confidence level overlaps with that of the $z=6.6$ Ly$\alpha$ LF but the $1\sigma$ confidence level contours of these LFs do not overlap. Hence, these two LFs are different at 1--2$\sigma$ level. Moreover, the $z=7$ error contour at $3\sigma$ confidence level does not overlap with that of the $z=5.7$ Ly$\alpha$ LF. Therefore, the Ly$\alpha$ LF evolves from $z=5.7$ to 7 at almost $3\sigma$ levels while it might evolve from $z=6.6$ to 7 at 1--2$\sigma$ level or might not. As whether the non-detections of LAEs in the faintest bins of the $z=7$ Ly$\alpha$ LF are due to the suppression of Ly$\alpha$ of faint LAEs by neutral IGM or the possible lack of the observation sensitivities are not clear, we cannot definitively conclude whether Ly$\alpha$ LF really evolves from $z=6.6$ to 7 at this moment. However, it should be noted that our $z=7$ Ly$\alpha$ LF is mostly based on the photometric LAE candidates (except for one $z=6.96$ LAE IOK-1), might include some contaminations and thus should be considered the upper limit. If we conduct spectroscopy of the photometric LAE candidates and clean the contaminations, we might confirm that the Ly$\alpha$ LF evolves from $z=6.6$ to 7.     

% Table 6
\begin{deluxetable*}{ccccccccl}
\tabletypesize{\scriptsize}
%\rotate
\tablecaption{Pure Luminosity and Pure Number Evolutions of the Schechter Parameters of the Ly$\alpha$ LF over $z=5.7$--7.3\label{PureL_PurePhi_Evolution}}
%\tablewidth{0pt}
%\tablewidth{510pt}
\tablehead{
 & & \multicolumn{3}{c}{\underline{Pure Luminosity Evolution}} & \multicolumn{3}{c}{\underline{Pure Number Evolution}} & \\
Redshift    & $\Delta t^{\rm a}$  & $L^*_{z_2}/L^*_{z_1}$$^{\rm b}$ & $\Delta L^*/\Delta t$$^{\rm c}$ & $\chi^2_{\rm red}$$^{\rm d}$  & $\phi^*_{z_2}/\phi^*_{z_1}$$^{\rm b}$ & $\Delta \phi^*/\Delta t$$^{\rm c}$ & $\chi^2_{\rm red}$$^{\rm d}$ & Ly$\alpha$ LF used to derive $L^*$ and $\phi^*$ $^{\rm e}$ \\         
$z_1 - z_2$  & [Myr] &  & [Gyr$^{-1}$] &  &  & [Gyr$^{-1}$] &  &
}
\startdata
5.7--6.6    & 160   & 0.7                &   1.9                & ---               & 0.5                      & 3.1                      & ---             &  O08 (5.7), O10 (6.6)\\
6.6--7.0    &  60   & 0.84               &   2.7                & 1.19              & 0.66                     & 5.7                      & 0.37            &  $z=7$ Differential LF$^{\rm f}$, O10 (6.6)\\
7.0--7.3    &  40   & 0.49               &  12.8                & 0.03              & 0.24                     & 19.0                     & 0.54            &  $z=7$ Differential LF$^{\rm f}$, K14 (7.3)\\
5.7--7.0    & 220   & 0.57               &   2.0                & 0.96              & 0.31                     & 3.1                      & 0.39            &  $z=7$ Differential LF$^{\rm f}$, O08 (5.7)\\
%5.7--7.0   & 220   & 0.59               &   1.9                & 0.15              & 0.39                     & 2.8                      & 2.88            &  This study ($z=7$ Differential Ly$\alpha$ LF), \citet{Kashikawa11} \\
%6.6--7.0   &  60   & 0.79               &   3.5                & 0.25              & 0.63                     & 6.2                      & 0.04            &  This study ($z=7$ Differential Ly$\alpha$ LF), \citet{Kashikawa11} \\
\hline
5.7--6.6    & 160   & 0.76               &   1.5                & ---               & 0.66                     & 2.1                      & ---             &  K11 (5.7, 6.6)\\
6.6--7.0    &  60   & 0.98               &   0.33               & 0.14              & 0.95                     & 0.83                     & 0.11   &  $z=7$ Cum LF excl 0 LAE bins$^{\rm g}$, K11 (6.6)\\
6.6--7.0    &  60   & 0.78               &   3.7                & 4.21              & 0.59                     & 6.8                      & 2.22   &  $z=7$ Cum LF incl 0 LAE bins$^{\rm h}$, K11 (6.6)\\
5.7--7.0    & 220   & 0.70               &   1.4                & 0.32              & 0.48                     & 2.3                      & 2.23   &  $z=7$ Cum LF excl 0 LAE bins$^{\rm g}$, K11 (5.7)\\
5.7--7.0    & 220   & 0.61               &   1.8                & 1.78              & 0.41                     & 2.7                      & 0.72   &  $z=7$ Cum LF incl 0 LAE bins$^{\rm h}$, K11 (5.7)
\enddata 
%% Text for table notes should follow after the \enddata but before
%% the \end{deluxetable}. Make sure there is at least one \tablenotemark
%% in the table for each \tablenotetext.
%\tablecomments{}
\tablenotetext{a}{Cosmic time interval in Myr corresponding to the redshift range $z_1 - z_2$ in column 1.}
\tablenotetext{b}{The ratio of the best-fit $L^*$ or $\phi^*$ values at the redshifts $z_2$ and $z_1$.}
\tablenotetext{c}{The rate of the decrease in $L^*$ or $\phi^*$ from $z_1$ to $z_2$ in the unit of Gyr$^{-1}$ defined as $\Delta L^*/\Delta t = (1 - L^*_{z_2}/L^*_{z_1})/\Delta t$ and $\Delta \phi^*/\Delta t = (1 - \phi^*_{z_2}/\phi^*_{z_1})/\Delta t$ where $\Delta t$ from column 2 is in the unit of Gyr instead of Myr.}
\tablenotetext{d}{The reduced $\chi^2$ value of the best-fit $L^*$ or $\phi^*$.}
\tablenotetext{e}{The number in parenthesis is redshift of the LF. The references from which the LFs come from are: O08 = \citet{Ouchi08}, O10 = \citet{Ouchi10}, K11 = \citet{Kashikawa11}, K14 = \citet{Konno14}.}
\tablenotetext{f}{The $z=7$ differential Ly$\alpha$ LF from the left panel in Figure \ref{plottingLyaLF}.}
\tablenotetext{g}{The $z=7$ cumulative Ly$\alpha$ LF excluding the faintest bins where no LAE is detected from the middle panel in Figure \ref{plottingLyaLF}.}
\tablenotetext{h}{The $z=7$ cumulative Ly$\alpha$ LF including the faintest bins where no LAE is detected from the right panel in Figure \ref{plottingLyaLF}.}
\end{deluxetable*}

% see also Madau plot top axis.
% z=5.7--6.6, dt = 0.16 Gyr
% z=6.6--7.0, dt = 0.06 Gyr
% z=7.0--7.3, dt = 0.04 Gyr
% z=6.6--7.3, dt = 0.10 Gyr
% z=5.6--7.0, dt = 0.22 Gyr

\subsubsection{Examining Rates of Number and Luminosity Evolution of the Ly$\alpha$ LF among $z=5.7$, 6.6, 7.0 and 7.3\label{LyaLF_LorPhi}}
Finally, the (possible) evolution of Ly$\alpha$ LF among $z=5.7$, 6.6, 7 and 7.3 seen in Figures \ref{LyaLFs_at_z5p7-7p3} and \ref{LyaLFs_at_z5p7-7_in_SDF} could reflect either the luminosity evolution, the number evolution or the combination of the both of the LAEs among these epochs. We can examine this quantitatively by using the best-fit Schechter function $L^*$ and $\phi^*$ of the $z=5.7$, 6.6, 7 and 7.3 Ly$\alpha$ LFs derived by us and taken from the literature \citep{Ouchi08,Ouchi10,Kashikawa11,Konno14} listed in Table \ref{Best-fitSchechter}. \citet{Ouchi10} and \citet{Kashikawa11} fixed either $L^*$ or $\phi^*$ of the Schechter function to the best-fit $L^*$ or $\phi^*$ values of their $z=5.7$ Ly$\alpha$ LFs and fitted the Schechter functions to their $z=6.6$ Ly$\alpha$ LFs by $\chi^2$ minimization. In this way, they examined which of the pure luminosity ($L^*$) and pure number ($\phi^*$) evolutions are more dominant by comparing their $\chi^2$ values and how much the $L^*$ or the $\phi^*$ changes from $z=5.7$ to 6.6. \citet{Ouchi10} found that for a fixed slope $\alpha=-1.5$, $L^*_{z=6.6} = 0.7 L^*_{z=5.7}$ for the pure luminosity evolution and $\phi^*_{z=6.6} = 0.5 \phi^*_{z=5.7}$ for the pure number evolution with the $\chi^2$ value of the pure luminosity evolution smaller than that of the pure number evolution. Also, \citet{Kashikawa11} confirmed that for a fixed slope $\alpha=-1.5$, $L^*_{z=6.6} = 0.76 L^*_{z=5.7}$ for the pure luminosity evolution and $\phi^*_{z=6.6} = 0.66 \phi^*_{z=5.7}$ for the pure number evolution. This means that the decrease in luminosity by 24--30\% is more favorable than that in number density by 34--50\% for the decline of the Ly$\alpha$ LF from $z=5.7$ to 6.6. 
%with the $\chi^2$ value of the pure luminosity evolution smaller than that of the pure number evolution

To further examine what the evolution of the Ly$\alpha$ LF looks like from $z=6.6$ to 7 and from $z=7$ to 7.3, we perform the similar fitting procedures to our $z=7$ Ly$\alpha$ LFs and the $z=7.3$ Ly$\alpha$ LF from \citet{Konno14}. More specifically, we fit the $\alpha=-1.5$ Schechter function with the fixed $L^* = L^*_{z=6.6}$ or $\phi^* = \phi^*_{z=6.6}$ of the differential (cumulative) $z=6.6$ Ly$\alpha$ LF from \citet{Ouchi10} (\citet{Kashikawa11}) to our differential (two types of cumulative) $z=7$ Ly$\alpha$ LF (LFs). Here, the two types of cumulative $z=7$ Ly$\alpha$ LFs are the ones that exclude or include the faintest bins where no LAE is detected and are plotted in the middle and the right panels of Figure \ref{plottingLyaLF}. Also, we fit the $\alpha=-1.5$ Schechter function with the fixed $L^* = L^*_{z=7}$ or $\phi^* = \phi^*_{z=7}$ of our differential $z=7$ Ly$\alpha$ LF to the differential $z=7.3$ Ly$\alpha$ LF from \citet{Konno14}. These $z=6.6$--7 and $z=7$--7.3 fitting results are summarized in Table \ref{PureL_PurePhi_Evolution} together with those for $z=5.7$--6.6 obtained by \citet{Ouchi10} and \citet{Kashikawa11}. We also perform the similar fittings for the $z=5.7$--7 case by using the $L^*_{z=5.7}$ or $\phi^*_{z=5.7}$ of the $z=5.7$ Ly$\alpha$ LFs from \citet{Ouchi10} and \citet{Kashikawa11} and our $z=7$ Ly$\alpha$ LFs and list the results in Table \ref{PureL_PurePhi_Evolution}. 

In the case of $z=6.6$--7, whether the case of the differential Ly$\alpha$ LF or the cumulative Ly$\alpha$ LFs, the reduced $\chi^2$ value of the pure luminosity evolution is larger than that of the pure number evolution, suggesting that the number evolution is more dominant. In the case of $z=7$--7.3, the reduced $\chi^2$ value of the pure luminosity evolution is smaller than that of the pure number evolution, suggesting that the luminosity evolution is more dominant. The evolution trend at $z=6.6$--7 is opposite to the case of the pure luminosity evolution favored at $z=5.7$--6.6 and $z=7$--7.3. In the case of pure number evolution, the decrease in $\phi^*$ from $z=6.6$ to 7 is smaller (larger) than that from $z=5.7$ to 6.6 if we exclude (include) the faintest bins from (in) our $z=7$ Ly$\alpha$ LF. In the case of pure luminosity evolution, the decrease in $L^*$ from $z=6.6$ to 7 is smaller than those from $z=5.7$ to 6.6 and $z=7$ to 7.3 in any $z=7$ Ly$\alpha$ LF cases. However, it should be noted that the cosmic time interval is quite different at $z=5.7$--6.6 (160 Myr), $z=6.6$--7 (60 Myr) and $z=7$--7.3 (40 Myr). In Table \ref{PureL_PurePhi_Evolution}, we also list the rates of the decrease in $L^*$ and $\phi^*$ in these cosmic time intervals (in units of Gyr$^{-1}$). We define them as $\Delta L^*/\Delta t = (1 - L^*_{z_2}/L^*_{z_1})/\Delta t$ and $\Delta \phi^*/\Delta t = (1 - \phi^*_{z_2}/\phi^*_{z_1})/\Delta t$ where $\Delta t$ in Gyr is a cosmic interval between redshifts $z_1$ and $z_2$. In the case of the differential Ly$\alpha$ LF, both $\Delta L^*/\Delta t$ and $\Delta \phi^*/\Delta t$ increase as redshift gets higher; a factor of 1.4 (1.8) increase in $\Delta L^*/\Delta t$ ($\Delta \phi^*/\Delta t$) from $z_1$ -- $z_2=5.7$--6.6 to $6.6$--7 and a factor of 4.7 (3.3) increase in $\Delta L^*/\Delta t$ ($\Delta \phi^*/\Delta t$) from $z_1$ -- $z_2=6.6$--7 to $7$--7.3. Both $\Delta L^*/\Delta t$ and $\Delta \phi^*/\Delta t$ increase with redshift also in the case of the cumulative $z=7$ LF including the faintest bins where no LAE is detected. However, both $\Delta L^*/\Delta t$ and $\Delta \phi^*/\Delta t$ decrease with redshift in the case of the cumulative $z=7$ LF excluding the faintest bins where no LAE is detected. If the large deficit of LAEs at $z=7.3$ suggested by the $z=7.3$ Ly$\alpha$ LF derived by \citet{Konno14} is real, it seems to be more natural to think that the deficit (non-detections) of $z=7$ LAEs we found at the faintest bins of our $z=7$ Ly$\alpha$ LF is not due to the possible lack of sensitivity but because of the luminosity or number evolution of faint LAEs.   
  
Eventually, our results suggest that the evolution of the Ly$\alpha$ LF at the epochs between $z=5.7$ and 7.3 is not always solely due to the change in luminosity of LAEs. The number density could also play a role especially at $z=6.6$--7. Moreover, there is a possibility that the luminosity and the number density of LAEs could evolve acceleratingly between $z=5.7$ and 7.3 including the short intervals $z=6.6$--7 and $z=7$--7.3, although we cannot completely rule out the possibility that the evolution might not be accelerating at $z=6.6$--7.       

% see also Madau plot top axis.
% z=5.7--6.6, dt = 0.16 Gyr
% z=6.6--7.0, dt = 0.06 Gyr
% z=7.0--7.3, dt = 0.04 Gyr
% z=6.6--7.3, dt = 0.10 Gyr

\subsection{Rest Frame UV Continuum Luminosity Function\label{UVLF}}
The significant and definitive decline of the Ly$\alpha$ LF from $z=5.7$ to 7 and possible decline from $z=6.6$ to 7 at its faint end might be due to the attenuation of Ly$\alpha$ emission of LAEs by higher fraction of neutral IGM at higher redshifts during the epoch of reionization. However, these declines can be also ascribed to the change in the detectability of LAEs due to galaxy evolution (their number and/or luminosity evolution). Hence, the decline of the Ly$\alpha$ LF can be caused by either IGM attenuation of Ly$\alpha$ alone, galaxy evolution alone or combination of the both. We can disentangle these if we can extract the galaxy evolution fraction alone contributing to the decline of the Ly$\alpha$ LF.

The rest frame UV continuum LF (UV LF) of LAEs can be used as a probe of evolution of LAEs because of the following reason. If there is no galaxy evolution (in both number and luminosity) between two epochs $z<6$ and $z>6$, the Ly$\alpha$ LF would shift to the fainter Ly$\alpha$ luminosity side from $z<6$ to $z>6$ if Ly$\alpha$ emissions of LAEs are attenuated by neutral IGM at $z>6$. UV-bright LAEs emitting stronger ionizing radiation can effectively ionize their surroundings, and thus their Ly$\alpha$ emissions are less attenuated by neutral IGM. Conversely, UV-faint LAEs emitting weaker ionizing radiation cannot effectively ionize their surroundings, and thus their Ly$\alpha$ emissions are more attenuated by neutral IGM. Therefore, only LAEs faint in both Ly$\alpha$ and the UV continuum around the faint end of the Ly$\alpha$ LF get fainter than a survey's Ly$\alpha$ luminosity limit at $z>6$ and would not be detected by a narrowband observation. Accordingly, the detection number of such LAEs is expected to decrease. Meanwhile, even in such a situation, the UV LF of LAEs will not shift to the fainter luminosity side as the rest frame UV continua of LAEs are not attenuated by neutral IGM at $z>6$. However, the number density of LAEs is expected to decrease only at the faint end of the UV LF due to the decrease in the detection number of LAEs faint in both Ly$\alpha$ and the UV continuum in the narrowband observation mentioned above. Hence, unless there is galaxy evolution, the bright side of the UV LF stays unchanged and only its faint end would change. This characteristic of the UV LF can be used to see if LAEs evolve between $z<6$ and $z>6$. Usually, we cannot accurately determine the faint end of the UV LF of LAEs to a sufficiently faint UV luminosity limit with high completeness due to the faintness of the UV continua of UV-faint LAEs and the difficulty in obtaining extremely deep imaging covering their UV-continua (especially for $z=5.7$--7 LAEs). Thus, we inevitably compare only the bright sides of the UV LFs of LAEs at $z<6$ and $z>6$.

Previously, \citet{Kashikawa11} estimated the UV continuum fluxes $f_c$ and then luminosities $M_{\rm UV}$ of the $z=5.7$ and 6.6 LAEs in SDF using their narrowband (NB816 and NB921 bands) and broadband ($z'$ band) magnitudes and the equation (\ref{Eqn_LyaUVLum}) and derived the UV LFs of the $z=5.7$ and 6.6 LAEs. They found that the LAE UV LF does not evolve much between $z=5.7$ and 6.6, suggesting no significant evolution of LAEs during this cosmic time interval. 

%%figure 14
\begin{figure}
\epsscale{1.17}
%\plotone{Cropped_Cumulative_UVLF_z5p7_z6p6_z7_adoptK11_inclCosmicVariance.eps}
\plotone{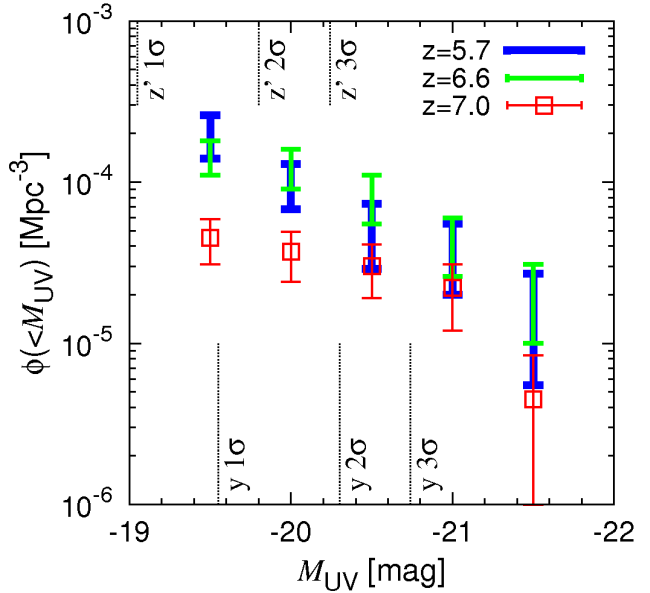}
\caption{The rest frame UV continuum LF of the $z=7$ LAE candidates in SDF we derived compared with those at $z=5.7$ and 6.6 in SDF taken from \citet{Kashikawa11}. The errors include Poisson error and cosmic variance. \citet{Kashikawa11} data originally included only Poisson error. Hence, we estimated cosmic variances of the $z=5.7$ and $z=6.6$ LAEs in SDF in the similar way as we did for our $z=7$ LAE candidates (see Section \ref{LyaLF_z7LAEs}) and include them in the error of the \citet{Kashikawa11} data. The vertical lines indicate the $3\sigma$, $2\sigma$ and $1\sigma$ limiting magnitudes in SDF in the $y$ ($z'$) band at $z=7.02$ (6.5) corresponding to $M_{\rm UV} = -20.74$ ($-20.24$), $-20.30$ ($-19.80$), and $-19.55$ ($-19.05$), respectively, given EW$_0 = 0$. The $M_{\rm UV}$'s at $z=5.7$ corresponding to the same SDF $z'$ band limiting magnitudes are fainter than those at $z=6.5$. The LAE number densities at the magnitudes fainter than $3\sigma$ suffers detection incompleteness and thus would probably be lower than actual number densities. This implies that at the $M_{\rm UV}=-20.5$ bin the LAE number density at $z=7$ could be higher and comparable to that at $z=5.7$ as it is already comparable to that at $z=6.6$ within the errors.\label{UVLFs_at_z5p7-7}}
\end{figure}

As we have also already estimated $M_{\rm UV}$'s of our $z=7$ LAE candidates in Section \ref{LyaUVLumSFRLimit} and Table \ref{Propertyz7LAECandidates} using their NB973 and $y$ band total magnitudes and the equation (\ref{Eqn_LyaUVLum}) in the same way as \citet{Kashikawa11}, we also derive the UV LF of the $z=7$ LAE candidates and compare it to those at $z=5.7$ and 6.6 derived by \citet{Kashikawa11}. To this end, we use only the $z=7$ LAE candidates in SDF, not including those in SXDS, as the $y$ band image of the SXDS is much shallower than that of the SDF. We count the number of the $z=7$ LAE candidates detected in the UV continuum (i.e., those with $f_c > 0$) in each $M_{\rm UV}$ bin, estimate its error including Poisson error and cosmic variance and correct the number and error for the detection completeness estimated in Section \ref{Completeness} and shown in Figure \ref{NB973Completeness} by number weighting according to the NB973 magnitude. Then, we divide the corrected numbers and errors by the SDF survey volume to convert them into the number densities. Note that this procedure is the same as the one used by \citet{Kashikawa11} who corrected the numbers of their $z=5.7$ and 6.6 LAEs using their detection completeness in the narrowbands NB816 and NB921. It is more ideal to use the detection completeness in the broadband covering the rest frame UV continuum of LAEs redward of Ly$\alpha$ (in the case of our $z=7$ LAEs, it is $y$ band). However, as we and \citet{Kashikawa11} used the narrowbands as the LAE detection images, it is impossible to use the detection completenesses in the broadbands. Thus, we and \citet{Kashikawa11} instead use the detection completenesses in the narrowbands as good approximation because they can detect not only Ly$\alpha$ emission but also UV continua of LAEs at the same time.    
 
In Figure \ref{UVLFs_at_z5p7-7}, we compare our $z=7$ LAE cumulative UV LF to those at $z=5.7$ and 6.6 derived by \citet{Kashikawa11}. As the errors in the \citet{Kashikawa11} data originally do not include cosmic variances at $z=5.7$ and 6.6, we estimate them in the same way as in \citet{Ota08,Ota10} and add them to the errors in quadrature for fair comparison. In Figure \ref{UVLFs_at_z5p7-7}, we also show the $M_{\rm UV}$'s corresponding to the $1\sigma$, $2\sigma$ and $3\sigma$ limiting magnitudes of the SDF $z'$ and $y$ band images which were used to derive $M_{\rm UV}$'s of the \citet{Kashikawa11}'s $z=5.7$ and 6.6 LAEs and our $z=7$ LAEs, respectively. As completeness is very low at magnitudes fainter than $3\sigma$, the $z=5.7$ and 6.6 UV LF at $M_{\rm UV}$ fainter than the $3\sigma$ $z'$ band magnitude and the $z=7$ UV LF at $M_{\rm UV}$ fainter than the $3\sigma$ $y$ band magnitude are highly uncertain and less reliable. We show the UV LFs at $M_{\rm UV}$ fainter than $3\sigma$ just for reference. 

Looking at the two most reliable brightest $M_{\rm UV}$ bins, we see that the UV LFs at the three epochs are consistent with each other within their errors. However, there is a sign of slight deficit in the $z=7$ UV LF compared to the $z=5.7$ and 6.6 ones. This implies possible galaxy evolution going on between $z=6.6$ and $z=7$. One possible origin of the galaxy evolution is the change in UV continuum luminosity due to dust extinction. \citet{Ono10} performed the SED-fitting of the stacking of the 401 $z=5.7$ and 207 $z=6.6$ LAEs detected by \citet{Ouchi08,Ouchi10} in the entire SXDS field and found that the $z=5.7$ and $z=6.6$ LAEs on average have negligibly low dust extinctions of $E(B-V)\sim 0.0$ and $\sim 0.1$, respectively. Also, \citet{Ono12} performed the SED-fitting of a $z=7.213$ LAE (but it was detected by $z$-band dropout technique, not by a narrowband) and found that it has negligibly low dust extinction $E(B-V)\sim 0.05$. Meanwhile, \citet{Jiang16} carried out the SED-fitting of the spectroscopically identified $z=5.7$--6.6 galaxies in SDF (including both narrowband-selected LAEs and $i'$-band dropouts) and the $z=6.96$ LAE IOK-1 (or NB973-SDF-85821 in Table \ref{z7LAECandidates} we detected in NB973). We notice that out of their galaxies, the narrowband-selected $z\sim5.7$ and $z\sim6.6$ LAEs have $E(B-V)\sim 0.0$--0.16 (0.06 on average) and $\sim 0.0$--0.36 (0.15 on average), respectively, and IOK-1 has $E(B-V)\sim 0.04$. These SED-fitting studies suggest that amount of dust extinction is very low and similar among $z=5.7$, 6.6 and 7 LAEs although the sample of $z\sim7$ LAEs in these studies is small. Hence, the deficit in the $z=7$ LAE UV LF compared to the $z=5.7$ and 6.6 ones may not be due to the difference in the amount of dust extinction between LAEs at these epochs. Whatever the origin of the deficit in the $z=7$ LAE UV LF is, it implies that galaxy evolution seems to partially contribute to the deficit in the $z=7$ Ly$\alpha$ LF. We will discuss this in more details and quantitatively in Section 5 when we obtain implications for reionization from the both $z=7$ LAE Ly$\alpha$ and UV LFs.    

\subsection{Ly$\alpha$ Equivalent Width Distribution\label{subsectionLyaEWdist}}
Another method of examining the possible attenuation of Ly$\alpha$ emission of LAEs by neutral IGM during the reionization epoch is to compare the Ly$\alpha$ EW distribution of LAEs at $z<6$ and $z>6$. \citet{Kashikawa11} found that the rest frame Ly$\alpha$ EW (EW$_0$) distributions of $z\sim3$--5.7 LAEs are very similar, but EW$_0$'s of $z=6.6$ LAEs are significantly lower. This implies that Ly$\alpha$ emission might be attenuated by neutral IGM at $z=6.6$. If this is the case, EW$_0$'s of our $z=7$ LAEs could be even lower than those of $z=6.6$ LAEs due to possibly stronger suppression of Ly$\alpha$ emission by potentially higher fraction of neutral IGM at higher redshift. 

\begin{figure}
\epsscale{1.17}
\plotone{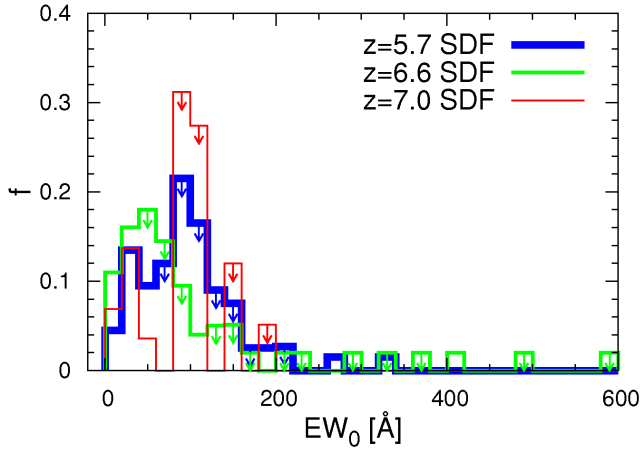}
%\plotone{Cropped_RestFrame_LyaEW_Distribution_z57_z66_z7_CompletenessCorrected_3LAECorr_z7SDF.eps}
\caption{The differential rest frame Ly$\alpha$ EW (EW$_0$) distribution of our $z=7$ LAE sample in SDF compared with those of the $z=5.7$ and $z=6.6$ LAE samples in SDF from \citet{Kashikawa11}. The bins with an arrow include the lower limits on EW$_0$'s of the LAEs not detected in the rest frame UV continuum, and thus their frequencies are upper limits as true values of their EW$_0$'s fall in the same bins or higher EW$_0$ bins.\label{EW_Distribution}}
\end{figure}

In Section \ref{LyaUVLumSFRLimit}, we derived EW$_0$'s (or lower limits on EW$_0$) of our $z=7$ LAE candidates from their Ly$\alpha$ fluxes $f_{\rm line}$ and UV continuum fluxes $f_c$ estimated by using the equation (\ref{Eqn_LyaUVLum}) and their NB973 and $y$-band total magnitudes in the same way as \citet{Kashikawa11}. We count the number of the $z=7$ LAEs in each EW$_0$ bin, correct it for the detection completeness estimated in Section \ref{Completeness} and shown in Figure \ref{NB973Completeness} by number weighting according to the NB973 magnitude and normalize it by the total completeness-corrected number of the $z=7$ LAEs. This procedure is also the same as the one taken by \citet{Kashikawa11}. Hence, we can compare the EW$_0$ distribution of our $z=7$ LAEs with those of the $z=5.7$ and 6.6 LAEs derived by \citet{Kashikawa11} with less or no bias/systematics. We make this comparison in Figure \ref{EW_Distribution}. As \citet{Kashikawa11} compiled EW$_0$'s of the $z=5.7$ and 6.6 LAEs only in SDF, we also use our $z=7$ LAEs only in SDF for the derivation of the $z=7$ LAE EW$_0$ distribution. As seen in Table \ref{LAESurveys}, we compare these $z=5.7$, 6.6 and 7 LAE samples in SDF to the fairly comparable EW$_0$ thresholds (10\AA, 7\AA~and 10\AA~for $z=5.7$, 6.6 and 7). \citet{Kashikawa11} estimated the lower limits on EW$_0$'s of the $z=5.7$ and 6.6 LAEs undetected ($<1\sigma$) in $z'$-band (i.e., undetected in the UV continuum) from the $1\sigma$ magnitude of the $z'$-band SDF image and included them in the EW$_0$ bins in the EW$_0$ distributions of the $z=5.7$ and 6.6 LAEs (the blue and the green bins with an arrow in Figure \ref{EW_Distribution}). Hence, for the fair comparison, in the EW$_0$ bins (the red bins with an arrow in Figure \ref{EW_Distribution}) we also include the lower limits on EW$_0$'s of the $z=7$ LAEs undetected ($<1\sigma$) in $y$-band estimated from the $1\sigma$ magnitude of the $y$-band SDF image (see Table \ref{Propertyz7LAECandidates} and Section \ref{LyaUVLumSFRLimit} for the EW$_0$ lower limits). The frequencies f of the bins with an arrow in Figure \ref{EW_Distribution} are upper limits as true values of included EW$_0$ lower limits fall in the same bins or higher EW$_0$ bins.

%%figure 15

Figure \ref{EW_Distribution} shows that the peak of the $z=7$ EW$_0$ distribution coincides with that of the $z=5.7$ distribution at the EW$_0 =$ 80--100\AA~bin. Also, the overall distributions look very similar between $z=7$ and $z=5.7$. However, the $z=7$ EW$_0$ distribution exhibits bimodality with another peak at the 20--40\AA~bin. The three consecutive (0--60\AA) bins around this peak consist of EW$_0$'s of only the $z=7$ LAEs detected in the UV continuum and do not include any EW$_0$ lower limits of the LAEs undetected in the UV continuum. Hence, there is no uncertainty due to including any EW$_0$ lower limits. Moreover, these three bins comprise two thirds of (4 out of the 6) $z=7$ LAEs detected in the UV continuum in SDF. The EW$_0$'s of these bins are lower than typical EW$_0$'s (and EW$_0$'s limits) of the $z=5.7$ and $z=6.6$ LAEs. Hence, there is a possibility that Ly$\alpha$ emission of a fraction of the $z=7$ LAEs might be more strongly attenuated by possibly higher fraction of neutral IGM at $z=7$ than $z=5.7$ and $z=6.6$ although our $z=7$ LAE sample size is much smaller than those of the $z=5.7$ and 6.6 LAEs. 

%Figure \ref{EW_Distribution} shows that except for the two higher EW$_0$ bins, EW$_0$'s of the $z=7$ LAEs (the lower EW$_0$ bins consisting of 4 out of the 6 LAEs detected in the UV continuum in SDF) are systematically lower than those of $z=5.7$ LAEs and even those of $z=6.6$ LAEs. We also perform the Kolmogorov-Smirnov test between the EW$_0$ distributions of the $z=5.7$ and $z=7$ LAEs as well as the $z=6.6$ and $z=7$ LAEs, and obtain the $p$-values of 0.302 and 0.029, respectively, suggesting that the EW$_0$ distributions of the $z=7$ LAEs are different from those of the $z=5.7$ and 6.6 LAEs. Although our $z=7$ LAE sample size is much smaller than those of the $z=5.7$ and 6.6 LAEs, this implies that Ly$\alpha$ emission might be more strongly attenuated by possibly higher fraction of neutral IGM at $z=7$ than $z=6.6$. 

%%figure 16
\begin{figure}
\epsscale{1.17}
%\plotone{Cropped_EW_vs_Muv_z7LAEs.eps}
\plotone{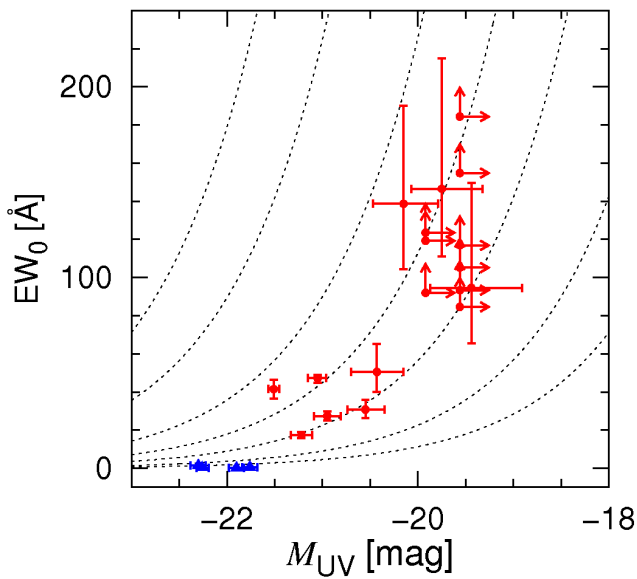}
\caption{The rest frame Ly$\alpha$ EW (EW$_0$) as a function of UV continuum luminosity ($M_{\rm UV}$). The arrows show lower limits on EW$_0$ and $M_{\rm UV}$. The red circles show the $z=7$ LAE candidates detected in the UV continuum in both SDF and SXDS while the blue triangles indicate the objects with an extremely faint or zero Ly$\alpha$ flux ($0\leq {\rm EW}_0 <10$\AA) and $z'-{\rm NB973}<0$ color (likely considered $z\sim7$ LBGs; see Section \ref{LyaFaintObjects} and Tables \ref{z7LAECandidates} and \ref{Propertyz7LAECandidates} for their details). The dotted curves are the EW$_0$'s at the fixed Ly$\alpha$ luminosities $L({\rm Ly}\alpha) =$ 10, 5, 2, 1, 0.5, 0.2, $0.1 \times 10^{43}$ erg s$^{-1}$ from top to bottom.\label{EW_vs_MUV}}
\end{figure}

\subsection{Ly$\alpha$ EW -- $M_{\rm UV}$ Relation\label{subsectionLyaEW-Muv}}
On the other hand, previous studies found that both $z=3.1$--6.6 LAEs and $z\sim3$--6 LBGs share the same trend in the EW$_0$--UV luminosity relation that there is an apparent deficit of high Ly$\alpha$ EW galaxies with a bright UV luminosity. Galaxies with a fainter UV continuum tend to exhibit higher Ly$\alpha$ EWs \citep{Shapley03,Ando06,Shimasaku06,Stanway07,Deharveng08,Ouchi08,Vanzella09,Stark10,Kashikawa11}. This trend has been clearly confirmed by the previous studies in the EW$_0$--$M_{\rm UV}$ relation in such a way that EW$_0$ systematically decreases as UV luminosity increases. To see if this also applies to $z=7$ LAEs, we plot the EW$_0$--$M_{\rm UV}$ diagram of our $z=7$ LAE candidates in SDF and SXDS in Figure \ref{EW_vs_MUV}. We find that our $z=7$ LAE candidates also lack the high EW bright UV luminosity objects and that those with fainter $M_{\rm UV}$ exhibit higher EWs following the same trend as that seen for lower redshift LAEs and LBGs. We also plot the objects with an extremely faint or zero Ly$\alpha$ flux ($0\leq {\rm EW}_0 <10$\AA) and a $z'-{\rm NB973}<0$ color (likely considered $z\sim7$ LBGs; see Section \ref{LyaFaintObjects}, Tables \ref{z7LAECandidates} and \ref{Propertyz7LAECandidates} and Figure \ref{SDF_4zLBGs_BlackWhite} for their details). They show brighter UV luminosities than the $z=7$ LAE candidates and are located at the low-EW$_0$, high-$M_{\rm UV}$ edge of the EW$_0$--$M_{\rm UV}$ trend.

The physical mechanism of the EW$_0$--$M_{\rm UV}$ trend of LAEs and LBGs has not been fully understood yet, but several studies suggested a few different explanations; e.g., higher metallicities in the UV-bright galaxies, older stellar populations in the UV-bright galaxies, enhancement of EWs of UV-faint galaxies at a pre-outflow phase by clumpy dust extinction and low EWs of low dust UV-bright galaxies at an outflow phase with a long history of starburst since its onset \citep{Ando06,Ouchi08,Kobayashi10}. Whatever the origin would be, the EW$_0$--$M_{\rm UV}$ trend also exists for the galaxies at the epoch as early as $z=7$. 
%
%The physical mechanism of the EW$_0$--$M_{\rm UV}$ trend of LAEs and LBGs has not been fully understood yet, but several studies suggested a few different explanations; e.g., higher metallicities in the UV-bright galaxies, older stellar populations in the UV-bright galaxies and enhancement of EW by clumpy dust extinction \citep{Ando06,Ouchi08,Kobayashi10}. Whatever the origin would be, the EW$_0$--$M_{\rm UV}$ trend also exists for the galaxies at the epoch as early as $z=7$. 

\section{Discussion\label{Discussion}}
In the previous sections, we have derived the Ly$\alpha$ LF, the UV LF and the EW$_0$ distribution of the $z=7$ LAEs from our observations. Then, we have found that the Ly$\alpha$ LF of LAEs evolves from $z=5.7$ and 6.6 to 7 and from $z=7$ to 7.3, that the UV LF of LAEs evolves from $z=5.7$--6.6 to 7 and that most of the EW$_0$'s of the $z=7$ LAEs detected in the UV continuum are lower than those of $z=5.7$ and 6.6 LAEs. All these results would suggest that LAEs themselves do evolve and neutral fraction of IGM could also increase from $z=5.7$--6.6 to 7 and from $z=7$ to 7.3. In this section, we further investigate and discuss this by extracting the number and luminosity densities of LAEs as well as comparing the observation results with theoretical models of galaxy evolution and reionization.

% Table 7
\begin{deluxetable*}{llcc}
\tabletypesize{\scriptsize}
%\rotate
\tablecaption{Number and UV Continuum Luminosity Densities of $z=5.7$, 6.6 and 7 LAEs\label{n_rho_UV}}
%\tablewidth{0pt}
%\tablewidth{510pt}
\tablehead{
$z$  & Reference          & $n_{\rm UV}^{\rm obs}$$^{\rm a}$ & $\rho_{\rm UV}^{\rm obs}$$^{\rm a}$ \\
     &                    & (10$^{-5}$ Mpc$^{-3}$)       & (10$^{24}$ erg s$^{-1}$ Hz$^{-1}$Mpc$^{-3}$)      
}
\startdata
5.7 & \citet{Ouchi08}     & 3.1$_{-2.6}^{+8.4}$           & 4.9$_{-4.1}^{+14.6}$ \\
5.7 & \citet{Kashikawa11} & 3.3$_{-2.7}^{+15.0}$          & 7.2$_{-5.8}^{+63.5}$  \\
6.6 & \citet{Kashikawa11} & 4.0$_{-2.5}^{+7.0}$           & 8.4$_{-5.7}^{+21.6}$  \\
7.0 & This study          & 2.2$_{-1.0}^{+0.9}$           & 2.8$_{-2.3}^{+6.4}$  
\enddata
%% Text for table notes should follow after the \enddata but before
%% the \end{deluxetable}. Make sure there is at least one \tablenotemark
%% in the table for each \tablenotetext.
%\tablecomments{}
\tablenotetext{a}{The number and UV continuum luminosity densities of LAEs. The densities of the $z=$5.7 and 6.6 LAEs were obtained by integrating the best-fit Schechter function derived by each author of the corresponding reference in column 2 down to the observed UV continuum luminosity limit $M_{\rm UV}=-21$ mag. The number (UV continuum luminosity) densities of the $z=7$ LAEs were derived by counting the number (summing the UV continuum luminosities) of the LAEs to the $z=7$ UV LF bin corresponding to the same limit.}
\end{deluxetable*}
%\clearpage
%\begin{turnpage}

\subsection{Evolution of Number and Luminosity Densities of LAEs\label{Number_Luminosity_Densities}}
If LAEs evolve and/or their detectability is affected by the attenuation of their Ly$\alpha$ emission by increasing neutral IGM towards higher redshift, the number, Ly$\alpha$ luminosity and UV continuum luminosity densities of LAEs would evolve with redshift. More specifically, the change in the number and Ly$\alpha$ luminosity densities reflects both galaxy evolution and attenuation of Ly$\alpha$ by neutral IGM while the UV luminosity density traces only galaxy evolution. Hence, comparing the redshift evolution of these three types of densities, we could obtain some implications for LAE evolution and reionization. 

The number and luminosity densities can be calculated by integrating the Ly$\alpha$ and UV LFs to certain Ly$\alpha$ and UV luminosity limits. In Section \ref{LyaLF}, we derived the Ly$\alpha$ LFs of our $z=7$ LAE candidates (differential one and two types of cumulative ones) and fitted the Schechter function (with a fixed slope $\alpha=-1.5$) to them. The best-fit results were presented in Table \ref{Best-fitSchechter} together with those of the $z=5.7$, 6.6 and 7.3 Ly$\alpha$ LFs obtained from the previous Subaru LAE surveys \citep{Ouchi08,Ouchi10,Kashikawa11,Konno14}. We can calculate the number and Ly$\alpha$ luminosity densities of the $z=5.7$--7.3 LAEs by integrating these best-fit Schechter functions. As for the Ly$\alpha$ luminosity limit, all these previous Subaru LAE surveys and our $z=7$ LAE one reached comparable depths of $L({\rm Ly}\alpha) \simeq 2 \times 10^{42}$ erg s$^{-1}$. \citet{Ouchi08,Ouchi10} and \citet{Konno14} already calculated the observed number and/or Ly$\alpha$ luminosity densities ($n_{{\rm Ly}\alpha}^{\rm obs}$ and/or $\rho_{{\rm Ly}\alpha}^{\rm obs}$) of their LAEs at $z=5.7$, 6.6 and 7.3, respectively, by integrating their best-fit Schechter functions to the common Ly$\alpha$ luminosity limit of $\log L({\rm Ly}\alpha)$ (erg s$^{-1}$) $=$ 42.4. To facilitate the comparison with them, we also integrate our best-fit Schechter functions of the three types of the $z=7$ Ly$\alpha$ LFs and those of the $z=5.7$ and 6.6 Ly$\alpha$ LFs derived by \citet{Kashikawa11} to $\log L({\rm Ly}\alpha)$ (erg s$^{-1}$) $=$ 42.4 to obtain $n_{{\rm Ly}\alpha}^{\rm obs}$ and $\rho_{{\rm Ly}\alpha}^{\rm obs}$ at $z=5.7$, 6.6 and 7. Moreover, similarly, we also calculate total Ly$\alpha$ luminosity densities $\rho_{{\rm Ly}\alpha}^{\rm tot}$ by integrating the best-fit Schechter functions to the zero luminosity $L({\rm Ly}\alpha)=0$. The $n_{{\rm Ly}\alpha}^{\rm obs}$, $\rho_{{\rm Ly}\alpha}^{\rm obs}$ and $\rho_{{\rm Ly}\alpha}^{\rm tot}$ we calculate and take from the previous Subaru LAE surveys are listed in Table \ref{Best-fitSchechter}. 

%%figure 17
\begin{figure*}
\epsscale{1.17}
%\plotone{z57_z66_z7_z73_LAEs_Madau-Plot_nLya_rhoLya_Ota16-II_pmLLphiErrors.eps}
%\plotone{z57_z66_z7_z73_LAEs_Madau-Plot_nLya_rhoLya_Ota16-III_pmLLphiErrors.eps}
\plotone{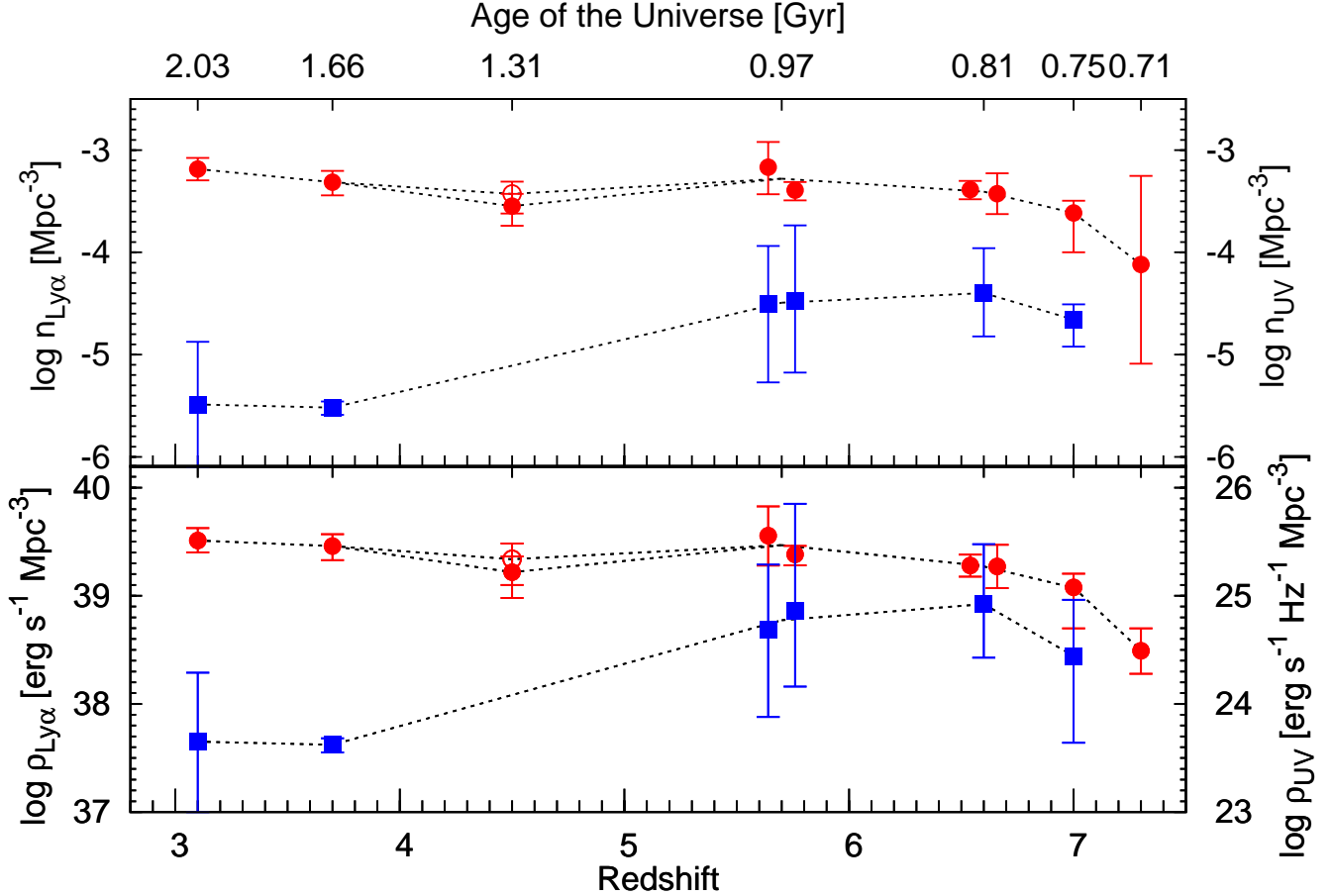}
\caption{Redshift evolution of number densities in Ly$\alpha$ and UV continuum detections ($n_{{\rm Ly}\alpha}$ and $n_{\rm UV}$ indicated by the circles and squares, respectively, in the upper panel) as well as Ly$\alpha$ and UV continuum luminosity densities ($\rho_{{\rm Ly}\alpha}$ and $\rho_{\rm UV}$ indicated by the circles and squares, respectively, in the lower panel) of LAEs over cosmic time of $z=3.1$--7.3 down to the Ly$\alpha$ and UV continuum luminosity limits $\log L({\rm Ly}\alpha)$ (erg s$^{-1}$) $=$ 42.4 and $M_{\rm UV} = -21$ mag. The densities are calculated by integrating the Schechter functions best-fitted to the Ly$\alpha$ and UV LFs of LAEs derived by \citet{Ouchi08} ($z=3.1$, 3.7 and 5.7 LAEs in SXDS), \citet{Dawson07} ($z=4.5$ LAEs from the Large Area Ly$\alpha$ (LALA) survey), \citet{Kashikawa11} ($z=5.7$ and 6.6 LAEs in SDF), \citet{Ouchi10} ($z=6.6$ LAEs in SXDS + SDF), this study ($z=7$ LAEs in SDF and/or SXDS; we here plot the densities derived from our $z=7$ differential Ly$\alpha$ LF in Figure \ref{plottingLyaLF}) and \citet{Konno14} ($z=7.3$ LAEs in SXDS and COSMOS). As for the $n_{{\rm Ly}\alpha}$ and $\rho_{{\rm Ly}\alpha}$ at $z=4.5$, the filled circles show the densities calculated from the \citet{Dawson07} $z=4.5$ Ly$\alpha$ LF based on the spectroscopically confirmed LAEs while the open circles indicate the densities corrected for the LAE selection reliability of $\sim 76$\% estimated by \citet{Dawson07} (see also text in Section \ref{Number_Luminosity_Densities} for more details). The two data points for each of the $n_{{\rm Ly}\alpha}$, $n_{\rm UV}$, $\rho_{{\rm Ly}\alpha}$ and $\rho_{\rm UV}$ at $z=5.7$ as well as $n_{{\rm Ly}\alpha}$ and $\rho_{{\rm Ly}\alpha}$ at $z=6.6$ are slightly horizontally shifted for clarity; i.e., the left and right ones correspond to the data based on \citet{Ouchi08,Ouchi10} and \citet{Kashikawa11}, respectively. The top axis shows the cosmic ages corresponding to the redshifts at which the densities are calculated. The dotted lines simply connect the data points to help elucidate how the densities vary with redshift. \label{Madau_Plot}}
\end{figure*}

On the other hand, we also calculate the number and UV luminosity densities, $n_{\rm UV}^{\rm obs}$ and $\rho_{\rm UV}^{\rm obs}$, of $z=5.7$, 6.6 and 7 LAEs and list them in Table \ref{n_rho_UV}. \citet{Ouchi08} and \citet{Kashikawa11} derived the best-fit Schechter functions of the UV LFs of the $z=5.7$ LAEs in SXDS and the $z=5.7$ and 6.6 LAEs in SDF, respectively. To calculate $n_{\rm UV}^{\rm obs}$ and $\rho_{\rm UV}^{\rm obs}$ of these LAEs, we integrate these Schechter functions to the UV luminosity limit of $M_{\rm UV}=-21$ mag. This corresponds to the UV luminosity of the second brightest bin of the UV LFs of LAEs shown in Figure \ref{UVLFs_at_z5p7-7} and the faintest bin just above the $3\sigma$ limit of our $z=7$ LAE UV LF. Meanwhile, we cannot accurately fit the Schechter function to our $z=7$ LAE UV LF with only its two brightest bins above the $3\sigma$ UV luminosity limit. Thus, we estimate $n_{\rm UV}^{\rm obs}$ and $\rho_{\rm UV}^{\rm obs}$ of the $z=7$ LAEs to $M_{\rm UV}=-21$ mag by counting the number of LAEs or summing their UV luminosities, correcting them for completeness and dividing them by our SDF survey volume.  

%%figure 18
\begin{figure}
\epsscale{1.17}
%\plotone{rhoLyaLAE_rhoUVLAE_rhoUVLBGs_vs_z-II_pmLphiErrors.eps}
\plotone{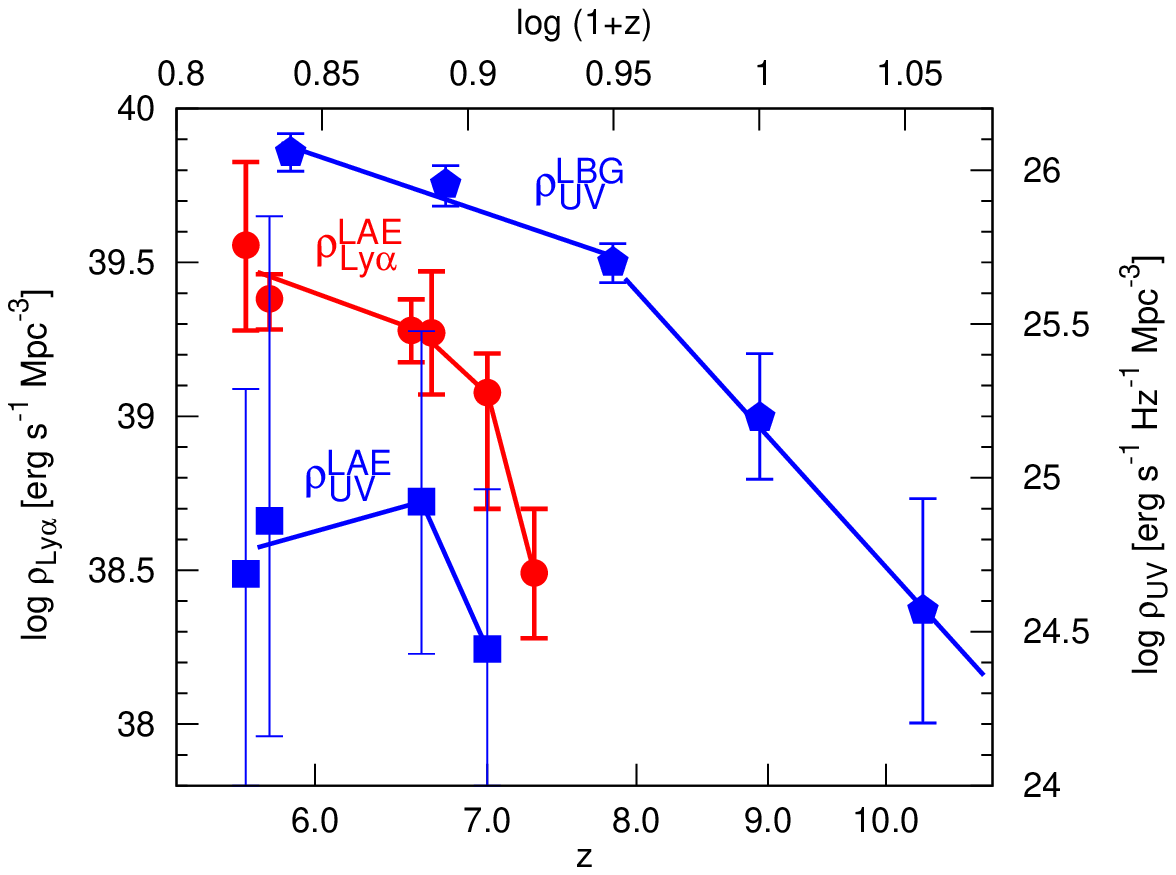}
\caption{The close-up of redshift evolution of the Ly$\alpha$ and UV luminosity densities of LAEs ($\rho_{{\rm Ly}\alpha}^{\rm LAE}$ and $\rho_{\rm UV}^{\rm LAE}$) among $z=5.7$, 6.6, 7 and 7.3 indicated by the red circles and the blue squares, respectively, as well as that of the UV luminosity densities of LBGs ($\rho_{\rm UV}^{\rm LBG}$) among $z\sim 5.9$, 6.8, 7.9, 9.0 and 10.4 shown by the blue pentagons. The $\rho_{{\rm Ly}\alpha}^{\rm LAE}$ and $\rho_{\rm UV}^{\rm LAE}$ data and the solid lines connecting them are the same as the data and the dotted lines plotted in Figure \ref{Madau_Plot} (see the caption of Figure \ref{Madau_Plot} and the references therein). The $\rho_{\rm UV}^{\rm LBG}$ data and the solid lines fitted to the data are taken from Figure 11 of \citet{Konno14}. They were originally taken from \citet{Bouwens15a} for $z=5.9$, 6.8, 7.9, and 10.4, and \citet{Ellis13} for $z=9.0$. \label{rhoLyaLAE_rhoUVLAE_rhoUVLBGs_vs_z}}
%The $\rho_{\rm UV}^{\rm LBG}$ data and the solid/extrapolated dotted lines fitted to the data are taken from Figure 11 of \citet{Konno14}.
\end{figure}

In Figure \ref{Madau_Plot}, we plot the number, Ly$\alpha$ luminosity and UV luminosity densities ($n_{{\rm Ly}\alpha}^{\rm obs}$, $n_{\rm UV}^{\rm obs}$, $\rho_{{\rm Ly}\alpha}^{\rm obs}$ and $\rho_{\rm UV}^{\rm obs}$) of the $z=5.7$, 6.6, 7 and 7.3 LAEs as a function of redshift and cosmic time (we plot only $n_{{\rm Ly}\alpha}^{\rm obs}$ and $\rho_{{\rm Ly}\alpha}^{\rm obs}$ in the case of the $z=7.3$ LAEs as the $z=7.3$ UV LF is not available). Furthermore, although not listed in Tables \ref{Best-fitSchechter} and \ref{n_rho_UV}, we also calculate $n_{{\rm Ly}\alpha}^{\rm obs}$ and $\rho_{{\rm Ly}\alpha}^{\rm obs}$ of LAEs at $z=3.1$, 3.7 and 4.5 to $\log L({\rm Ly}\alpha)$ (erg s$^{-1}$) $=$ 42.4 as well as $n_{\rm UV}^{\rm obs}$ and $\rho_{\rm UV}^{\rm obs}$ of LAEs at $z=3.1$ and 3.7 to $M_{\rm UV}=-21$ mag by integrating the best-fit Schechter functions of the $z=3.1$, 3.7 and 4.5 LAE Ly$\alpha$ LFs and the $z=3.1$ and 3.7 LAE UV LFs derived by \citet{Ouchi08} and \citet{Dawson07}. We also plot these densities in Figure \ref{Madau_Plot} to trace the evolution of the densities over the wide redshift range of $z=3.1$--7.3. 

Figure \ref{Madau_Plot} shows that $n_{{\rm Ly}\alpha}^{\rm obs}$ and $\rho_{{\rm Ly}\alpha}^{\rm obs}$ do not vary much at $z=3.1$--5.7 (except for the decrease at $z=4.5$; see below for more details), mildly decrease at $z=5.7$--6.6, slightly more rapidly decrease at $z=6.6$--7 and even more rapidly decrease at $z=7$--7.3. The rates of the decrease in $n_{{\rm Ly}\alpha}^{\rm obs}$ and $\rho_{{\rm Ly}\alpha}^{\rm obs}$ increase as redshift gets higher at $z=5.7$--7.3. Meanwhile, Figure \ref{Madau_Plot} also shows that $n_{\rm UV}^{\rm obs}$ and $\rho_{\rm UV}^{\rm obs}$ do not change much at $z=3.1$--3.7, increase at $z=3.7$--5.7, apparently very slightly increase or possibly stay constant (given large errors) at $z=5.7$--6.6 and then decrease at $z=6.6$--7. 

The possible origins of the decreases in $n_{{\rm Ly}\alpha}^{\rm obs}$ and $\rho_{{\rm Ly}\alpha}^{\rm obs}$ (the filled circles in Figure \ref{Madau_Plot}) at $z=4.5$ might be (1) the difference in telescope/instruments/broadband filters used for the observations and most likely (2) the $z=4.5$ Ly$\alpha$ LF based on only the spectroscopically confirmed LAEs in the \citet{Dawson07} sample. As for (1), the \citet{Dawson07} LAE sample is based on observations using the Mosaic CCD cameras on the 4 m Mayall Telescope at Kitt Peak National Observatory and on the 4 m Blanco Telescope at Cerro Tololo Inter-American Observatory while all the other data in Figure \ref{Madau_Plot} are based on Subaru Suprime-Cam observations with its narrowband and broadband filter set. This difference might cause some systematic difference in selecting LAEs. As to the most likely origin (2), by using only spectroscopically confirmed LAEs, both contaminations and photometric LAE candidates not yet spectroscopically confirmed are removed from the $z=4.5$ Ly$\alpha$ LF. Thus, the $n_{{\rm Ly}\alpha}^{\rm obs}$ and $\rho_{{\rm Ly}\alpha}^{\rm obs}$ of LAEs at $z=4.5$ could be lower than those estimated from the \citet{Ouchi08} $z=3.1$, 3.7 and 5.7 Ly$\alpha$ LFs based on LAE samples including both large fraction of photometric candidates and smaller fraction of spectroscopically confirmed LAEs. For example, the difference in densities caused by spectroscopy fraction can be seen for the $z=5.7$ and 6.6 $n_{{\rm Ly}\alpha}^{\rm obs}$ and $\rho_{{\rm Ly}\alpha}^{\rm obs}$ derived by using the \citet{Ouchi08,Ouchi10} $z=5.7$ and 6.6 Ly$\alpha$ LFs based on large fraction of photometric LAE candidates (the left data points at $z=5.7$ and 6.6 in Figure \ref{Madau_Plot}) and the \citet{Kashikawa11} $z=5.7$ and 6.6 Ly$\alpha$ LFs based on large fraction of spectroscopically confirmed LAEs (the right data points at $z=5.7$ and 6.6 in Figure \ref{Madau_Plot}). The  $z=5.7$ and 6.6 $n_{{\rm Ly}\alpha}^{\rm obs}$ and $\rho_{{\rm Ly}\alpha}^{\rm obs}$ based on the \citet{Kashikawa11} LFs are lower than those based on the \citet{Ouchi08,Ouchi10} LFs. Moreover, based on their spectroscopy, \citet{Dawson07} estimated their LAE selection reliability to be $\sim 76$\%. If we correct the $z=4.5$ $n_{{\rm Ly}\alpha}^{\rm obs}$ and $\rho_{{\rm Ly}\alpha}^{\rm obs}$ (the filled circles at $z=4.5$ in Figure \ref{Madau_Plot}) for this reliability to obtain the densities based on the ``photometric'' $z=4.5$ LAE sample, they become closer to and consistent within errors with those at $z=3.7$ as shown by the open circles at $z=4.5$ in Figure \ref{Madau_Plot}. Hence, we consider that the $n_{{\rm Ly}\alpha}^{\rm obs}$ and $\rho_{{\rm Ly}\alpha}^{\rm obs}$ do not change much among $z=3.1$, 3.7, 4.5 and 5.7.
%If we correct the $z=4.5$ $n_{{\rm Ly}\alpha}^{\rm obs}$ and $\rho_{{\rm Ly}\alpha}^{\rm obs}$ for this reliability, they are $\log n_{{\rm Ly}\alpha}^{\rm obs} = -3.43_{-0.19}^{+0.12}$ and $\rho_{{\rm Ly}\alpha}^{\rm obs}= 39.34_{-0.24}^{+0.15}$ shown by the open circles in Figure  \ref{Madau_Plot}. 

%Table 8 
\begin{turnpage}
\begin{deluxetable*}{llcccccc}
\tabletypesize{\scriptsize}
%\rotate
\tablecaption{Changes from Lower Redshifts in Number and Ly$\alpha$ Luminosity Densities ($n_{{\rm Ly}\alpha}^{{\rm obs}}$, $\rho_{{\rm Ly}\alpha}^{{\rm obs}}$, $\rho_{{\rm Ly}\alpha}^{{\rm tot}}$) of LAEs at $z=6.6$, 7 and 7.3\label{n_rho_Lya}}
\tablewidth{0pt}
%\tablewidth{490pt}
\tablehead{
%\multicolumn{8}{c}{Decreases in $n_{{\rm Ly}\alpha}^{{\rm obs}}$, $\rho_{{\rm Ly}\alpha}^{{\rm obs}}$ and $\rho_{{\rm Ly}\alpha}^{{\rm tot}}$ from $z=5.7$ to 6.6, 7 and 7.3}\\
\multicolumn{8}{c}{Changes from $z=5.7$ to 6.6, 7 and 7.3}\\ 
\hline
$z$ & LF$^{\rm a}$ for $n_{{\rm Ly}\alpha}^{{\rm obs},z}$, $\rho_{{\rm Ly}\alpha}^{{\rm obs},z}$, $\rho_{{\rm Ly}\alpha}^{{\rm tot},z}$ & \multicolumn{3}{c}{\underline{\citet{Ouchi08} $z=5.7$ Ly$\alpha$ LF$^{\rm e}$}} & \multicolumn{3}{c}{\underline{\citet{Kashikawa11} $z=5.7$ Ly$\alpha$ LF$^{\rm e}$}}\\                           
%\cline{3-5}\cline{6-8}
&  & $n_{{\rm Ly}\alpha}^{{\rm obs},z}$/$n_{{\rm Ly}\alpha}^{{\rm obs},z=5.7}$ & $\rho_{{\rm Ly}\alpha}^{{\rm obs},z}$/$\rho_{{\rm Ly}\alpha}^{{\rm obs},z=5.7}$ & $\rho_{{\rm Ly}\alpha}^{{\rm tot},z}$/$\rho_{{\rm Ly}\alpha}^{{\rm tot},z=5.7}$ & $n_{{\rm Ly}\alpha}^{{\rm obs},z}$/$n_{{\rm Ly}\alpha}^{{\rm obs},z=5.7}$ & $\rho_{{\rm Ly}\alpha}^{{\rm obs},z}$/$\rho_{{\rm Ly}\alpha}^{{\rm obs},z=5.7}$ & $\rho_{{\rm Ly}\alpha}^{{\rm tot},z}$/$\rho_{{\rm Ly}\alpha}^{{\rm tot},z=5.7}$
}
\startdata
6.6 & \citet{Ouchi10}                               & 0.60$_{-0.33}^{+0.75}$ & 0.53$_{-0.30}^{+0.74}$ & 0.72$_{-0.35}^{+0.66}$ & 1.01$_{-0.33}^{+0.54}$ & 0.79$_{-0.27}^{+0.47}$ & 1.33$_{-0.36}^{+0.60}$ \\
6.6 & \citet{Kashikawa11}                           & 0.55$_{-0.35}^{+1.05}$ & 0.52$_{-0.34}^{+1.04}$ & 0.57$_{-0.36}^{+0.94}$ & 0.92$_{-0.44}^{+0.92}$ & 0.77$_{-0.37}^{+0.77}$ & 1.06$_{-0.50}^{+1.05}$ \\
%\hline
7.0 & Differential LF$^{\rm b}$ & 0.36$_{-0.27}^{+0.53}$ & 0.33$_{-0.26}^{+0.52}$ & 0.38$_{-0.28}^{+0.47}$ & 0.60$_{-0.38}^{+0.42}$ & 0.50$_{-0.32}^{+0.35}$ & 0.70$_{-0.45}^{+0.49}$ \\
7.0 & Cum LF excl.~0 bins$^{\rm c}$  & 0.48$_{-0.36}^{+1.80}$ & 0.47$_{-0.36}^{+1.81}$ & 0.49$_{-0.36}^{+1.60}$ & 0.81$_{-0.51}^{+1.82}$ & 0.70$_{-0.44}^{+1.56}$ & 0.90$_{-0.57}^{+2.01}$ \\
7.0 & Cum LF incl.~0 bins$ ^{\rm d}$ & 0.27$_{-0.19}^{+0.72}$ & 0.29$_{-0.21}^{+0.82}$ & 0.25$_{-0.17}^{+0.58}$ & 0.45$_{-0.26}^{+0.68}$ & 0.44$_{-0.26}^{+0.66}$ & 0.46$_{-0.27}^{+0.69}$ \\
%\hline
7.3 & \citet{Konno14}                               & 0.11$_{-0.11}^{+1.40}$ & 0.09$_{-0.06}^{+0.18}$ & 0.20$_{-0.15}^{+0.82}$ & 0.19$_{-0.17}^{+1.55}$ & 0.13$_{-0.06}^{+0.13}$ & 0.36$_{-0.25}^{+1.06}$ \\
\hline
%\multicolumn{8}{c}{Decreases in $n_{{\rm Ly}\alpha}^{{\rm obs}}$, $\rho_{{\rm Ly}\alpha}^{{\rm obs}}$ and $\rho_{{\rm Ly}\alpha}^{{\rm tot}}$ from $z=6.6$ to 7 and 7.3}\\ 
\multicolumn{8}{c}{Changes from $z=6.6$ to 7 and 7.3}\\ 
\hline
$z$ & LF$^{\rm a}$ for $n_{{\rm Ly}\alpha}^{{\rm obs},z}$, $\rho_{{\rm Ly}\alpha}^{{\rm obs},z}$, $\rho_{{\rm Ly}\alpha}^{{\rm tot},z}$ & \multicolumn{3}{c}{\underline{\citet{Ouchi10} $z=6.6$ Ly$\alpha$ LF$^{\rm e}$}} & \multicolumn{3}{c}{\underline{\citet{Kashikawa11} $z=6.6$ Ly$\alpha$ LF$^{\rm e}$}}\\                           
%\cline{3-5}\cline{6-8}
&  & $n_{{\rm Ly}\alpha}^{{\rm obs},z}$/$n_{{\rm Ly}\alpha}^{{\rm obs},z=6.6}$ & $\rho_{{\rm Ly}\alpha}^{{\rm obs},z}$/$\rho_{{\rm Ly}\alpha}^{{\rm obs},z=6.6}$ & $\rho_{{\rm Ly}\alpha}^{{\rm tot},z}$/$\rho_{{\rm Ly}\alpha}^{{\rm tot},z=6.6}$ & $n_{{\rm Ly}\alpha}^{{\rm obs},z}$/$n_{{\rm Ly}\alpha}^{{\rm obs},z=6.6}$ & $\rho_{{\rm Ly}\alpha}^{{\rm obs},z}$/$\rho_{{\rm Ly}\alpha}^{{\rm obs},z=6.6}$ & $\rho_{{\rm Ly}\alpha}^{{\rm tot},z}$/$\rho_{{\rm Ly}\alpha}^{{\rm tot},z=6.6}$\\
\hline
7.0 & Differential LF$^{\rm b}$ & 0.59$_{-0.38}^{+0.40}$ & 0.63$_{-0.42}^{+0.45}$ & 0.53$_{-0.33}^{+0.28}$ & 0.65$_{-0.47}^{+0.74}$ & 0.64$_{-0.47}^{+0.73}$ & 0.66$_{-0.48}^{+0.75}$ \\
7.0 & Cum LF excl.~0 bins$^{\rm c}$  & 0.80$_{-0.51}^{+1.76}$ & 0.89$_{-0.57}^{+2.00}$ & 0.68$_{-0.41}^{+1.30}$ & 0.88$_{-0.63}^{+2.70}$ & 0.90$_{-0.65}^{+2.78}$ & 0.85$_{-0.61}^{+2.62}$ \\
7.0 & Cum LF incl.~0 bins$ ^{\rm d}$ & 0.45$_{-0.26}^{+0.66}$ & 0.56$_{-0.34}^{+0.85}$ & 0.34$_{-0.19}^{+0.44}$ & 0.49$_{-0.33}^{+1.06}$ & 0.57$_{-0.39}^{+1.22}$ & 0.43$_{-0.30}^{+0.94}$ \\
%\hline
7.3 & \citet{Konno14}                               & 0.19$_{-0.17}^{+1.51}$ & 0.16$_{-0.08}^{+0.17}$ & 0.27$_{-0.18}^{+0.69}$ & 0.20$_{-0.19}^{+2.17}$ & 0.17$_{-0.10}^{+0.26}$ & 0.34$_{-0.26}^{+1.35}$ \\
\hline
%\multicolumn{8}{c}{Decreases in $n_{{\rm Ly}\alpha}^{{\rm obs}}$, $\rho_{{\rm Ly}\alpha}^{{\rm obs}}$ and $\rho_{{\rm Ly}\alpha}^{{\rm tot}}$ from $z=7$ to 7.3}\\ 
\multicolumn{8}{c}{Changes from $z=7$ to 7.3}\\ 
\hline
$z$ & LF$^{\rm a}$ for $n_{{\rm Ly}\alpha}^{{\rm obs},z}$, $\rho_{{\rm Ly}\alpha}^{{\rm obs},z}$, $\rho_{{\rm Ly}\alpha}^{{\rm tot},z}$ & \multicolumn{3}{c}{\underline{Differential $z=7$ Ly$\alpha$ LF (this study)$^{\rm e}$}} & \multicolumn{3}{c}{} \\                           
%\cline{3-5}
&  & $n_{{\rm Ly}\alpha}^{{\rm obs},z}$/$n_{{\rm Ly}\alpha}^{{\rm obs},z=7}$ & $\rho_{{\rm Ly}\alpha}^{{\rm obs},z}$/$\rho_{{\rm Ly}\alpha}^{{\rm obs},z=7}$ & $\rho_{{\rm Ly}\alpha}^{{\rm tot},z}$/$\rho_{{\rm Ly}\alpha}^{{\rm tot},z=7}$ & & & \\
\hline
7.3 & \citet{Konno14}                               & 0.31$_{-0.29}^{+4.96}$ & 0.26$_{-0.14}^{+0.72}$ & 0.52$_{-0.37}^{+3.18}$ & & & 
\enddata
%\tablecomments{}
\tablenotetext{a}{The Ly$\alpha$ LFs from the references listed in this column are used to calculate the numerators of the ratios (i.e., $n_{{\rm Ly}\alpha}^{{\rm obs}, z}$, $\rho_{{\rm Ly}\alpha}^{{\rm obs}, z}$ and $\rho_{{\rm Ly}\alpha}^{{\rm tot}, z}$ at the redshifts $z$ in column 1).}
\tablenotetext{b}{The best-fit Schechter function of the differential $z=7$ Ly$\alpha$ LF (see the left panel of Figure \ref{plottingLyaLF}) is used to calculate the numerators of the ratios (i.e., $n_{{\rm Ly}\alpha}^{{\rm obs},z}$, $\rho_{{\rm Ly}\alpha}^{{\rm obs},z}$, $\rho_{{\rm Ly}\alpha}^{{\rm tot},z}$).}
\tablenotetext{c}{The best-fit Schechter function of the cumulative $z=7$ Ly$\alpha$ LF excluding the faintest bins where no LAE is detected (see the middle panel of Figure \ref{plottingLyaLF}) is used to calculate the numerators of the ratios.}
\tablenotetext{d}{The best-fit Schechter function of the cumulative $z=7$ Ly$\alpha$ LF including the faintest bins (see the right panel of Figure \ref{plottingLyaLF}) is used to calculate the numerators of the ratios.}
\tablenotetext{e}{The Ly$\alpha$ LFs used to calculate the denominators of the ratios (i.e., $n_{{\rm Ly}\alpha}^{{\rm obs}}$, $\rho_{{\rm Ly}\alpha}^{{\rm obs}}$ and $\rho_{{\rm Ly}\alpha}^{{\rm tot}}$ at $z=5.7$, 6.6 and 7).}
\end{deluxetable*}

% Table 9
\begin{deluxetable*}{llcccccc}
\tabletypesize{\scriptsize}
%\rotate
\tablecaption{Changes from Lower Redshifts in Number and UV Continuum Luminosity Densities ($n_{\rm UV}^{\rm obs}$, $\rho_{\rm UV}^{\rm obs}$) of LAEs at $z=6.6$ and 7\label{Ratios_nUV_rhoUV}}
\tablewidth{0pt}
%\tablewidth{510pt}
\tablehead{
\multicolumn{2}{c}{} & \multicolumn{4}{c}{Changes from $z=5.7$ to 6.6 and 7} & \multicolumn{2}{c}{Changes from $z=6.6$ to 7}\\ 
\hline
$z$  & UV LF$^{\rm a}$ for $n_{\rm UV}^{\rm obs, z}$, $\rho_{\rm UV}^{\rm obs, z}$ & \multicolumn{2}{c}{\underline{\citet{Ouchi08} $z=5.7$ UV LF$^{\rm b}$}} & \multicolumn{2}{c}{\underline{\citet{Kashikawa11} $z=5.7$ UV LF$^{\rm b}$}}& \multicolumn{2}{c}{\underline{\citet{Kashikawa11} $z=6.6$ UV LF$^{\rm b}$}}\\         
     &           & $n_{\rm UV}^{\rm obs,z}$/$n_{\rm UV}^{\rm obs,z=5.7}$ & $\rho_{\rm UV}^{\rm obs,z}$/$\rho_{\rm UV}^{\rm obs,z=5.7}$ & $n_{\rm UV}^{\rm obs,z}$/$n_{\rm UV}^{\rm obs,z=5.7}$ & $\rho_{\rm UV}^{\rm obs,z}$/$\rho_{\rm UV}^{\rm obs,z=5.7}$ & $n_{\rm UV}^{\rm obs,z}$/$n_{\rm UV}^{\rm obs,z=6.6}$ & $\rho_{\rm UV}^{\rm obs,z}$/$\rho_{\rm UV}^{\rm obs,z=6.6}$   
}
\startdata
6.6 & \citet{Kashikawa11} & 1.3$_{-1.2}^{+20.7}$          & 1.7$_{-1.6}^{+35.8}$  & 1.2$_{-1.1}^{+15.2}$  & 1.2$_{-1.1}^{+19.6}$  & ---                & --- \\
7.0 & This study          & 0.71$_{-0.61}^{+5.5}$         & 0.57$_{-0.55}^{+10.9}$  & 0.66$_{-0.59}^{+3.99}$ & 0.38$_{-0.37}^{+5.97}$ & 0.55$_{-0.44}^{+1.51}$ & 0.33$_{-0.32}^{+3.09}$ 
\enddata
%% Text for table notes should follow after the \enddata but before
%% the \end{deluxetable}. Make sure there is at least one \tablenotemark
%% in the table for each \tablenotetext.
%\tablecomments{}
\tablenotetext{a}{The UV LFs from the references listed in this column are used to calculate the numerators of the ratios (i.e., $n_{\rm UV}^{\rm obs,z}$ and $\rho_{\rm UV}^{\rm obs,z}$ at the redshifts $z$ in column 1).}
\tablenotetext{b}{The UV LFs used to calculate the denominators of the ratios (i.e., $n_{\rm UV}^{\rm obs}$ and $\rho_{\rm UV}^{\rm obs}$ at $z=5.7$ and 6.6).}
\end{deluxetable*}
\end{turnpage}

While $n_{{\rm Ly}\alpha}^{\rm obs}$ and $\rho_{{\rm Ly}\alpha}^{\rm obs}$ of LAEs do not evolve much at $z=3.1$--5.7, they seem to decrease with increasing rate at $z=5.7$--7.3 in Figure \ref{Madau_Plot}. Meanwhile, $n_{\rm UV}^{\rm obs}$ and $\rho_{\rm UV}^{\rm obs}$ of LAEs slightly increase or almost stay unchanged at $z=5.7$--6.6 (given large errors) and decrease at $z=6.6$--7. Figure \ref{rhoLyaLAE_rhoUVLAE_rhoUVLBGs_vs_z} shows the close-up of the change in Ly$\alpha$ and UV luminosity densities $\rho_{{\rm Ly}\alpha}^{\rm obs}$ and $\rho_{\rm UV}^{\rm obs}$ of LAEs at $z=5.7$--7.3 and at $z=5.7$--7, respectively (labeled $\rho_{{\rm Ly}\alpha}^{\rm LAE}$ and $\rho_{\rm UV}^{\rm LAE}$ in the figure). The rate of decrease in $\rho_{{\rm Ly}\alpha}^{\rm LAE}$ seems to increase as redshift gets higher. On the other hand, the $\rho_{\rm UV}^{\rm LAE}$ seems to decrease with a similar or slightly higher rate than that of $\rho_{{\rm Ly}\alpha}^{\rm LAE}$. To look into this quantitatively, we calculate and list the ratios of the number and Ly$\alpha$ luminosity densities at $z$ ($=$ 6.6, 7 or 7.3) to those at $z=5.7$, $n_{{\rm Ly}\alpha}^{\rm obs, z}/n_{{\rm Ly}\alpha}^{\rm obs, z=5.7}$, $\rho_{{\rm Ly}\alpha}^{\rm obs, z}/\rho_{{\rm Ly}\alpha}^{\rm obs, z=5.7}$ and $\rho_{{\rm Ly}\alpha}^{\rm tot, z}/\rho_{{\rm Ly}\alpha}^{\rm tot, z=5.7}$, in Table \ref{n_rho_Lya}. The ratios of the densities at $z$ ($=$ 7 or 7.3) to those at $z=6.6$ as well as the densities at $z$ ($=$ 7.3) to those at $z=7$ are also listed in Table \ref{n_rho_Lya}. Similarly, we also calculate and list the ratios of the number and UV luminosity densities at $z$ ($=$ 6.6 or 7) to those at $z=5.7$, $n_{\rm UV}^{\rm obs, z}/n_{\rm UV}^{\rm obs, z=5.7}$ and $\rho_{\rm UV}^{\rm obs, z}/\rho_{\rm UV}^{\rm obs, z=5.7}$, in Table \ref{Ratios_nUV_rhoUV}. The ratios of the UV number and luminosity densities at $z$ ($=$ 7) to those at $z=6.6$, $n_{\rm UV}^{\rm obs, z}/n_{\rm UV}^{\rm obs, z=6.6}$ and $\rho_{\rm UV}^{\rm obs, z}/\rho_{\rm UV}^{\rm obs, z=6.6}$, are also listed in Table \ref{Ratios_nUV_rhoUV}. As seen in Tables \ref{n_rho_Lya} and \ref{Ratios_nUV_rhoUV}, we calculate the ratios using combinations of densities obtained from different Ly$\alpha$/UV LFs derived by different authors or methods to see whether the trend of change in the ratio with redshift depends on how the Ly$\alpha$/UV LFs were derived. 

The ``Changes from $z=5.7$ to 6.6, 7 and 7.3'' section of Table \ref{n_rho_Lya} shows that the ratios $n_{{\rm Ly}\alpha}^{\rm obs, z}/n_{{\rm Ly}\alpha}^{\rm obs, z=5.7}$, $\rho_{{\rm Ly}\alpha}^{\rm obs, z}/\rho_{{\rm Ly}\alpha}^{\rm obs, z=5.7}$ and $\rho_{{\rm Ly}\alpha}^{\rm tot, z}/\rho_{{\rm Ly}\alpha}^{\rm tot, z=5.7}$ all continue to decrease as redshift increases from $z=6.6$ via 7 to 7.3 (i.e. $n_{{\rm Ly}\alpha}^{\rm obs, z=6.6}/n_{{\rm Ly}\alpha}^{\rm obs, z=5.7} > n_{{\rm Ly}\alpha}^{\rm obs, z=7}/n_{{\rm Ly}\alpha}^{\rm obs, z=5.7} > n_{{\rm Ly}\alpha}^{\rm obs, z=7.3}/n_{{\rm Ly}\alpha}^{\rm obs, z=5.7}$, $\rho_{{\rm Ly}\alpha}^{\rm obs, z=6.6}/\rho_{{\rm Ly}\alpha}^{\rm obs, z=5.7} > \rho_{{\rm Ly}\alpha}^{\rm obs, z=7}/\rho_{{\rm Ly}\alpha}^{\rm obs, z=5.7} > \rho_{{\rm Ly}\alpha}^{\rm obs, z=7.3}/\rho_{{\rm Ly}\alpha}^{\rm obs, z=5.7}$ and $\rho_{{\rm Ly}\alpha}^{\rm tot, z=6.6}/\rho_{{\rm Ly}\alpha}^{\rm tot, z=5.7} > \rho_{{\rm Ly}\alpha}^{\rm tot, z=7}/\rho_{{\rm Ly}\alpha}^{\rm tot, z=5.7} > \rho_{{\rm Ly}\alpha}^{\rm tot, z=7.3}/\rho_{{\rm Ly}\alpha}^{\rm tot, z=5.7}$) no matter what combination of densities obtained from different Ly$\alpha$ LFs derived by different authors or methods is considered. On the other hand, Table \ref{Ratios_nUV_rhoUV} shows that $n_{\rm UV}^{\rm obs, z}/n_{\rm UV}^{\rm obs, z=5.7}$ and $\rho_{\rm UV}^{\rm obs, z}/\rho_{\rm UV}^{\rm obs, z=5.7}$ very modestly increase from $z=5.7$ to 6.6 and decrease from $z=6.6$ to 7. Comparison of the ``Changes from $z=6.6$ to 7 and 7.3'' section of Table \ref{n_rho_Lya} and the ``Changes from $z=6.6$ to 7'' section of Table \ref{Ratios_nUV_rhoUV} shows that the decreases in the number and luminosity densities in UV continuum from $z=6.6$ to 7 are comparable or more rapid than those in the number and luminosity densities in Ly$\alpha$ from $z=6.6$ to 7 (i.e. $n_{\rm UV}^{\rm obs, z=7}/n_{\rm UV}^{\rm obs, z=6.6}  \lesssim n_{{\rm Ly}\alpha}^{\rm obs, z=7}/n_{{\rm Ly}\alpha}^{\rm obs, z=6.6}$ and $\rho_{\rm UV}^{\rm obs, z=7}/\rho_{\rm UV}^{\rm obs, z=6.6}  \lesssim \rho_{{\rm Ly}\alpha}^{\rm obs, z=7}/\rho_{{\rm Ly}\alpha}^{\rm obs, z=6.6}$).

% Table 10
\begin{deluxetable*}{ccccccll}
\tabletypesize{\scriptsize}
%\rotate
\tablecaption{Time Evolution of Rates of Decrease in the Number, Ly$\alpha$ Luminosity and UV Luminosity Densities of LAEs at $z=5.7$--7.3 \label{Density_Rate_Evolution}}
%\tablewidth{0pt}
%\tablewidth{510pt}
\tablehead{
Redshift    & $\Delta t^{\rm a}$  & $\frac{\Delta n_{{\rm Ly}\alpha}^{\rm obs}}{\Delta t}$ & $\frac{\Delta n_{\rm UV}^{\rm obs}}{\Delta t}$ & $\frac{\Delta \rho_{{\rm Ly}\alpha}^{\rm obs}}{\Delta t}$ & $\frac{\Delta \rho_{\rm UV}^{\rm obs}}{\Delta t}$ & Ly$\alpha$ LF$^{\rm b}$ used for $n_{{\rm Ly}\alpha}^{\rm obs}$, $\rho_{{\rm Ly}\alpha}^{\rm obs}$ & UV LF$^{\rm b}$ used for $n_{\rm UV}^{\rm obs}$, $\rho_{\rm UV}^{\rm obs}$\\   
$z_1 - z_2$  & [Myr] & [Gyr$^{-1}$] & [Gyr$^{-1}$] & [Gyr$^{-1}$] & [Gyr$^{-1}$] & &
%\hline  
%\multicolumn{8}{l}{Using \citet{Ouchi08,Ouchi10}'s $z=5.7$, 6.6 Ly$\alpha$ LFs for $n_{{\rm Ly}\alpha}^{\rm obs}$, $\rho_{{\rm Ly}\alpha}^{\rm obs}$}
}
\startdata
5.7--6.6    & 160   & $2.5_{-4.7}^{+2.1}$    & $-1.3_{-95.0}^{+6.9}$ & $2.9_{-4.6}^{+1.9}$  & $-1.3_{-122.5}^{+6.9}$ & O08 (5.7), O10 (6.6) & K11 (5.7, 6.6)\\
6.6--7.0    &  60   & $6.8_{-6.7}^{+6.3}$   & $7.5_{-25.2}^{+7.3}$  & $6.2_{-7.5}^{+7.0}$ & $11.2_{-51.5}^{+5.3}$ & O10 (6.6), This study (7.0)$^{\rm c}$ & K11 (6.6), This study (7.0)\\
%\hline
%\multicolumn{8}{l}{Using \citet{Kashikawa11}'s $z=5.7$, 6.6 Ly$\alpha$ LFs for $n_{{\rm Ly}\alpha}^{\rm obs}$, $\rho_{{\rm Ly}\alpha}^{\rm obs}$}\\
\hline
5.7--6.6    & 160   & $0.5_{-5.8}^{+2.8}$    & $-1.3_{-95.0}^{+6.9}$ & $1.4_{-4.8}^{+2.3}$   & $-1.3_{-122.5}^{+6.9}$ & K11 (5.7, 6.6) & K11 (5.7, 6.6)\\
6.6--7.0    &  60   & $2.0_{-45.0}^{+10.5}$  & $7.5_{-25.2}^{+7.3}$  & $1.7_{-46.3}^{+10.8}$ & $11.2_{-51.5}^{+5.3}$ & K11 (6.6), This study (7.0)$^{\rm d}$ & K11 (6.6), This study (7.0)\\
6.6--7.0    &  60   & $8.5_{-17.7}^{+5.5}$   & $7.5_{-25.2}^{+7.3}$  & $7.2_{-20.3}^{+6.5}$  & $11.2_{-51.5}^{+5.3}$ & K11 (6.6), This study (7.0)$^{\rm f}$ & K11 (6.6), This study (7.0)\\
\hline
7.0--7.3    &  40   & $17.3_{-124.0}^{+7.3}$ & ---                   & $18.5_{-18.0}^{+3.5}$ & ---                   & This study (7.0)$^{\rm c}$, K14 (7.3) & ---
\enddata 
%% Text for table notes should follow after the \enddata but before
%% the \end{deluxetable}. Make sure there is at least one \tablenotemark
%% in the table for each \tablenotetext.
\tablecomments{See text in Section \ref{Number_Luminosity_Densities} for the definitions of the decrease rates in densities $\Delta n_{{\rm Ly}\alpha}^{\rm obs}/\Delta t$, $\Delta n_{\rm UV}^{\rm obs}/\Delta t$, $\Delta \rho_{{\rm Ly}\alpha}^{\rm obs}/\Delta t$ and $\Delta \rho_{\rm UV}^{\rm obs}/\Delta t$. A plus value means decrease in a density while a minus value means increase in a density.}
\tablenotetext{a}{Cosmic time interval in Myr corresponding to the redshift range $z_1 - z_2$ in column 1.}
\tablenotetext{b}{Ly$\alpha$ and UV LFs used to compute the densities. The number in parenthesis is redshift of the LF. The references from which the LFs come from are: O08 = \citet{Ouchi08}, O10 = \citet{Ouchi10}, K11 = \citet{Kashikawa11}, K14 = \citet{Konno14}.}
\tablenotetext{c}{The $z=7$ differential Ly$\alpha$ LF from the left panel in Figure \ref{plottingLyaLF}.}
\tablenotetext{d}{The $z=7$ cumulative Ly$\alpha$ LF excluding the faintest bins where no LAE is detected from the middle panel in Figure \ref{plottingLyaLF}.}
\tablenotetext{f}{The $z=7$ cumulative Ly$\alpha$ LF including the faintest bins where no LAE is detected from the right panel in Figure \ref{plottingLyaLF}.}
\end{deluxetable*}

Furthermore, we also calculate the rates of the decrease in the densities, $\Delta n_{{\rm Ly}\alpha}^{\rm obs}/\Delta t$, $\Delta n_{\rm UV}^{\rm obs}/\Delta t$, $\Delta \rho_{{\rm Ly}\alpha}^{\rm obs}/\Delta t$ and $\Delta \rho_{\rm UV}^{\rm obs}/\Delta t$ in Gyr$^{-1}$ at $z=5.7$--6.6, 6.6--7 and 7--7.3 and list them in Table \ref{Density_Rate_Evolution}. We define these decrease rates at a redshift range $z=z_1$ -- $z_2$ corresponding to a cosmic time interval $t=t_1$ -- $t_2$ Gyr as $\Delta n^{\rm obs}/\Delta t = (1 - n_{z_2}^{\rm obs}/n_{z_1}^{\rm obs})/(t_2 - t_1)$ and $\Delta \rho^{\rm obs}/\Delta t = (1 - \rho_{z_2}^{\rm obs}/\rho_{z_1}^{\rm obs})/(t_2 - t_1)$. Here, $n_{z_1}^{\rm obs}$, $n_{z_2}^{\rm obs}$, $\rho_{z_1}^{\rm obs}$ and $\rho_{z_2}^{\rm obs}$ are the number and luminosity densities at the redshifts $z_1$ and $z_2$. Also, the time intervals are $\Delta t = t_2 - t_1 =$ 0.16, 0.06 and 0.04 Gyr for $z=5.7$--6.6, 6.6--7 and 7--7.3, respectively. Figures \ref{Madau_Plot} and \ref{rhoLyaLAE_rhoUVLAE_rhoUVLBGs_vs_z} show only the $n_{{\rm Ly}\alpha}^{\rm obs}$ and $\rho_{{\rm Ly}\alpha}^{\rm obs}$ derived from our $z=7$ differential Ly$\alpha$ LF in the left panel of Figure \ref{plottingLyaLF}. In Table \ref{Density_Rate_Evolution}, we also calculate $\Delta n_{{\rm Ly}\alpha}^{\rm obs}/\Delta t$ and $\Delta \rho_{{\rm Ly}\alpha}^{\rm obs}/\Delta t$ at $z=6.6$--7 by using the $z=7$ LAE $n_{{\rm Ly}\alpha}^{\rm obs}$ and $\rho_{{\rm Ly}\alpha}^{\rm obs}$ derived from our $z=7$ cumulative Ly$\alpha$ LFs excluding (including) the faintest bins where no LAE is detected shown in the middle (right) panel of Figure \ref{plottingLyaLF}. 

Table \ref{Density_Rate_Evolution} shows that $\Delta n_{{\rm Ly}\alpha}^{\rm obs}/\Delta t$ and $\Delta \rho_{{\rm Ly}\alpha}^{\rm obs}/\Delta t$ at $z=5.7$--6.6 are $\sim 2.5$ (0.5) and 2.9 (1.4) Gyr$^{-1}$, respectively, if we use $n_{{\rm Ly}\alpha}^{\rm obs}$ and $\rho_{{\rm Ly}\alpha}^{\rm obs}$ derived from the \citet{Ouchi08,Ouchi10} \citep{Kashikawa11} $z=5.7$ and 6.6 Ly$\alpha$ LFs. The $\Delta n_{{\rm Ly}\alpha}^{\rm obs}/\Delta t$ and $\Delta \rho_{{\rm Ly}\alpha}^{\rm obs}/\Delta t$ at $z=6.6$--7 are $\sim 6.8$ and 6.2 Gyr$^{-1}$, respectively, if we use $n_{{\rm Ly}\alpha}^{\rm obs}$ and $\rho_{{\rm Ly}\alpha}^{\rm obs}$ derived from the \citet{Ouchi10} $z=6.6$ differential Ly$\alpha$ LF and our $z=7$ differential Ly$\alpha$ LF. Meanwhile, $\Delta n_{{\rm Ly}\alpha}^{\rm obs}/\Delta t$ and $\Delta \rho_{{\rm Ly}\alpha}^{\rm obs}/\Delta t$ at $z=6.6$--7 are $\sim 2.0$ (8.5) and 1.7 (7.2) Gyr$^{-1}$, respectively, if we use $n_{{\rm Ly}\alpha}^{\rm obs}$ and $\rho_{{\rm Ly}\alpha}^{\rm obs}$ derived from the \citet{Kashikawa11} $z=6.6$ cumulative Ly$\alpha$ LF and our $z=7$ cumulative Ly$\alpha$ LFs excluding (including) the faintest bins. Finally, $\Delta n_{{\rm Ly}\alpha}^{\rm obs}/\Delta t$ and $\Delta \rho_{{\rm Ly}\alpha}^{\rm obs}/\Delta t$ at $z=7$--7.3 are $\sim 17.3$ and 18.5 Gyr$^{-1}$, respectively, if we use $n_{{\rm Ly}\alpha}^{\rm obs}$ and $\rho_{{\rm Ly}\alpha}^{\rm obs}$ derived from our $z=7$ differential Ly$\alpha$ LF and the \citet{Konno14} $z=7.3$ differential Ly$\alpha$ LF. On the other hand, the $\Delta n_{\rm UV}^{\rm obs}/\Delta t$ and $\Delta \rho_{\rm UV}^{\rm obs}/\Delta t$ at $z=5.7$--6.6 are $\sim -1.3$ and $-1.3$ Gyr$^{-1}$ (a negative value means that the density increases), respectively. The $\Delta n_{\rm UV}^{\rm obs}/\Delta t$ and $\Delta \rho_{\rm UV}^{\rm obs}/\Delta t$ at $z=6.6$--7 are $\sim 7.5$ and 11.2 Gyr$^{-1}$, respectively. These rates are derived using the $n_{\rm UV}^{\rm obs}$ and $\rho_{\rm UV}^{\rm obs}$ derived from the \citet{Kashikawa11} $z=5.7$ and 6.6 LAE UV LFs and our $z=7$ LAE UV LF. 

These results presented in Table \ref{Density_Rate_Evolution} suggest that rates of the decrease in $n_{{\rm Ly}\alpha}^{\rm obs}$ and $\rho_{{\rm Ly}\alpha}^{\rm obs}$ of LAEs increase as redshift gets higher at $z=5.7$--7.3. This means that the $n_{{\rm Ly}\alpha}^{\rm obs}$ and $\rho_{{\rm Ly}\alpha}^{\rm obs}$ of LAEs decrease acceleratingly at $z=5.7$--7.3. On the other hand, the $n_{\rm UV}^{\rm obs}$ and $\rho_{\rm UV}^{\rm obs}$ of LAEs very modestly increase or almost remain unchanged at $z=5.7$--6.6 and decrease at $z=6.6$--7 with a comparable rate or more rapidly than the decrease in $n_{{\rm Ly}\alpha}^{\rm obs}$ and $\rho_{{\rm Ly}\alpha}^{\rm obs}$ at $z=6.6$--7. This implies that there is almost no LAE evolution at $z=5.7$--6.6 in terms of number and Ly$\alpha$ and UV luminosities while LAEs evolve at $z=6.6$--7. Hence, the changes in $n_{{\rm Ly}\alpha}^{\rm obs}$ and $\rho_{{\rm Ly}\alpha}^{\rm obs}$ at $z=5.7$--6.6 are not due to galaxy evolution while those at $z=6.6$--7 are partly due to galaxy evolution. Moreover, the accelerating decreases in $n_{{\rm Ly}\alpha}^{\rm obs}$ and $\rho_{{\rm Ly}\alpha}^{\rm obs}$ found here would be consistent with the accelerating decrease in the best-fit Schechter function $L^*$ and $\phi^*$ parameters (i.e. increases in $\Delta L^*/\Delta t$ and $\Delta \phi^*/\Delta t$) found earlier in Section \ref{LyaLF_LorPhi} and Table \ref{PureL_PurePhi_Evolution}.

\citet{Konno14} compared the redshift evolution of $\rho_{{\rm Ly}\alpha}^{\rm obs}$ among $z=5.7$, 6.6 and 7.3 with that of $\rho_{\rm UV}$ of LBGs (dropout galaxies) at $z\sim6$--10 by assuming that the redshift evolutions of $\rho_{\rm UV}$ of LAEs and LBGs are the same. We take these LBG UV luminosity densities data from \citet{Konno14} \citep[originally from][]{Ellis13,Bouwens15a} and plot them in Figure \ref{rhoLyaLAE_rhoUVLAE_rhoUVLBGs_vs_z} (labeled $\rho_{\rm UV}^{\rm LBG}$). As seen in this figure \citep[and Figure 11 in][]{Konno14}, \citet{Konno14} argued that the rates of the decrease in $\rho_{{\rm Ly}\alpha}^{\rm LAE}$ and $\rho_{\rm UV}^{\rm LBG}$ are almost the same at $z\sim6$--6.6, but $\rho_{{\rm Ly}\alpha}^{\rm LAE}$ starts to decrease more rapidly from $z=6.6$ while $\rho_{\rm UV}^{\rm LBG}$ begins to decrease more sharply from $z\sim8$. They called these redshifts $\rho_{{\rm Ly}\alpha}^{\rm LAE}$ knee and $\rho_{\rm UV}^{\rm LBG}$ knee. Hence, as $\rho_{{\rm Ly}\alpha}^{\rm LAE}$ and $\rho_{\rm UV}^{\rm LBG}$ knees are not the same, they concluded that the rapid decrease in $\rho_{{\rm Ly}\alpha}^{\rm LAE}$ is not due to galaxy evolution but possibly because of attenuation of Ly$\alpha$ emission of LAEs by neutral IGM. 

However, strictly speaking, the redshift evolutions of $\rho_{\rm UV}$ of LAEs and LBGs ($\rho_{\rm UV}^{\rm LAE}$ and $\rho_{\rm UV}^{\rm LBG}$ versus $z$) may not necessarily be the same. This is because galaxies detected by a narrowband excess (LAEs) and those by a broadband dropout method (we call dropout galaxies LBGs for simplicity but some fraction of LAEs are included) are quite different despite some degree of overlap. A narrowband excess method detects mostly galaxies with a faint to bright Ly$\alpha$ emission and a faint UV continuum as well as small fraction of galaxies with a very bright Ly$\alpha$ emission and a bright UV continuum. Meanwhile, a dropout method detects mainly galaxies with no Ly$\alpha$ emission or a faint to bright Ly$\alpha$ emission and a bright UV continuum as well as small fraction of galaxies with a very bright Ly$\alpha$ emission and a faint UV continuum. Thus, a narrowband excess can detect galaxies with a faint Ly$\alpha$ emission and a faint UV continuum while a dropout method cannot. Moreover, a typical narrowband excess criterion does not detect galaxies with zero/faint Ly$\alpha$ emission and a bright UV continuum but a dropout method does\footnote[8]{In the case of our $z=7$ LAE selection, we detected such galaxies by the $z'-{\rm NB973}>1$ excess because the NB973 bandpass is located at the red edge of the $z'$ bandpass and because the NB973 bandpass (200\AA) is wider than that of a typical narrowband filter ($\lesssim 100$\AA). However, we removed them from our LAE sample by imposing another narrowband excess criterion $y-{\rm NB973}>0$ as the NB973 bandpass is located in the middle of the $y$ bandpass.}. 

Hence, though limited to only redshifts $z=5.7$, 6.6 and 7 and the large errors in $\rho_{\rm UV}^{\rm LAE}$, we compare redshift evolutions of the Ly$\alpha$ and UV luminosity densities of LAEs ($\rho_{{\rm Ly}\alpha}^{\rm LAE}$ and $\rho_{\rm UV}^{\rm LAE}$) along with that of the UV luminosity density of LBGs ($\rho_{\rm UV}^{\rm LBG}$) in Figure \ref{rhoLyaLAE_rhoUVLAE_rhoUVLBGs_vs_z}. As mentioned earlier, both $\rho_{{\rm Ly}\alpha}^{\rm LAE}$ and $\rho_{\rm UV}^{\rm LAE}$ start to decrease rapidly at $z=6.6$ with the comparable rate or the higher rate in $\rho_{\rm UV}^{\rm obs}$ ($\Delta \rho_{{\rm Ly}\alpha}^{\rm obs}/\Delta t \lesssim \Delta \rho_{\rm UV}^{\rm obs}/\Delta t$) at $z=6.6$--7 seen in Table \ref{Density_Rate_Evolution} (i.e. $\rho_{{\rm Ly}\alpha}^{\rm LAE}$ and $\rho_{\rm UV}^{\rm LAE}$ knees are the same) while $\rho_{\rm UV}^{\rm LBG}$ begins to decrease at $z\sim8$. As the $\rho_{\rm UV}^{\rm LAE}$ knee and the $\rho_{\rm UV}^{\rm LBG}$ knee are different, LAEs and LBGs evolve in different way at $z \geq 6.6$ in terms of global luminosity or SFR density. Hence, as mentioned earlier, the decrease in $\rho_{{\rm Ly}\alpha}^{\rm LAE}$ (and also $n_{{\rm Ly}\alpha}^{\rm LAE}$) from $z=6.6$ to $z=7$ (and also possibly from $z=7$ to $z=7.3$) would be partly due to the evolution of LAE population (as implied by the decrease in $\rho_{\rm UV}^{\rm LAE}$ at $z=6.6$--7).
%This is because these two galaxy populations are quite different despite some degree of overlap as LAEs are galaxies with brighter Ly$\alpha$ emission and fainter UV continuum detected/selected by narrowband excess while LBGs fainter Ly$\alpha$ and brighter UV continuum by broadband dropout technique. 
%%%%
%As the $\Delta n_{\rm UV}^{\rm obs}/\Delta t$ and $\Delta \rho_{\rm UV}^{\rm obs}/\Delta t$ at $z=5.7$--6.6 are only $\sim -1.3$ and -1.3 Gyr$^{-1}$ and the errors in $n_{{\rm Ly}\alpha}^{\rm obs}$ and $\rho_{{\rm Ly}\alpha}^{\rm obs}$ at $z=5.7$ and 6.6 are large,    

\subsection{Implications for Galaxy Evolution and Cosmic Reionization \label{IforCR}}
In the previous sections, we have found that the Ly$\alpha$ LF and the Ly$\alpha$ luminosity density of LAEs decline from $z=5.7$ and 6.6 to 7. We have also found that the UV LF and the UV luminosity density of LAEs decline from $z=5.7$ and 6.6 to 7. Thus, we have considered that the declines of the Ly$\alpha$ LF and the Ly$\alpha$ luminosity density can be partly ascribed to galaxy evolution from $z=5.7$ and 6.6 to 7. Meanwhile, the previous Subaru Suprime-Cam LAE surveys at $z=6.6$, 7 and 7.3 suggested that the declines of the Ly$\alpha$ LF and the Ly$\alpha$ luminosity density can be attributed to both galaxy evolution and attenuation of Ly$\alpha$ emission of LAEs by neutral IGM whose fraction increases as redshift gets higher \citep{Iye06,Kashikawa06,Kashikawa11,Ota08,Ota10,Ouchi10,Shibuya12,Konno14}. Hence, in this section, we compare our Ly$\alpha$ and UV LFs and luminosity densities of the $z=7$ LAE candidates with galaxy evolution and reionization models to examine if the declines of the Ly$\alpha$ LF and the Ly$\alpha$ luminosity density at $z=7$ can be also partly ascribed to attenuation of Ly$\alpha$ by neutral IGM.  

%%figure 19
\begin{figure*}
\epsscale{1.17}
%\plottwo{Differential_z7LyaLF_KTN10model_Ota16.eps}{z7LyaLF_KTN10_XHI_TEST_Ota16-II.eps}
\plottwo{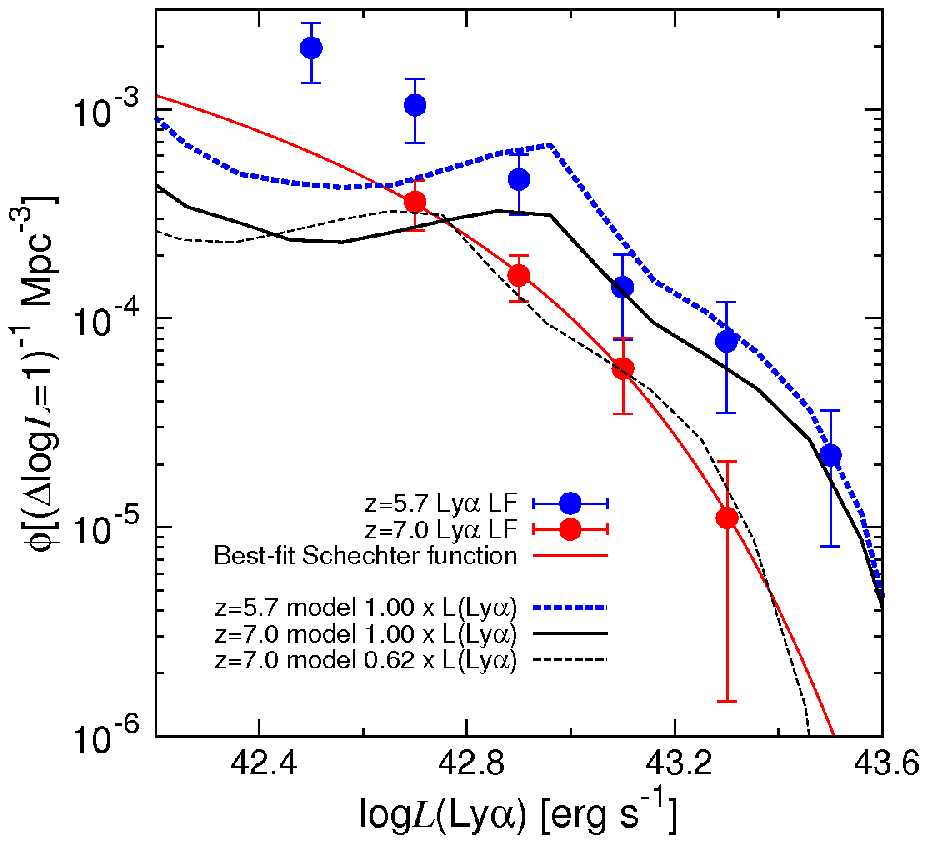}{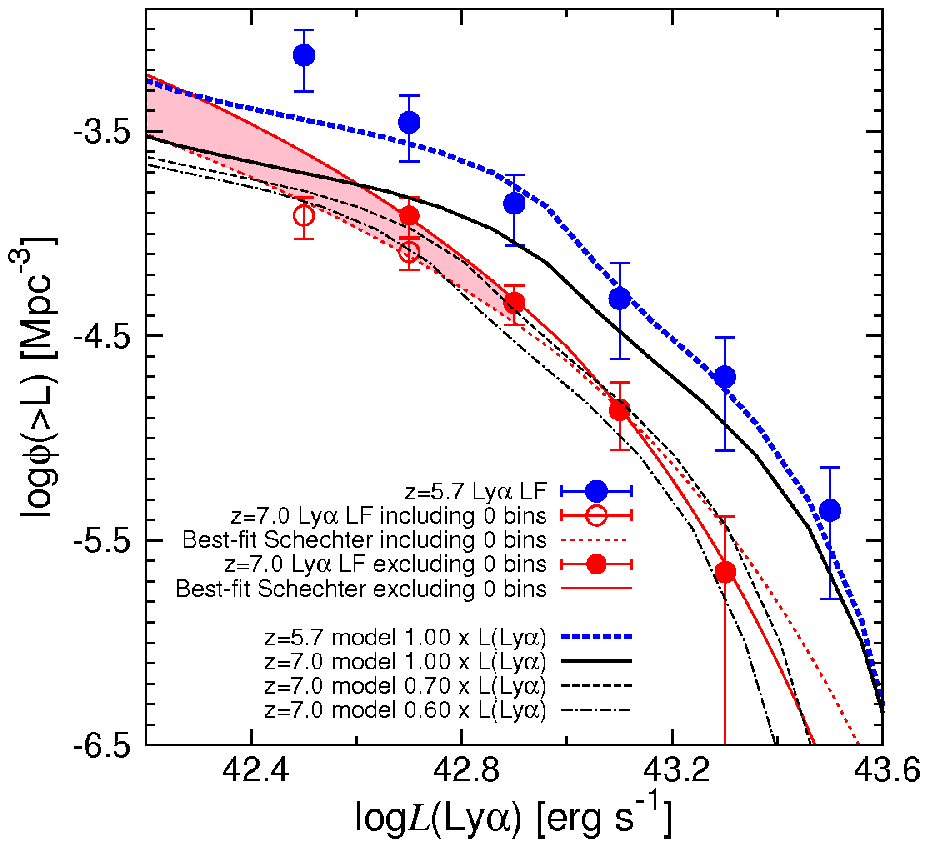}
\caption{(Left) Comparison of the observed differential $z=7$ Ly$\alpha$ LF (red filled circles) with the $z=5.7$ and $z=7$ Ly$\alpha$ LFs in the case of $T_{{\rm Ly}\alpha}^{\rm IGM}=1$ (or $x_{\rm HI}=0$) predicted by the KTN10 model (the thick blue dashed and the thick black solid lines, respectively). The $z=5.7$ KTN10 model Ly$\alpha$ LF is calibrated by the observed $z=5.7$ Ly$\alpha$ LF (blue filled circles) derived by \citet{Ouchi08}. The error bars include both Poisson errors and cosmic variance. The red solid line is the Schechter function best fitted to the observed $z=7$ Ly$\alpha$ LF. The thin black dashed line is the KTN10 $z=7$ Ly$\alpha$ LF with its Ly$\alpha$ luminosities attenuated by a factor of 0.62, which best fits the observed $z=7$ Ly$\alpha$ LF by $\chi^2$ minimization. (Right) The same comparison as the left panel but for the observed cumulative $z=7$ Ly$\alpha$ LFs excluding (including) the faintest bins where no LAE is detected and their best-fit Schechter function is shown by the filled (open) circles and the red solid (dotted) line, respectively. The pink shaded region shows the difference between the two $z=7$ Ly$\alpha$ LFs at their faint ends which can be considered the possible range of the $z=7$ Ly$\alpha$ LF at the faint end. The thin black dashed (dot-dashed) line is the KTN10 $z=7$ Ly$\alpha$ LF with its Ly$\alpha$ luminosities attenuated by a factor of 0.7 (0.6), which best fits the observed cumulative $z=7$ Ly$\alpha$ LF excluding (including) the faintest bins by $\chi^2$ minimization. The three types of observed $z=7$ Ly$\alpha$ LFs and their best-fit Schechter functions in the both panels here are taken from the three panels in Figure \ref{plottingLyaLF}. \label{KTN10_z7LyaLF}}
\end{figure*}
%The red solid (dotted) line is the Schechter function best fitted to the observed cumulative $z=7$ Ly$\alpha$ LFs excluding (including) the faintest bins.

\subsubsection{Comparison with Galaxy Evolution Models and Constraint on IGM Transmission for Ly$\alpha$ Photons\label{GalEvoModel}}
First, in Figure \ref{KTN10_z7LyaLF}, we compare our three types of $z=7$ Ly$\alpha$ LFs (one differential and two cumulative ones from the left, middle and right panels of Figure \ref{plottingLyaLF}) with the Ly$\alpha$ LFs theoretically predicted by using the \citet[][hereafter KTN10]{Kobayashi07,Kobayashi10} LAE evolution model. KTN10 constructed this LAE model by incorporating new modeling for an escape fraction of Ly$\alpha$ photons from galaxies into a recent hierarchical clustering model of \citet{Nagashima04}, physically considering dust extinction of Ly$\alpha$ photons and the effect of galaxy-scale outflows. The KTN10 model was empirically calibrated to fit the observed Ly$\alpha$ LFs of $z=5.7$ LAEs in SDF and SXDS derived by \citet{Shimasaku06} and \citet{Ouchi08} (see Figure 2 in the KTN10 paper; also see the observed $z=5.7$ Ly$\alpha$ LF and the one predicted by the KTN10 model in Figure \ref{KTN10_z7LyaLF}). It is worth noting that, with the consistent set of model parameters, the model naturally reproduces observed data of LAEs (i.e., Ly$\alpha$ LF, UV LF, and EW$_0$ distribution) in the redshift range of $z\sim3$--6 under the standard scenario of hierarchical galaxy formation.

In the left panel of Figure \ref{KTN10_z7LyaLF}, we plot our observed differential $z=7$ LAE Ly$\alpha$ LF (red circles), its best-fit Schechter function (red solid line), Ly$\alpha$ LFs of LAEs at $z=5.7$ and 7 expected in the case of the IGM transmission for Ly$\alpha$ photons $T_{{\rm Ly}\alpha}^{\rm IGM} = 1$ (or equivalently neutral fraction $x_{\rm HI} = 0$), which we calculated by using the KTN10 model (thick blue dashed and black solid lines, respectively). The KTN10 model $z=5.7$ Ly$\alpha$ LF is calibrated by the observed $z=5.7$ Ly$\alpha$ LF derived by \citet{Ouchi08}. Note that the KTN10 model apparently underpredicts the $z=5.7$ Ly$\alpha$ LF at the fainter luminosities $\log L({\rm Ly}\alpha) \lesssim 42.8$ by a factor of $\sim2$--3 (considering the errors of the observed LF). \citet{Ouchi08} estimated that the detection completeness in their faintest narrowband magnitude bin (NB816 $=$ 25.5--26.0 mag) is 50\%--60\% in their $z=5.7$ LAE survey. They also estimated that the contamination rate of their $z=5.7$ LAE photometric sample is 0\%--25\%. Thus, the fainter side of the observed $z=5.7$ Ly$\alpha$ LF has an uncertainty of up to a factor of $\sim 3$, and we cannot tell if the observed and predicted LFs are consistent at the faint luminosities. Therefore, we compare the observed and model Ly$\alpha$ LFs at the brighter luminosities $\log L({\rm Ly}\alpha) \gtrsim 42.8$.

We can see that the KTN10 model Ly$\alpha$ LF declines from $z=5.7$ to $z=7$ due to galaxy evolution. However, our observed $z=7$ Ly$\alpha$ LF is even lower than the model $z=7$ Ly$\alpha$ LF beyond the errors including Poisson errors and cosmic variance. As the model $z=7$ Ly$\alpha$ LF has already evolved from $z=5.7$, this discrepancy between the observed and predicted Ly$\alpha$ LFs cannot be explained by galaxy evolution. Although there are a lot of uncertainties in theoretical modeling of LAEs \citep{Cai14,Dayal12,Dijkstra12,Garel15,Jensen13,Jose13,Laursen13,Nagamine10,Orsi12,Shimizu11,Tilvi09,Yajima14}, in general it is theoretically unlikely that an unknown effect changes only Ly$\alpha$ luminosity compared with UV continuum luminosity in this short redshift range of $z=5.7$--7.0. Thus, it could be due to the Ly$\alpha$ attenuation by neutral IGM. Actually, if we attenuate the Ly$\alpha$ luminosity of the KTN10 model $z=7$ Ly$\alpha$ LF by a factor of 0.62 (that is, $T_{{\rm Ly}\alpha, z=7}^{\rm IGM}/T_{{\rm Ly}\alpha, z=5.7}^{\rm IGM} = 0.62$), it best fits the observed $z=7$ Ly$\alpha$ LF as shown by the thin black dashed line in the left panel of Figure \ref{KTN10_z7LyaLF} (we perform the fitting by $\chi^2$ minimization by treating the attenuation factor as a free parameter).

On the other hand, in the right panel of Figure \ref{KTN10_z7LyaLF}, we also plot our two observed cumulative $z=7$ LAE Ly$\alpha$ LFs excluding (including) the faintest bins where no LAE is detected (filled (open) circles), their best-fit Schechter functions (red solid (dotted) line), the $z=5.7$ and 7 KTN10 model Ly$\alpha$ LFs in the case of $T_{{\rm Ly}\alpha}^{\rm IGM} = 1$ (thick blue dashed and black solid lines, respectively). Again, the both observed $z=7$ LFs are lower than the $z=7$ KTN10 model Ly$\alpha$ LF. If we assume that this is due to the Ly$\alpha$ attenuation by neutral IGM and attenuate the Ly$\alpha$ luminosity of the model $z=7$ Ly$\alpha$ LF by a factor of $T_{{\rm Ly}\alpha, z=7}^{\rm IGM}/T_{{\rm Ly}\alpha, z=5.7}^{\rm IGM} = 0.70$ (0.60), it best fits the observed $z=7$ cumulative Ly$\alpha$ LFs excluding (including) the faintest bins as shown by the thin black dashed (dot-dashed) line in the right panel of Figure \ref{KTN10_z7LyaLF}.           

Meanwhile, based on their Subaru Suprime-Cam $z=6.6$ and $z=7.3$ LAE surveys and using a different method, \citet{Ouchi10} and \citet{Konno14} estimated the IGM transmission for Ly$\alpha$ photons at $z=6.6$ and $z=7.3$ relative to $z=5.7$ to be $T_{{\rm Ly}\alpha, z=6.6}^{\rm IGM}/T_{{\rm Ly}\alpha, z=5.7}^{\rm IGM} = 0.80 \pm 0.18$ and $T_{{\rm Ly}\alpha, z=7.3}^{\rm IGM}/T_{{\rm Ly}\alpha, z=5.7}^{\rm IGM} = 0.29$, respectively. To estimate these, they used the equation they derived,
\begin{equation}
\frac{T_{{\rm Ly}\alpha, z}^{\rm IGM}}{T_{{\rm Ly}\alpha, z=5.7}^{\rm IGM}} = \frac{\rho_{{\rm Ly}\alpha}^{{\rm tot}, z}/\rho_{{\rm Ly}\alpha}^{{\rm tot}, z=5.7}}{\rho_{\rm UV}^z/\rho_{\rm UV}^{z=5.7}}
\label{Eqn_TLya}
\end{equation}
by assuming that stellar population and escape fraction of Ly$\alpha$ photons through interstellar medium (galactic neutral hydrogen and dust) of LAEs do not evolve between $z=5.7$ and the redshift of interest $z$ \citep[see][for more details]{Ouchi10,Konno14}. The numerator is the ratio of the total Ly$\alpha$ luminosity densities of LAEs at $z=5.7$ and $z$. This represents the attenuation of Ly$\alpha$ emission of LAEs by neutral IGM at $z$ relative to $z=5.7$ where there is no such attenuation. The denominator is the ratio of the UV luminosity densities of LAEs at $z=5.7$ and $z$. This ratio corrects the IGM transmission of Ly$\alpha$ photons for the galaxy evolution factor that contributes to the Ly$\alpha$ luminosity density evolution between $z=5.7$ and $z$. 

For example, \citet{Konno14} used the $\rho_{\rm UV}^{z=7.3}/\rho_{\rm UV}^{z=5.7}$ of LBGs instead of that of LAEs as the UV LF of $z=7.3$ LAEs is not available. In our case, though limited to the UV luminosity of $M_{\rm UV} \leq -21$ mag, we have calculated the $\rho_{\rm UV}$'s of LAEs at $z=5.7$ and $z=7$ from their UV LFs and listed the ratios in Table \ref{Ratios_nUV_rhoUV}. We have $\rho_{\rm UV}^{z=7}/\rho_{\rm UV}^{z=5.7} = 0.57$ if we use $\rho_{\rm UV}^{z=5.7}$ calculated by using the \citet{Ouchi08} $z=5.7$ LAE UV LF. From Table \ref{n_rho_Lya}, we have $\rho_{{\rm Ly}\alpha}^{{\rm tot}, z=7}/\rho_{{\rm Ly}\alpha}^{{\rm tot}, z=5.7} = 0.38$ if we use the Ly$\alpha$ luminosity densities calculated by using the \citet{Ouchi08} $z=5.7$ LAE Ly$\alpha$ LF and our differential $z=7$ LAE Ly$\alpha$ LF. Hence, we obtain $T_{{\rm Ly}\alpha, z=7}^{\rm IGM}/T_{{\rm Ly}\alpha, z=5.7}^{\rm IGM} = 0.67$. Instead, we can also estimate $\rho_{\rm UV}^{z=7}/\rho_{\rm UV}^{z=5.7}$ of LBGs to be 0.56 from the interpolated line of the redshift evolution of $\rho_{\rm UV}^{\rm LBG}$ in Figure \ref{rhoLyaLAE_rhoUVLAE_rhoUVLBGs_vs_z}. This gives $T_{{\rm Ly}\alpha, z=7}^{\rm IGM}/T_{{\rm Ly}\alpha, z=5.7}^{\rm IGM} = 0.68$, almost same as the one estimated by using $\rho_{\rm UV}^{z=7}/\rho_{\rm UV}^{z=5.7}$ of LAEs. $T_{{\rm Ly}\alpha, z=7}^{\rm IGM}/T_{{\rm Ly}\alpha, z=5.7}^{\rm IGM} = 0.67$--0.68 estimated here are consistent with $T_{{\rm Ly}\alpha, z=7}^{\rm IGM}/T_{{\rm Ly}\alpha, z=5.7}^{\rm IGM} = 0.6$--0.7 estimated earlier by using the KTN10 model and our observed $z=7$ Ly$\alpha$ LFs. These values are comparable or lower than the IGM transmission of Ly$\alpha$ photons at $z=6.6$ and higher than that at $z=7.3$, which were estimated by \citet{Ouchi10} and \citet{Konno14}, respectively (see text above). Hence, the IGM transmission of Ly$\alpha$ photons becomes lower at higher redshifts at $z>5.7$.         

%%figure 20
\begin{figure*}
\epsscale{1.17}
\plottwo{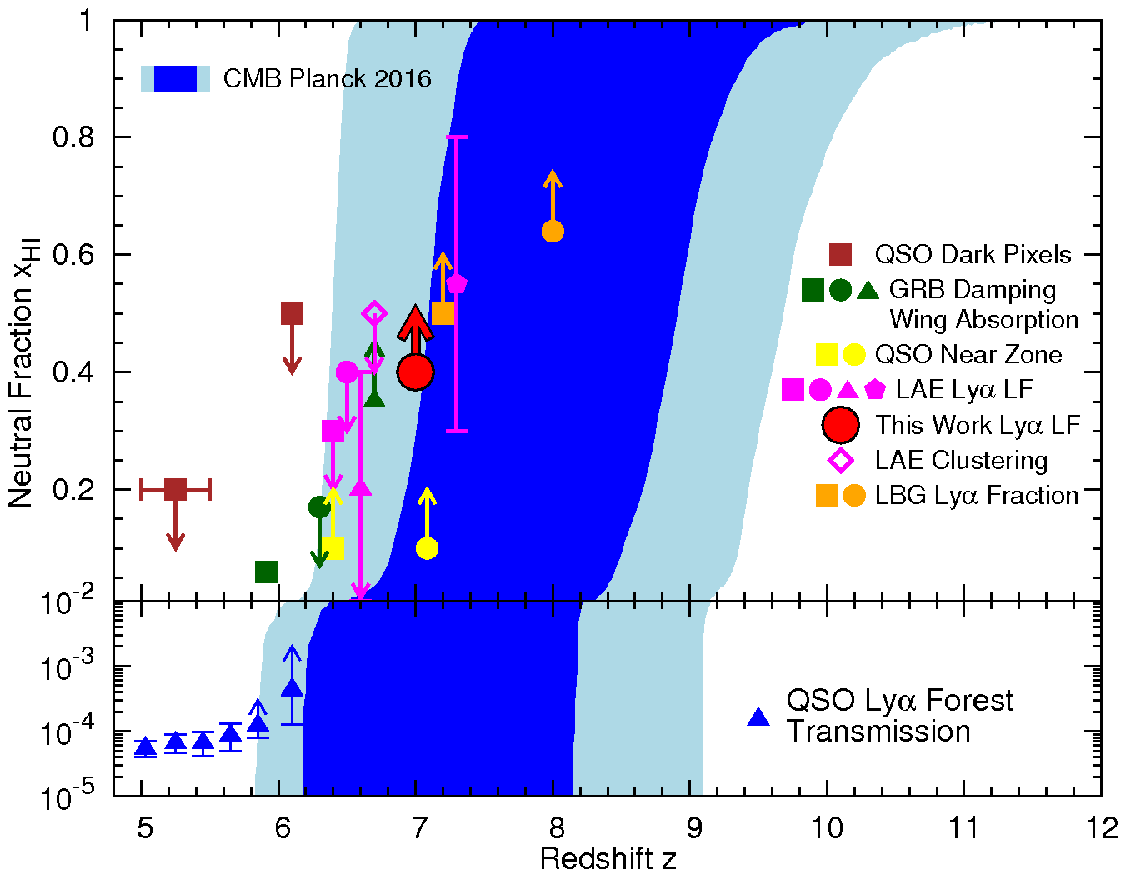}{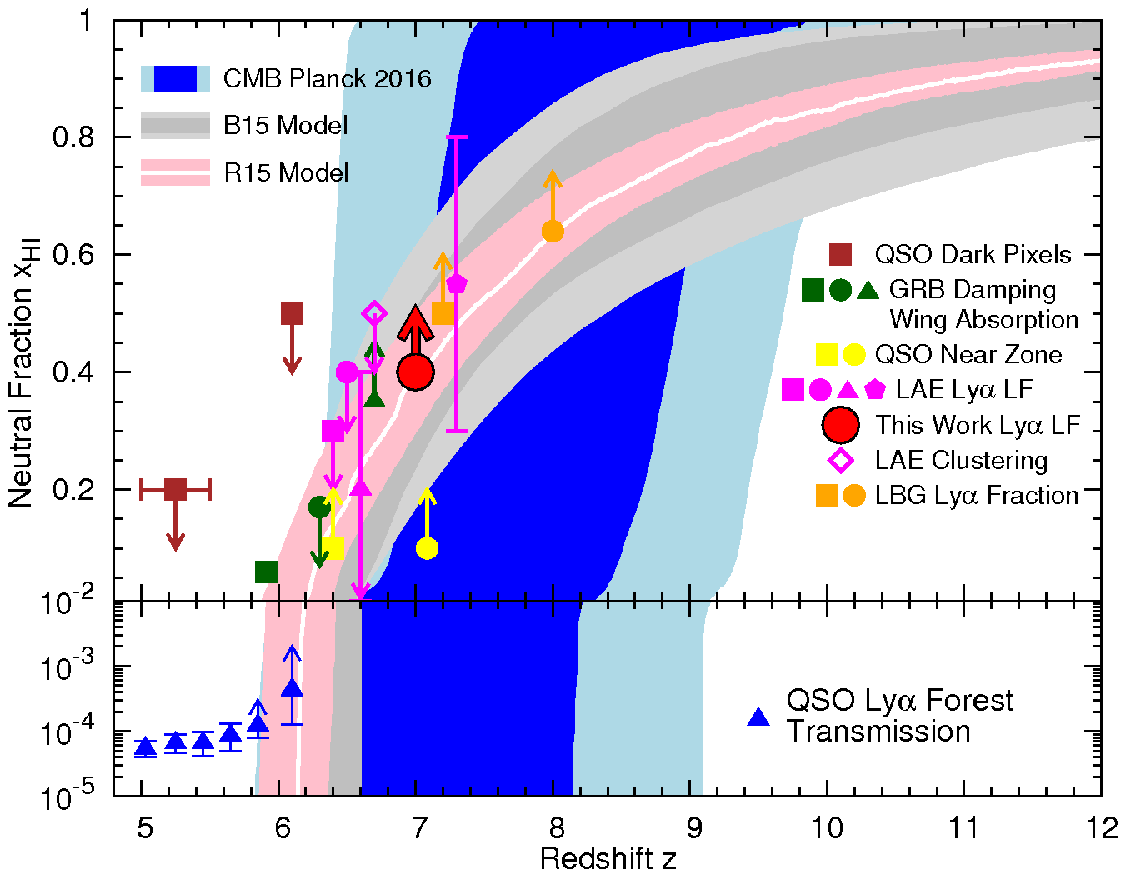}
\caption{(Left) Cosmic reionization history (neutral fraction $x_{\rm HI}$ as a function of redshift) constrained by our $z=7$ LAE survey and previous studies using various probes. The lower limit on $x_{\rm HI}$ at $z=7$ obtained by our study based on the Ly$\alpha$ LF is indicated by the large red filled circle. Meanwhile, the filled magenta square, circle, triangle and pentagon denote the neutral fractions estimated from the Ly$\alpha$ LFs at $z=6.6$ and 7.3 by \citet{Malhotra04}, \citet{Kashikawa11}, \citet{Ouchi10} and \citet{Konno14}, respectively. The three $z=6.6$ data points are slightly horizontally shifted from each other for clarity. The open diamond is the $x_{\rm HI}$ constraint from the analysis of the clustering of $z=6.6$ LAEs by \citet{Ouchi10}. The blue filled triangle shows $x_{\rm HI}$ constrained by QSO GP test by \citet{Fan06}. The brown filled square denotes the constraints from the QSO dark Ly$\alpha$ forest pixels obtained by \citet{McGreer11}. The green filled square, circle and triangle are the constraints based on GRB damping wing absorption by \citet{Totani16}, \citet{Totani06} and \citet{Greiner09}, respectively. The yellow filled square and circle indicate $x_{\rm HI}$ estimated from sizes of QSO near zones by \citet{Schroeder13} and \citet{Bolton11}, respectively. The orange filled square and circle are $x_{\rm HI}$ constrained from the fraction of Ly$\alpha$ emitting LBGs at $z\sim7$ \citep[the combined constraint from][]{Stark10,Pentericci11,Pentericci14,Schenker12,Schenker14,Ono12,Treu12,Caruana12,Caruana14,Furusawa16} and $z\sim8$ \citep{Schenker14}. The blue and light-blue shaded regions show the 68\% and 95\% allowed intervals of reionization history, respectively, constrained by the redshift-symmetric reionization model and the analysis of the {\it Planck} pre-2016 CMB observations data by \citet{Planck16b}. (Right) The same figure as the left panel that also plots reionization histories inferred from the contribution from star-forming galaxies derived by the \citet{Robertson15} (R15) model and the cosmic ionizing emissivity by the \citet{Bouwens15b} (B15) model. The white line and the pink shaded regions indicates the R15 model and its 68\% confidence interval. The gray and light-gray areas are the 68\% and 95\% confidence intervals of the B15 model.\label{ReionizationHistory}}
\end{figure*}

\subsubsection{Comparison with Reionization Models and Constraint on Neutral Fraction at $z=7$\label{ReionizationModels}}
Eventually, the observed decline of the $z=7$ Ly$\alpha$ LF or the Ly$\alpha$ luminosity density from $z=5.7$ can be only partially explained by galaxy evolution alone, and the Ly$\alpha$ attenuation by neutral IGM can explain the rest. This implies that the IGM neutral fraction significantly evolves between $z=5.7$ and $z=7$ corresponding to $T_{{\rm Ly}\alpha, z=7}^{\rm IGM}/T_{{\rm Ly}\alpha, z=5.7}^{\rm IGM} = 0.6$--0.7 estimated from the observed decline of the Ly$\alpha$ LF or the Ly$\alpha$ luminosity density at $z=7$. If we can translate this IGM transmission of Ly$\alpha$ photons into neutral IGM fraction $x_{\rm HI}$ at $z=7$ and compare it to $x_{\rm HI}$ at different redshifts estimated by previous studies, we can obtain implication for evolution of reionization state over cosmic time. However, note that this procedure is not simple and could be highly uncertain as the conversion depends on currently proposed models that quantify Ly$\alpha$ attenuation by neutral IGM. To facilitate comparison with the estimates of $x_{\rm HI}$ at $z=6.6$ and 7.3 obtained by the previous Subaru Suprime-Cam LAE surveys \citep{Kashikawa06,Kashikawa11,Ouchi10,Konno14}, we use the same and as many reionization models as possible used by these surveys \citep[i.e., models of][]{Santos04,Furlanetto06,Dijkstra07,McQuinn07}. 

We first use the \citet{Santos04} model that has been most frequently used by the previous LAE studies to convert $T_{{\rm Ly}\alpha}^{\rm IGM}$ to $x_{\rm HI}$. Their dynamical model assumes no velocity shift or a velocity shift of the Ly$\alpha$ line by 360 km s$^{-1}$ redward of the systemic velocity and provides $T_{{\rm Ly}\alpha}^{\rm IGM}$ as a function of $x_{\rm HI}$. We adopt the latter assumption, as recent studies revealed that there are shifts of a few hundreds km s$^{-1}$ between Ly$\alpha$ and systemic velocities in low redshift LAEs \citep[e.g.,][]{McLinden11,Hashimoto13,Shibuya14}. Applying $T_{{\rm Ly}\alpha, z=7}^{\rm IGM}/T_{{\rm Ly}\alpha, z=5.7}^{\rm IGM} \leq 0.6$--0.7 we estimated from our observed $z=7$ Ly$\alpha$ LF and the KTN10 model to the \citet{Santos04} model (Figure 25 in his paper), we obtain $x_{\rm HI}^{z=7} \gtrsim 0.3$--0.4. This is the lower limit as our observed $z=7$ Ly$\alpha$ LF is based on the photometric LAE candidates (except for the $z=6.96$ LAE IOK-1) and might include some contaminations. If it does and if we remove the contaminations, the actual $z=7$ Ly$\alpha$ LF will exhibit more deficit, implying lower $T_{{\rm Ly}\alpha, z=7}^{\rm IGM}/T_{{\rm Ly}\alpha, z=5.7}^{\rm IGM}$ and thus higher $x_{\rm HI}^{z=7}$.  

On the other hand, \citet{Ouchi10} and \citet{Konno14} used the \citet{Dijkstra07} model that predicted a radius of ionized bubbles $R_{\rm HII}$ at $z=6.5$ as a function of $T_{{\rm Ly}\alpha, z=6.5}^{\rm IGM}/T_{{\rm Ly}\alpha, z=5.7}^{\rm IGM}$ and the \citet{Furlanetto06} model that relates the characteristic radius of ionized bubbles at $z=6.5$ to ionized fraction $x_{\rm i}$ (i.e., $x_{\rm HI} = 1-x_{\rm i}$) in order to translate their estimates of $T_{{\rm Ly}\alpha, z}^{\rm IGM}/T_{{\rm Ly}\alpha, z=5.7}^{\rm IGM}$ into $x_{\rm HI}$ at $z=6.6$ and 7.3. The \citet{Dijkstra07} model provides their predictions with two cases where the ionizing background is (or is not) boosted by undetected surrounding sources. If we assume that the characteristic size of ionized bubbles does not change between $z=6.5$ and $z=7$ at a fixed $x_{\rm i}$ and also apply our estimate of $T_{{\rm Ly}\alpha, z=7}^{\rm IGM}/T_{{\rm Ly}\alpha, z=5.7}^{\rm IGM} \leq 0.6$--0.7 to the \citet{Dijkstra07} model (Figure 6 of their paper or Figure 20 of \citet{Ouchi10}), the typical radius of ionized bubbles at $z=7$ would be $R_{\rm HII} \lesssim 13$--24 comoving Mpc and $R_{\rm HII} \lesssim 24$--80 comoving Mpc for the boost and the non-boost cases, respectively. According to the \citet{Furlanetto06} model (the top panel of Figure 1 in their paper), these ionized bubble radii convert to $x_{\rm HI}^{z=7} \gtrsim 0.14$--0.22 and $x_{\rm HI}^{z=7} \gtrsim 0.04$--0.14, respectively. Here, we use their model with $z=6.5$ and the halo mass threshold corresponding to a virial temperature $10^4$ K where hydrogen line cooling becomes efficient.
   
Finally, another reionization model frequently used by the previous LAE studies is the \citet{McQuinn07} model that predicts the cumulative Ly$\alpha$ LFs in the cases of several different $x_{\rm i}$. Comparing our two types of cumulative $z=7$ Ly$\alpha$ LFs excluding (including) the faintest bins from the middle (right) panel of Figure \ref{plottingLyaLF} with the predicted Ly$\alpha$ LFs in Figure 4 of \citet{McQuinn07}, we obtain $x_{\rm HI}^{z=7} \geq 0.0$--0.38. 

Combining all the $x_{\rm HI}^{z=7}$ estimates above based on different reionization models, we conclude that the neutral IGM fraction at $z=7$ would be $x_{\rm HI}^{z=7} \gtrsim 0.4$. As mentioned earlier in Section 1, \cite{Zheng17} also conducted their $z=6.9$ LAE survey and constrained the neutral fraction at $z=6.9$ to be $x_{\rm HI}^{z=6.9} \sim 0.4$--0.6 in the similar way; i.e., comparing the $z=6.9$ Ly$\alpha$ LF and the decline of the Ly$\alpha$ luminosity density from $z=5.7$ to 6.9 with the same reionization models we used \citep{Santos04,Furlanetto06,Dijkstra07,McQuinn07}. As their $z=6.9$ Ly$\alpha$ LF and luminosity density are based on photometric LAE candidates, which could include some contaminations, their constraint on $x_{\rm HI}$ would be the lower limit and is consistent with ours. 

Figure \ref{ReionizationHistory} shows cosmic reionization history ($x_{\rm HI}$ as a function of redshift) obtained by combining our $x_{\rm HI}$ estimate at $z=7$ and constraints on $x_{\rm HI}$ at $z\sim5$--8 from previous studies of LAEs at $z=6.6$ and 7.3 (Ly$\alpha$ LF and clustering), QSOs (GP optical depth, dark pixels and near zone), GRB damping wing absorptions and LBG Ly$\alpha$ fractions. It suggests that $x_{\rm HI}$ increases rapidly from $\sim 10^{-4}$ to $\gtrsim 0.6$ at $z\sim6$--8. Figure \ref{ReionizationHistory} also overplots the most recent constraint on reionization history (the 68\% and 95\% confidence intervals) from the analysis of the Planck 2016 CMB observations intermediate results assuming the redshift-symmetric reionization model \citep{Planck16b}. Our $x_{\rm HI}$ estimate at $z=7$ together with the compilation of $x_{\rm HI}$'s at $z\sim6$--8 from previous studies are consistent with the reionization history inferred from the Planck 2016 result. Moreover, in the right panel of Figure \ref{ReionizationHistory}, we overlay the reionization histories inferred from the contribution from star-forming galaxies derived by the \citet{Robertson15} model and the cosmic ionizing emissivity by the \citet{Bouwens15b} model. Our constraint on $x_{\rm HI}$ at $z=7$ is also consistent with their models.

\subsubsection{Possible Impact on Sky Distribution of $z=7$ LAE Candidates by Reionization\label{SkyDist}}
In Section \ref{ReionizationModels}, we have constrained the neutral fraction at $z=7$ to be $x_{\rm HI}^{z=7} \gtrsim 0.4$. If the universe is really partly neutral at $z=7$, this may affect the visibility and sky distribution of LAEs. LAEs in ionized bubbles near ionizing sources would be more easily seen than those in the regions of locally higher neutral fraction. 

\begin{figure}
%\hspace*{-2.5cm}
\epsscale{1.3}
%\plotone{14z7LAEs_SDF_SkyDist_MaskRegions_4LBGs.eps}
%\plotone{14z7LAEs_SDF_SkyDist_MaskRegions_4LBGs_LyaLuminosityBin.eps}
%\plottwo{14z7LAEs_SDF_SkyDist_MaskRegions_4LBGs_LyaLuminosityBin_LabelUVbrightLAEs_RHII.eps}{6z7LAEs_SXDS_SkyDist_MaskRegions_LyaLumbin_LabelUVbrightLAEs_RHII.eps}
\plotone{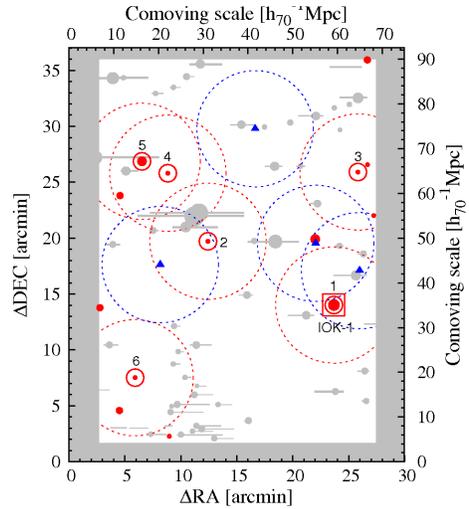}
\caption{Sky distribution of the 14 $z=7$ LAE candidates in SDF. They are indicated by the filled red circles whose sizes are proportional to the Ly$\alpha$ luminosity range corresponding to each bin of the $z=7$ Ly$\alpha$ LF in Figure \ref{LyaLFs_at_z5p7-7p3} (huge: 1 LAE in log$L$(Ly$\alpha$) $=$ 43.2--43.4, large: 2 LAEs in log$L$(Ly$\alpha$) $=$ 43.0--43.2, medium: 4 LAEs in log$L$(Ly$\alpha$) $=$ 42.8--43.0, small: 7 LAEs in log$L$(Ly$\alpha$) $=$ 42.6--42.8). We also encircle and number the six LAE candidates detected in the UV continuum in the order of decreasing UV continuum luminosity ($M_{\rm UV} \sim -21.6$ to $-19.44$ mag; also see Table \ref{Propertyz7LAECandidates}). These LAEs can be moderate ionizing sources. The one brightest in both Ly$\alpha$ and the UV continuum, NB973-SDF-85821, previously spectroscopically confirmed as a $z=6.96$ LAE, IOK-1, by \citet{Iye06}, \citet{Ota08} and \citet{Ono12} is marked with the red square. The four objects with a $y-{\rm NB973}<0$ color and an extremely faint or zero Ly$\alpha$ flux ($0\leq {\rm EW}_0 <10$\AA; likely considered $z\sim7$ LBGs with a very bright UV continuum $M_{\rm UV} \lesssim -21.8$ mag) finally removed from our $z=7$ LAE sample are also shown by the blue triangles (see Tables \ref{z7LAECandidates} and \ref{Propertyz7LAECandidates} and Section \ref{LyaFaintObjects} for their details). They can be stronger ionizing sources. We also plot the circles (dashed lines) of a radius $R_{\rm HII} = 13$ comoving Mpc (the most stringent constraint on the typical radius of ionized bubbles at $z=7$; see Section \ref{ReionizationModels}) around the UV-continuum-detected LAE candidates and the LBG candidates. The regions we masked when we selected the LAE candidates are shown by the shades. North is up and east to the left. The scale in the left and bottom axes are in arcmin while the right and top ones in comoving Mpc at $z=7$.\label{SDF_SkyDist}}
\end{figure}

\begin{figure}
%\hspace*{-2.5cm}
\epsscale{1.1}
\plotone{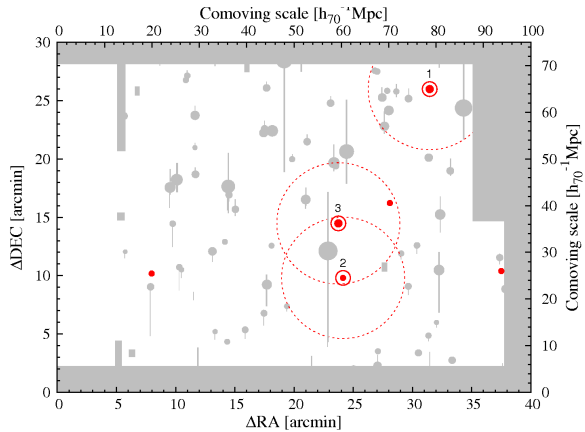}
\caption{The same as Figure \ref{SDF_SkyDist} but for sky distribution of the 6 $z=7$ LAE candidates in SXDS. Two are in the log$L$(Ly$\alpha$) $=$ 43.0--43.2 bin (large filled circles) and four in the log$L$(Ly$\alpha$) $=$ 42.8--43.0 bin (medium filled circles). Three of them are detected in the UV continuum, encircled and numbered in the order of decreasing UV continuum luminosity ($M_{\rm UV} \sim -21.05$ to $-20.15$ mag; also see Table \ref{Propertyz7LAECandidates}). They are also encircled with $R_{\rm HII} = 13$ circles (dashed lines). The shades are the masked regions. North is up and east to the left.\label{SXDS_SkyDist}}
\end{figure}

To examine this, we plot sky distributions of the 14 and 6 $z=7$ LAE candidates in SDF and SXDS in Figures \ref{SDF_SkyDist} and \ref{SXDS_SkyDist}, respectively. They are indicated by the filled red circles whose sizes are proportional to the Ly$\alpha$ luminosity range corresponding to each bin of the $z=7$ Ly$\alpha$ LFs in Figures \ref{plottingLyaLF} and \ref{LyaLFs_at_z5p7-7p3} (huge: 1 SDF LAE in log$L$(Ly$\alpha$) $=$ 43.2--43.4, large: 2 SDF and 2 SXDS LAEs in log$L$(Ly$\alpha$) $=$ 43.0--43.2, medium: 4 SDF and 4 SXDS LAEs in log$L$(Ly$\alpha$) $=$ 42.8--43.0, small: 7 SDF LAEs in log$L$(Ly$\alpha$) $=$ 42.6--42.8). We also encircle and number the six and three LAE candidates in SDF and SXDS detected in the UV continuum in the order of decreasing UV continuum luminosity ($M_{\rm UV} \sim -21.6$ to $-19.44$ mag and $-21.05$ to $-20.15$ mag for the LAEs in SDF and SXDS, respectively; also see Table \ref{Propertyz7LAECandidates}). These LAEs can be moderate ionizing sources. The brightest source NB973-SDF-85821 in SDF in the brightest Ly$\alpha$ LF bin log$L$(Ly$\alpha$) $=$ 43.2--43.4 (the largest filled red circle marked by a red square in Figure \ref{SDF_SkyDist}) is the $z=6.96$ LAE IOK-1. It is also the brightest in the UV continuum of all the UV-continuum-detected LAEs in SDF and SXDS. The four objects in SDF with a $y-{\rm NB973}<0$ color and an extremely faint or zero Ly$\alpha$ flux finally removed from our $z=7$ LAE sample are also shown by the blue triangles (see Tables \ref{z7LAECandidates} and \ref{Propertyz7LAECandidates} and Section \ref{LyaFaintObjects}). As discussed earlier, we consider them $z\sim7$ LBG candidates with a very bright UV continuum luminosity ($M_{\rm UV} \lesssim -21.8$ mag), and thus they can be stronger ionizing sources. 

In Section \ref{ReionizationModels}, we have estimated the typical radius of ionized bubbles at $z=7$ to be $R_{\rm HII} \lesssim 13$--24 (24--80) comoving Mpc by using the boost (non-boost) case of the \citet{Dijkstra07} reionization model. To see how many observed LAE candidates exist near the ionizing sources within the typical radius of ionized bubbles, we also plot the circles (dashed lines) of a radius $R_{\rm HII} = 13$ comoving Mpc (the most stringent constraint on $R_{\rm HII}$) around the UV-continuum-detected LAE candidates and the LBG candidates in Figures \ref{SDF_SkyDist} and \ref{SXDS_SkyDist}. Note that this radius $R_{\rm HII}$ is not a typical minimum size of ionized bubbles created by one isolated ionizing source. Here, we try to see if LAE(s) and ionizing source(s) are located within a typical ionized bubble of radius $R_{\rm HII}$. We see that most of the LAE candidates are either located within $R_{\rm HII}$ from any ionizing sources or are themselves moderate ionizing sources (i.e. UV-continuum-detected) that can ionize their surroundings. Some other LAEs (three in SDF and two in SXDS) are neither located at the distances within $R_{\rm HII}$ from any ionizing sources nor themselves moderate ionizing sources. However, most of them (four out of the five) are located close to the edges of the SDF and SXDS images, and thus we do not know whether they have neighboring ionizing sources within $R_{\rm HII}$. Also, if we slightly loosen the constraint on $R_{\rm HII}$, some of these LAEs are located at the distances within $R_{\rm HII}$ from the ionizing sources. Eventually, sky distributions of the LAEs and the ionizing sources imply that the LAEs near the ionizing sources within typical ionized bubbles and/or the UV-bright LAEs that are themselves moderate ionizing sources would be preferentially seen by our narrowband NB973 observations. However, it should be noted that we cannot rule out the possibility that the observed $z=7$ LAE sky distributions are just product of chance, either.
 
%%figure 21

%%figure 22

On the other hand, the sky distributions of the LAE and LBG candidates in SDF and SXDS as well as brightness in the UV continuum of the LAEs could explain the non-evolution of the Ly$\alpha$ LF between $z=6.6$ and $z=7$ at the bright end (the two brightest bins log$L$(Ly$\alpha$) $=$ 43.2--43.4 and 43.0--43.2) seen in Figure \ref{LyaLFs_at_z5p7-7p3}. The brightest Ly$\alpha$ LF bin includes only IOK-1 in SDF as seen in Figure \ref{SDF_SkyDist}. IOK-1 is detected in the UV continuum and has the brightest UV continuum luminosity of all the UV-continuum-detected LAE candidates. Hence, it may be able to effectively ionize its surroundings by itself, and this could reduce the neutral fraction locally, allow the higher transmission of its Ly$\alpha$ photons and make IOK-1 very bright in Ly$\alpha$ emission, too. Meanwhile, the second brightest Ly$\alpha$ LF bin contains two LAEs in SDF and two LAEs in SXDS as seen in Figures \ref{SDF_SkyDist} and \ref{SXDS_SkyDist}. The one in SDF and the two in SXDS are detected in the UV continuum and thus UV-bright galaxies that could ionize their surroundings by themselves. The remaining one LAE in SDF is not detected in the UV continuum and hence a UV-faint galaxy that might not be able to ionize its surroundings well. However, as seen in Figure \ref{SDF_SkyDist}, there is a $z\sim7$ LBG candidate very close (within $\lesssim 2$ comoving Mpc) to this LAE candidates. This LBG could ionize the neutral IGM around the LAE. All these situations may allow the locally higher transmission of Ly$\alpha$ photons from these four LAEs in SDF and SXDS, making them bright in Ly$\alpha$ emission. As a result, the two brightest Ly$\alpha$ LF bins include Ly$\alpha$-bright LAEs and could contribute to the apparent non-evolution of the Ly$\alpha$ LF between $z=6.6$ and $z=7$ at the bright end in possible conspiracy with the field-to-field variance of the number of LAEs and the difference in the method of estimating the narrowband detection completeness between our and previous LAE studies discussed earlier.

\vspace{0.5cm}
\section{Summary and Conclusion\label{Conclusion}}
We carried out very deep Subaru Suprime-Cam NB973 imaging of SDF and SXDS to conduct a census of $z=7$ LAEs to the Ly$\alpha$ luminosity limit as comparable as possible to the depths of the previous Subaru surveys of LAEs at $z=5.7$, 6.6 and 7.3 and to investigate the $z=7$ Ly$\alpha$ LF to its faint end. Our observations led to the Ly$\alpha$ flux limits of $f$(Ly$\alpha$)$_{\rm lim} = 3.4 \times 10^{-18}$ and $4.7 \times 10^{-18}$ erg s$^{-1}$ cm$^{-2}$ ($4\sigma$) or the Ly$\alpha$ luminosity limits of $L$(Ly$\alpha$)$_{\rm lim} = 2.0 \times 10^{42}$ and $2.7 \times 10^{42}$ erg s$^{-1}$ ($4\sigma$) for SDF and SXDS, respectively. These limits correspond to 0.22--0.36 $L^*_{z=7}$ (SDF) and 0.29--0.49 $L^*_{z=7}$ (SXDS) and probe the $z=7$ Ly$\alpha$ LF to its fainter end. The $L^*_{z=7}$ are derived by fitting the Schechter function to the $z=7$ Ly$\alpha$ LFs (see below). 

We also newly obtained fairly deep Suprime-Cam $y$-band imaging of SXDS and the very deep $y$-band image of SDF \citep[][and our observations]{Ouchi09,Tadaki12}. The $y$-band covers both $z=7$ Ly$\alpha$ emission and the rest frame UV continuum redward of it. Hence, we simulate the colors of $z=7$ LAEs including the $y$-band and established a new robust color selection criteria of them. Application of the criteria to our photometric catalog led to the detections of 14 and 6 $z=7$ LAE candidates (including one $z=6.96$ LAE previously spectroscopically confirmed) in SDF and SXDS, respectively. This is considerably smaller number than the 89 $z=5.7$ and 58 $z=6.6$ LAE photometric candidates previously detected in SDF to the comparable depths but 1.7 ($z=5.7)$ and 1.4 ($z=6.6$) times smaller survey volumes with Subaru Suprime-Cam \citep{Taniguchi05,Shimasaku06,Kashikawa06,Kashikawa11}. Using NB973 and $y$-band total magnitudes (or limits) of the $z=7$ LAE candidates, we estimate their Ly$\alpha$ fluxes, luminosities and EWs as well as UV continuum magnitudes, fluxes and luminosities. Then, we derive the Ly$\alpha$ and UV LFs and EW distribution of the $z=7$ LAE candidates. We summarize main results of our study below. 

\begin{enumerate}
\item We derive both differential and cumulative $z=7$ Ly$\alpha$ LFs to facilitate the comparison with the $z=5.7$, 6.6 and 7.3 Ly$\alpha$ LFs derived differentially or cumulatively by the previous Subaru LAE studies. We do not detect any $z=7$ LAE candidates at the Ly$\alpha$ luminosity ranges of $\log L({\rm Ly}\alpha)$ (erg s$^{-1}$) = 42.3--42.6 in SDF and 42.43--42.8 in SXDS although these luminosity ranges are close to but still within our survey limits. Though actual sensitivities could be somewhat shallower than our estimates, it is also possible that the very faint end of the $z=7$ Ly$\alpha$ LF might be suppressed as Ly$\alpha$ emissions of fainter LAEs are more preferentially attenuated by neutral IGM as suggested by \citet{Matthee15}. However, we cannot distinguish between these two possibilities from the current data alone. Hence, we derive the two different cumulative Ly$\alpha$ LFs of $z=7$ LAEs: (1) the LF excluding the faintest Ly$\alpha$ luminosity bins within our survey limits where no LAE candidate is detected and (2) the LF including these faintest bins. 
%They are presented in the middle and the right panels of Figure \ref{plottingLyaLF}, respectively.  
 
\item We compare the differential Ly$\alpha$ LF of our $z=7$ LAE candidates to those of LAEs at $z=5.7$, 6.6 and 7.3 in SXDS and COSMOS fields mostly based on photometric LAE candidates derived from the previous Subaru Suprime-Cam LAE surveys conducted by \citet{Ouchi08}, \citet{Ouchi10} and \citet{Konno14}. The $z=7$ Ly$\alpha$ LF exhibits a significant deficit from the $z=5.7$ one from the bright to faint end. Also, the $z=7$ Ly$\alpha$ LF shows a significant deficit from the $z=6.6$ one at the fainter end $\log L({\rm Ly}\alpha)$ (erg s$^{-1}$) $<43.0$ and the both LFs are almost same at the bright end $\log L({\rm Ly}\alpha)$ (erg s$^{-1}$) $>43.0$. Moreover, the $z=7.3$ Ly$\alpha$ LF shows a significant deficit from our $z=7$ Ly$\alpha$ LF from the bright to faint end.  

\item We also compare the two cumulative Ly$\alpha$ LFs of our $z=7$ LAE candidates to those of LAEs at $z=5.7$ and 6.6 in SDF mostly based on spectroscopically confirmed LAEs derived by the \citet{Kashikawa11}. Their LAEs were also originally detected by using Subaru Suprime-Cam. The both cumulative $z=7$ Ly$\alpha$ LFs show significant deficits from the $z=5.7$ LF from bright to faint end. On the other hand, the cumulative $z=7$ Ly$\alpha$ LF excluding the faintest bins does not exhibit any deficit from the $z=6.6$ LF. However, the cumulative $z=7$ Ly$\alpha$ LF including the faintest bins shows a deficit from the $z=6.6$ LF at the faint end $\log L({\rm Ly}\alpha)$ (erg s$^{-1}$) $<42.7$.

\item We fit the Schechter function to our $z=7$ LAE Ly$\alpha$ LFs fixing the slope to $\alpha=-1.5$ (to facilitate the comparison with previous LAE studies) and derive their characteristic luminosities and number densities, $L^*$ and $\phi^*$. Also, we fit the Schechter functions with their $L^*$ or $\phi^*$ fixed to those of the $z=5.7$ or 6.6 Ly$\alpha$ LF to our $z=7$ LAE Ly$\alpha$ LFs in order to see which of the luminosity evolution or the number evolution is more dominant factor in the evolution of the Ly$\alpha$ LF at $z=5.7$--7 and $z=6.6$--7. We find that the number evolution is more dominant. Combining these results with the $L^*$ and $\phi^*$ evolutions at $z=5.7$--6.6 and $z=7$--7.3 derived from similar Schechter function fittings, we estimate the rates of decrease in $L^*$ and $\phi^*$ ($\Delta L^*/\Delta t$ and $\Delta \phi^*/\Delta t$) at the redshift ranges of $z=5.7$--6.6 ($\Delta t=0.16$ Gyr), 6.6--7 ($\Delta t=0.06$ Gyr) and 7--7.3 ($\Delta t=0.04$ Gyr) in the cases of the pure luminosity and pure number evolutions. We find that $L^*$ and $\phi^*$ decrease acceleratingly as redshift increases.

\item Meanwhile, we compare the UV LF of our $z=7$ LAE candidates in SDF with those of the $z=5.7$ and 6.6 LAEs in SDF derived by the same method by \citet{Kashikawa11}. \citet{Kashikawa11} already found that the UV LF does not evolve much between $z=5.7$ and 6.6. We newly find that the $z=7$ LAE UV LF shows a deficit from $z=5.7$ and 6.6 ones although these three LFs are consistent within statistical errors and cosmic variance. This implies that LAEs evolve between $z=6.6$ and 7, and this partially contributes to the decline of the Ly$\alpha$ LF from $z=6.6$ to 7.     

\item We also compare the rest frame Ly$\alpha$ EW (EW$_0$) distribution of our $z=7$ LAE candidates in SDF with those of the $z=5.7$ and 6.6 LAEs in SDF derived by the same method by \citet{Kashikawa11}. \citet{Kashikawa11} already found that the EW$_0$'s of the $z=6.6$ LAEs are systematically lower than those of the $z=5.7$ LAEs. We further find that two thirds of our $z=7$ LAE candidates detected in the UV continuum exhibit EW$_0$'s lower than those of the $z=6.6$ LAEs. This implies that Ly$\alpha$ emission of LAEs could be more strongly suppressed at $z=7$ than $z=6.6$ by possibly higher fraction of neutral IGM at $z=7$.   

\item Furthermore, we combine the number, Ly$\alpha$ luminosity, UV luminosity densities ($n_{{\rm Ly}\alpha}$, $n_{\rm UV}$, $\rho_{{\rm Ly}\alpha}$ and $\rho_{\rm UV}$) of our $z=7$ LAEs with those of the $z=3.1$, 3.7, 4.5, 5.7, 6.6 and 7.3 LAEs calculated by integrating the Ly$\alpha$ and UV LFs to $\log L({\rm Ly}\alpha)$ (erg s$^{-1}$) $=$ 42.4 and $M_{\rm UV}=-21$ mag to trace the redshift evolution of these densities which could reflect the change in neutral IGM fraction with redshift. We find that the $n_{{\rm Ly}\alpha}$ and $\rho_{{\rm Ly}\alpha}$ do not change much at $z=3.1$--5.7 but modestly decrease at $z=5.7$--6.6, slightly more rapidly decrease at $z=6.6$--7 and more rapidly decrease at $z=7$--7.3. Meanwhile, the $n_{\rm UV}$ and $\rho_{\rm UV}$ increase at $z=3.1$--5.7, very modestly increase or stay constant at $z=5.7$--6.6 and decrease at $z=6.6$--7. Moreover, we estimate the rates of decrease in the densities ($\Delta n_{{\rm Ly}\alpha}/\Delta t$, $\Delta n_{\rm UV}/\Delta t$, $\Delta \rho_{{\rm Ly}\alpha}/\Delta t$ and $\Delta \rho_{\rm UV}/\Delta t$) at the redshift ranges $z=5.7$--6.6, 6.6--7, and 7--7.3. We find that $\Delta n_{{\rm Ly}\alpha}/\Delta t$ and $\Delta \rho_{{\rm Ly}\alpha}/\Delta t$ get higher as redshift increases, implying accelerating decrease in $n_{{\rm Ly}\alpha}$ and $\rho_{{\rm Ly}\alpha}$ at $z>5.7$. Also, $\Delta n_{\rm UV}/\Delta t$ and $\Delta \rho_{\rm UV}/\Delta t$ are close to zero at $z=5.7$--6.6 but comparable to or even higher than $\Delta n_{{\rm Ly}\alpha}/\Delta t$ and $\Delta \rho_{{\rm Ly}\alpha}/\Delta t$ at $z=6.6$--7. All these results imply that neutral IGM fraction could acceleratingly get higher at $z>5.7$ and suppress Ly$\alpha$ emission of LAEs while LAEs evolve between $z=6.6$ and 7, partially contributing to the decrease in $n_{{\rm Ly}\alpha}$ and $\rho_{{\rm Ly}\alpha}$ from $z=6.6$ to 7.    

\item We also compare our three types of observed $z=7$ LAE Ly$\alpha$ LFs (one differential LF and two cumulative LFs; see above) with the $z=5.7$ and $z=7$ Ly$\alpha$ LFs in the case of $x_{\rm HI}=0$ theoretically predicted by the KTN10 LAE evolution model. Although the KTN10 $z=7$ Ly$\alpha$ LF already includes effect of the LAE evolution from the KTN10 $z=5.7$ Ly$\alpha$ LF matched to the observed $z=5.7$ Ly$\alpha$ LF, our observed $z=7$ Ly$\alpha$ LFs show significant deficits from the KTN10 $z=7$ Ly$\alpha$ LF. These discrepancies can be reconciled if we attenuate the Ly$\alpha$ luminosities of the KTN10 $z=7$ Ly$\alpha$ LF by a factor of 0.6--0.7. This factor corresponds to the upper limit on the IGM transmission of Ly$\alpha$ photons at $z=7$, $T_{{\rm Ly}\alpha, z=7}^{\rm IGM}/T_{{\rm Ly}\alpha, z=5.7}^{\rm IGM} \leq 0.6$--0.7 as our $z=7$ Ly$\alpha$ LFs are based on photometric LAE candidates that may include some contaminations. We also independently estimate this factor to be 0.67 by using the $\rho_{{\rm Ly}\alpha}$'s and the $\rho_{\rm UV}$'s of LAEs at $z=7$ and 5.7 and the equation (\ref{Eqn_TLya}). We convert $T_{{\rm Ly}\alpha, z=7}^{\rm IGM}/T_{{\rm Ly}\alpha, z=5.7}^{\rm IGM}$ into the neutral IGM fraction at $z=7$, $x_{\rm HI}^{z=7}$ by using several different reionization models. We obtain $x_{\rm HI}^{z=7} \gtrsim 0.4$, suggesting that reionization was not complete at $z=7$. Our result combined with estimates of $x_{\rm HI}$ from previous studies of LAEs at other epochs, QSOs, GRBs and LBG Ly$\alpha$ fraction also suggests that $x_{\rm HI}$ increases rapidly from $\sim 10^{-4}$ to $\gtrsim 0.6$ at $z\sim6$--8. This is consistent with the most recent constraint on reionization history from the Planck 2016 CMB observations intermediate results as well as the contribution from star-forming galaxies derived by the \citet{Robertson15} model and the cosmic ionizing emissivity by the \citet{Bouwens15b} model.
  
\item Finally, we examine the sky distribution of our $z=7$ LAE candidates in SDF and SXDS. We find that LAEs near ionizing sources (either UV-bright LAEs or LBGs) within typical ionized bubbles ($R_{\rm HII} < 13$ comoving Mpc at $z=7$) and/or the UV-bright LAEs that are themselves moderate ionizing sources would be preferentially seen by our narrowband NB973 observations. However, we cannot rule out the possibility that the sky distribution is just the product of chance. Meanwhile, the sky distributions of the LAE and LBG candidates in SDF and SXDS as well as brightness in the UV continuum of the LAEs could also explain the non-evolution of the Ly$\alpha$ LF between $z=6.6$ and $z=7$ at the bright end (the two brightest bins log$L$(Ly$\alpha$) $=$ 43.2--43.4 and 43.0--43.2). These bins contain four UV-bright LAEs that could effectively ionize their surroundings by themselves and one UV-faint LAE that might not be able to ionize its surroundings well but has a UV-bright LBG candidate as an immediate neighbor that could instead ionize the LAE's surroundings. These situations could allow the locally higher transmission of Ly$\alpha$ photons from these five LAEs, making them bright enough in Ly$\alpha$ emission to form the brightest $z=7$ Ly$\alpha$ LF bins that apparently show no evolution from $z=6.6$. Again, the possible field-to-field variance of the number of LAEs and the difference in the method of estimating the narrowband detection completeness between our and previous LAE studies could also partly contribute to this non-evolution of the Ly$\alpha$ LF at the bright end.
\end{enumerate}

The present study is based on our $z=7$ Ly$\alpha$ LF consisting of the photometric $z=7$ LAE candidates (except for one $z=6.96$ LAE spectroscopically confirmed). If the candidates include some contaminations, the actual deficit in the $z=7$ Ly$\alpha$ LF from $z=5.7$ and 6.6 are more significant and the estimated neutral IGM fraction at $z=7$ will be higher. To confirm this, the follow-up spectroscopy of all the $z=7$ LAE candidates is essential as a next step. Also, this study is the last one of the series of the Subaru Suprime-Cam narrowband LAE surveys in SDF and SXDS. These LAE surveys have successfully revealed galaxy evolution and cosmic reionization at $z=5.7$--7.3 taking full advantage of the large field-of-view (FoV) and the fully depleted red-sensitive CCDs of the Suprime-Cam. Currently, the strategic survey program of the Hyper Suprime-Cam (HSC), which has a seven times larger FoV, is now in progress by investing 300 nights of Subaru time for five years. These observations include the narrowband imaging targeting $z=5.7$--7.3 LAEs and will significantly improve the statistics of the LAE studies. This will enable to obtain more robust constraints on the reionization state.


%% Included in this acknowledgments section are examples of the
%% AASTeX hypertext markup commands. Use \url without the optional [HREF]
%% argument when you want to print the url directly in the text. Otherwise,
%% use either \url or \anchor, with the HREF as the first argument and the
%% text to be printed in the second.

\acknowledgments
We are grateful to the staff at the Subaru Telescope for their great support during our observations. We thank Masami Ouchi for providing his Ly$\alpha$ LF data and SDF $y$-band image and for useful discussion, Tomoe Takeuchi and Hiroko Tamazawa for helping us conduct parts of our observations, Tomoki Morokuma and Linhua Jiang for providing the deep $R$, $i'$ and $z'$ images of the SDF, Tadayuki Kodama and Ken-ichi Tadaki for providing the information about their $y$-band imaging observations of SXDS. We also thank our referee for the useful comments and suggestions that helped us improve this paper. This research has benefitted from the SpeX Prism Spectral Libraries, maintained by Adam Burgasser at http://pono.ucsd.edu/{\textasciitilde}adam/browndwarfs/spexprism. IRAF is distributed by the National Optical Astronomy Observatory, which is operated by the Association of Universities for Research in Astronomy (AURA) under a cooperative agreement with the National Science Foundation. K.O. acknowledges the Kavli Institute Fellowship at the Kavli Institute for Cosmology in the University of Cambridge supported by the Kavli Foundation. The long term appointment by this fellowship allowed him to complete this time-consuming project. N.K. acknowledges supports from the JSPS grant 15H03645. The authors recognize and acknowledge the very significant cultural role and reverence that the summit of Mauna Kea has always had within the indigenous Hawaiian community. We are most fortunate to have the opportunity to conduct observations from this mountain.

%We thank XX, XX and XX for their comments on the paper. We thank our referee for useful comments that helped us to improve this paper.
%This work is based in part on data collected at the Subaru Telescope, which is operated by the National Astronomical Observatory of Japan (NAOJ).
%X.X. was supported with a fellowship from the Japan Society for the Promotion of Science. 


%% in parentheses, after the keyword, but they will not be verified.

{\it Facilities:} \facility{Subaru (Suprime-Cam)}

\end{document}